\documentclass[twocolumn]{aastex63}
\usepackage{amsmath}
\newcommand{\ba}{\begin{eqnarray}}
\newcommand{\ea}{\end{eqnarray}}

\begin{document}
% Title
\title{Chemistry and Isotope Ratios of Substellar Atmospheres in the $\beta$ Pictoris Young Moving Group and Vicinity}

\author[0009-0007-9211-2884]{Yurou Liu} 
\affiliation{Department of Astronomy, Yale University, New Haven, CT 06511, USA}

\author[0000-0003-0097-4414]{Yapeng Zhang}
\altaffiliation{51 Pegasi b Fellow}
\affiliation{Department of Astronomy, California Institute of Technology, Pasadena, CA 91125, USA}

\author[0000-0002-6618-1137]{Jerry W. Xuan}
\altaffiliation{51 Pegasi b Fellow}
\affiliation{Department of Astronomy, California Institute of Technology, Pasadena, CA 91125, USA}
\affiliation{Department of Earth, Planetary, and Space Sciences, University of California, Los Angeles, CA 90095, USA}

\author[0000-0002-8895-4735]{Dimitri Mawet}
\affiliation{Department of Astronomy, California Institute of Technology, Pasadena, CA 91125, USA}

\author[0000-0003-1624-3667]{Ignas Snellen}
\affiliation{Leiden Observatory, Leiden University, Postbus 9513, 2300 RA Leiden, The Netherlands}

\author[0000-0002-7261-8083]{Rico Landman}
\affiliation{Leiden Observatory, Leiden University, Postbus 9513, 2300 RA Leiden, The Netherlands}

\author[0000-0002-5823-3072]{Tomas Stolker}
\affiliation{Leiden Observatory, Leiden University, Postbus 9513, 2300 RA Leiden, The Netherlands}

\author[0000-0003-4760-6168]{Sam de Regt}
\affiliation{Leiden Observatory, Leiden University, Postbus 9513, 2300 RA Leiden, The Netherlands}

\author[0000-0002-3239-5989]{Aurora Kesseli}
\affiliation{NASA Exoplanet Science Institute, IPAC, MS100-22, Caltech, 1200 E. California Blvd., Pasadena, CA 91125, USA}

\author[0000-0002-7670-670X]{Malena Rice}
\affiliation{Department of Astronomy, Yale University, New Haven, CT 06511, USA}

%Add authors

\begin{abstract}
  Measuring the chemical and isotopic compositions of gas giants and brown dwarfs provides insights into their formation pathways and birth environments. 2MASS J0249-0557 c is an L2-type planetary mass companion ($\sim 12 M_{\mathrm{Jup}}$) orbiting a pair of brown dwarfs in the $\beta$ Pic young moving group and vicinity \footnote{While \cite{Shkolnik2017bpmg} and \cite{Dupuy2018hawaii} have attributed 2MASS J0249-0557AB to the $\beta$ Pic young moving group, work by \cite{Luhman2024bpmg} attributed 2MASS J0249-0557AB to the 32 Ori young moving group, which spatially overlaps with the $\beta$ Pic young moving group. The two groups have been proposed to be physically connected \citep{Gagnne2021movinggroup}.}. Its mass places it at the intersection of planets and brown dwarfs, making it an interesting target for constraining formation pathways at the planet-brown-dwarf boundary. Using high-resolution spectroscopic data of the planet acquired with CRIRES+ mounted on VLT, we conduct atmospheric retrieval with the radiative transfer code \texttt{petitRADTRANS} and the nested sampling tool PyMultiNest. We retrieve a C/O ratio of $0.57\pm0.01$, a metallicity of [M/H] = $0.18\pm0.05$, and a $^{12}$CO/$^{13}$CO ratio of $95^{+23}_{-17}$. We also retrieve atmospheric compositions for two benchmark brown dwarfs in the $\beta$ Pic YMG, 2MASSI J0443+0002 and SIPS J2000-7523, using CRIRES+ data and find consistent compositions. Together with 2MASS J0249-0557 c's wide separation from its host, its compositional consistency with benchmark brown dwarfs supports gravitational collapse in a star-like manner as its most likely formation mechanism. These results deliver a homogeneous comparison of three substellar members in the $\beta$ Pic YMG and vicinity. Their solar-like abundances provide a baseline for exoplanet members in the same moving group, such as $\beta$ Pic b, 51 Eri b, and AF Lep b, whose host stellar compositions are difficult to measure. Future comparisons of atmospheric compositions among this moving group offer the potential to distinguish between formation mechanisms for its planetary members.

\end{abstract}

\section{Introduction}

The chemical composition of gas giant atmospheres can provide insight into their formation mechanisms and evolutionary histories. Fundamental questions about gas giant formation remain unanswered. Do they form via the bottom-up core accretion process \citep{Pollack1996accretion, Lambrechts2012accretion} or top-down gravitational collapse \citep{Boss1997instability, Kratter2016instability}? What determines which evolutionary pathways dominate for gas giants across mass and orbital separation regimes? 

One way to investigate the formation pathways of gas giants is by analyzing their atmospheric composition through direct spectroscopy. Different chemical compositions are suggestive of different formation scenarios, with some degeneracies. For example, a C/O ratio similar to that of the host star could indicate formation through gravitational collapse or enhanced solid accretion beyond the CO snowline, whereas a  superstellar C/O and C/H indicate either gas accretion close to the CO or CO$_2$ snowlines or
substantially polluted from carbon-grains  \citep{Oberg2011CO}. Additionally, a low ${}^{12}$C/${}^{13}$C ratio has been suggested to be an indicator of ice-accretion outside of the snowline, whereas a stellar ${}^{12}$C/${}^{13}$C ratio is indicative of formation inside the CO snowline or gravitational collapse \citep{Zhang2021C13}. 

We can obtain direct spectroscopy for self-luminous, young giant planets at large orbital distances \citep{Madhusudhan2019atmospheres, Snellen2025review}. Using ground-based telescopes, high spectral resolution and high signal-to-noise (S/N) spectroscopy of gas giant planets can advance our understanding of giant planet formation.
Previous high-resolution spectroscopy studies have yielded significant progress in characterizing the atmospheres of directly imaged substellar objects. The Keck Planet Imager and Characterizer (KPIC) survey has been able to detect molecular species such as CO and H${}_2$O; measure properties such as radial velocities and spins; retrieve C/O ratio and isotopic abundance ratios; detect clouds; and search for exomoons around notable targets such as the planets in the HR 8799 system, PDS 70b, HR 7672 B, HD 4747 B, and HIP 55507 B. \citep[e.g.,][]{Mawet2016kpic, Wang2021hr8799, Wang2022hr7672, Xuan2022cloud, Ruffio2023exomoon, Xuan2024isotope, Hsu2024pds70b, Wang2026_cd35}. There have been complementary efforts in the Southern Hemisphere with the ESO SupJup survey \citep{Regt2024eso1}. This survey has likewise detected molecular species and clouds, and measured radial and rotational velocities as well as chemical abundances for targets such as DENIS J0255-4700, the YSES 1 system, GQ Lup B, AB Pictoris b, and the Luhman 16 system \citep{Regt2024eso1, Gonalezz2024eso2, Zhang2024yses1, Gonalezz2025eso4, Gandhi2025eso5, Mulder2025eso6, Regt2025eso7, Grasser2025eso8}.

However, these studies are often limited by sample heterogeneity. The targets span a wide range of ages and formation environments, making it difficult to isolate intrinsic trends from environmental or evolutionary effects. Moreover, for many planetary-mass companions, the chemical abundances of their host stars remain poorly constrained or entirely unknown. The absence of reliable baseline abundances hinders our ability to constrain formation pathways through direct comparison of planetary atmospheric properties such as C/O and ${}^{12}$C/${}^{13}$C ratios with their natal environments.

Studying multiple substellar objects within the same Young Moving Group (YMG) using the same instrument helps mitigate the limitations described above. Members of a YMG share a common age and formation environment. By comparing these homogeneous members, we can robustly infer the formation pathway of these objects. In particular, isolated substellar members of the YMG, which are unaffected by disk interactions and stellar processes, offer a direct window into the natal composition of the group and can be used to establish a baseline abundance.

The $\beta$ Pictoris YMG is one of the most well-studied moving groups. It comprises several hundred candidate members and has a well-constrained age of $18.5_{-2.4}^{+2.0}$ Myr \citep{Miret2020betapic}. Owing to its proximity of $\sim$50 pc and young age, the $\beta$ Pic YMG provides many ideal targets for direct imaging studies of substellar objects.
Notable planetary-mass members with well-studied atmospheres include $\beta$ Pic b, 51 Eri b, and AF Lep b \citep[e.g.,][]{Collaboration2020bpicb, Brown2023Erib, DeRosa2023aflepb, Whiteford2023Erib, Zhang2023cloud, Franson2024aflepb, Kammerer2024bpicb, Landman2024bpicb, Palma2024aflepb, Parker2024bpicb, Reggiani2024betapic, Worthen2024bpicb, Balmer2025aflefb, Balmer2025Erib, Bonse2025aflepb, Denis2025aflepb, Hayoz2025aflepb, Janson2025bpicb}. However, directly comparing the atmospheric composition of these planets with their host stars is difficult because the host stars are typically fast-rotating or chemically peculiar, making it difficult to constrain their abundances \citep[e.g.,][]{Maldonado2022starrvaflep, Bowler2023starrv51eri, Borisov2023starrvbpic}. This analysis is particularly challenging for the star $\beta$ Pic, as its current abundances poorly reflect its natal environment due to its classification as an A-type star \citep{Reggiani2024betapic}. 

To address this issue, we present an initial chemical abundance baseline for the $\beta$ Pic YMG by studying three substellar atmospheres within the group and its vicinity. While the current sample size is limited, it provides a useful starting point for future work. In cases where host star abundances are unavailable or poorly constrained, planets in the $\beta$ Pic YMG can be compared directly to this baseline to infer their formation histories. \cite{Reggiani2024betapic} used an F dwarf in the $\beta$ Pic YMG, HD 181327, as a proxy for the natal environment of the YMG. In this work, we offer complementary benchmarks by characterizing the chemical composition of two benchmark isolated brown dwarfs from the same group. These late-type substellar objects open up a homogeneous way to measure abundance ratios through molecular features, as done for gas giant exoplanets. We also analyze a super-Jovian gas giant that is in the vicinity of the $\beta$ Pic YMG, 2MASS J02495436-0558015---named as 2MASS J0249-0557 c by \cite{Dupuy2018hawaii}). The membership of the 2MASS J0249-0557 system has been debated. It was originally attributed to the $\beta$ Pic YMG \citep{Shkolnik2017bpmg, Dupuy2018hawaii}, and later to the 32 Ori YMG \citep{Luhman2024bpmg}. The two moving groups spatially overlap, and the $\beta$ Pic YMG has been proposed as a leading tidal tail of the 32 Ori YMG \citep{Gagnne2021movinggroup}. They likely share similar natal environments. The 2MASS J0249-0557 system is found in the spatially overlapping region of the two moving groups, and therefore is challenging to assign membership.
All three targets were observed with the near-infrared high-resolution spectrograph VLT/CRIRES+ \citep{Dorn2014crires, Dorn2023crires}, ensuring a homogeneous dataset. 

In this work, we present an analysis of 2MASS J0249-0557 c's formation mechanism to show how comparisons with benchmark brown dwarfs sharing the same natal environment can shed light on its formation history. We begin by describing the properties of the targets in Section \ref{sec:targets}. Then, in Section \ref{sec:observations}, we discuss the observation details and data reduction process. In Section \ref{sec:retrieval}, we describe the atmospheric retrieval process, and in Section \ref{sec:results} we include the results of the retrievals and modifications to the retrieval process we made for each target. We discuss our results in the context of planet formation and the $\beta$ Pic YMG as a whole in Section \ref{sec:discussion}, and in Section \ref{sec:conclusion} we conclude with our main results.

\section{Targets}
\label{sec:targets}
2MASS J0249-0557 c has a spectral type of L$2.5\pm0.5$ in the optical and L$3\pm1$ in the infrared \citep{Chinchilla2021Halpha}. It has a mass of around $11.6_{-1.0}^{+1.3}M_{\mathrm{Jup}}$ and orbits at a separation of $1950\pm200$ AU from a pair of closely situated binary M dwarfs with spectral types of M6. The two M dwarfs (2MASS J0249-0557AB) have masses of around $48M_{\mathrm{Jup}}$ and $44M_{\mathrm{Jup}}$  and are $2.17 \pm 0.22$ AU apart \citep{Dupuy2018hawaii}. 

\cite{Chinchilla2021Halpha} studied the atmosphere of 2MASS J0249-0557 c with low-intermediate resolution spectra and found that the planet has strong H$\alpha$ emission either due to accretion or chromospheric activity. \cite{Bryan2020vsini} constrained the $v\sin{i}$ of 2MASS J0249-0557 c to be $15.3^{+0.4}_{-0.7}$ km/s. We acquired high signal-to-noise K-band spectra of this object through CRIRES+ (The CRyogenic InfraRed Echelle Spectrograph Upgrade Project) mounted on the Very Large Telescope \citep[VLT;][]{Dorn2014crires, Dorn2023crires}. Using this data, we can estimate the C/O ratio, CO isotopologue ratio ($^{12}$CO/$^{13}$CO), and metallicity ([C/H]) of 2M0249-0557 c and analyze its possible formation mechanisms.

% \textbf{Though previous studies by \cite{Shkolnik2017bpmg} and \citep{Dupuy2018hawaii} have attributed the 2MASS J0249-0557 system to the $\beta$ Pic YMG. More recent work by \cite{Luhman2024bpmg} has attributed it to the 32 Ori YMG. The two moving groups have similar ages and kinematics, and the $\beta$ Pic YMG has been suggested to be a leading tidal tail of the 32 Ori YMG \citep{Gagnne2021movinggroup}. These two moving groups are close to where clustering methods become inefficient in direct observable space, so attributing membership for certain co-moving objects is challenging. Due to their coeval nature, the two moving groups are expected to share similar chemical compositions. Therefore, we analyze the chemical composition of 2MASS J0249-0557 c along with isolated brown dwarfs in the $\beta$ Pic Moving Group.}

2MASSI J0443376+000205 (also known as 2MASSI J0443+0002) and SIPS J2000-7523 are two isolated M9 brown dwarfs in the $\beta$ Pic Moving Group \citep{Gagne2015banyan, Filippazzo2015mass}. The $v\sin{i}$ of 2MASSI J0443+0002 has been constrained to be $13.1\pm2.0$ km/s \citep{Deshpande2012MIJvsini}. \cite{Gagne2015banyan} and \cite{Filippazzo2015mass} used low-resolution spectra of the two objects to assign spectral types and derive the expected mass, radius, $\log(g)$, and $T_{\mathrm{eff}}$. For 2MASSI J0443+0002, \cite{Filippazzo2015mass} found a mass of $21.99\pm5.76 M_{\mathrm{Jup}}$ and a radius of $1.78\pm0.12 R_{\mathrm{Jup}}$. \cite{Gagne2015banyan} also constrained the mass of 2MASSI J0443+0002 to be $20.6^{+5.9}_{-3.8} M_{\mathrm{Jup}}$ and the radius to be $1.86^{+0.06}_{-0.05} R_{\mathrm{Jup}}$.
For SIPS J2000-7523, \cite{Filippazzo2015mass} found a mass of $24.75\pm6.06 M_{\mathrm{Jup}}$ and a radius of $1.88\pm0.11R_{\mathrm{Jup}}$. We acquired high signal-to-noise K-band spectra of these two objects through CRIRES+ as well.

We derive the expected masses and radii of the three targets ourselves with updated parallax values from Gaia DR3. These can be constrained from substellar evolutionary models \citep[e.g.][]{Dupuy2017evolution}. We use the age of the $\beta$ Pic YMG and the $L_{\mathrm{bol}}$ of the three targets from \cite{Dupuy2018hawaii} and \cite{Filippazzo2015mass}. Following the procedures in \cite{Xuan2024c13}, we consider four evolutionary models to mitigate model uncertainty: ATMO 2020 \citep{Phillips2020atmo, Chabrier2023atmo}, SM08 \citep{Saumon2008sm08}, AMES-Dusty \citep{Allard2001ames-dusty}, and BHAC15 \citep{Baraffe2015bhac15}. The predictions of these models for mass, radius, $\log(g)$, and $T_{\mathrm{eff}}$ of the three targets are shown in Figure \ref{fig:evolutionarymodels}.
The four models predict slightly different masses and radii, so for each variable, we estimate the central value of all four model predictions.
% visually estimate a Gaussian distribution that encompasses all four model predictions. The Gaussian distributions, their means, and their standard deviations are shown in Appendix Figure \ref{fig:evolutionarymodels}. 
Our predictions are consistent with those from the literature that we described above. 
We adapt these central values as priors for atmospheric retrieval as described in Sections \ref{sec:retrieval}, \ref{sec:2MASS J0249-0557c}, and \ref{sec:bds}. The priors used are summarized in Table \ref{tab:prior}.

\section{Observations and data reduction}
\label{sec:observations}
We observed 2MASS J0249-0557 c on September 21, 2022 and September 22, 2022 with CRIRES+. We chose the wavelength setting of K2166 and a $0.4'' \times 10''$ slit, giving us a high resolution of $\mathcal{R} \sim 50,000$. The observations were taken using the standard ABBA nodding scheme with a 600 s exposure time. We took 10 exposures per night, amounting to 3.3 hr of total integration time. The seeing was $\sim 0.51''-1.43''$ both nights. The target was at airmass $1.056-1.073$ and $1.056-1.099$ during the first and second night, respectively.

% Owing to the lack of standard star observations, we model the telluric lines in the spectra of 2MASS J0249-0557 c with ESO SkyCalc \citep{Jones2013skycalc, Noll2012skycalc}. The telluric line model is highly dependent on the variable precipitable water vapor (pwv). To find the value of pwv closest to the real conditions of the nights of observation, we use a range of pwv to generate different models. Then, we remove telluric lines from the spectra by dividing the data with each telluric line model while masking out the regions with strong telluric features. We assess the effectiveness of different telluric models by calculating the cross-correlation function (CCF) of each model and its corresponding telluric-removed data. We find that when pwv is larger than 1.5, there are peaks in the ccf figure at the radial velocity of the object. Therefore, we estimate that pwv is 1.5 (pwv less than 1.5 is less representative of normal weather). Using the telluric line model with this pwv value, we remove the effect of the Earth's atmosphere from the data. The wavelength range where the telluric features are too strong and where the telluric features are not strong enough for good wavelength resolution has been omitted.

We observed SIPS J2000-7523 on May 18th, 2023 and 2MASSI J0443+0002 on December 5, 2023 with CRIRES+. We chose the wavelength setting of K2166. We observed this target with a $0.2'' \times 10''$ slit, giving us a resolution of $\mathcal{R} \sim 100,000$. The observation was taken with the standard ABBA nodding scheme with a 300 s exposure time. We took 10 exposures per targets, amounting to a total integration time of 0.83 hr for each target. The seeing was $0.37'' - 0.99''$ for both targets, and the targets were at airmass 1.576-1.579 and 1.232-1.404, respectively.

Using the data reduction pipeline \texttt{excalibuhr} \citep{Zhang2024yses1}, we extracted spectra from the observations. 
We used the ESO sky tool \texttt{Molecfit} to correct for the telluric transmission in each spectrum.
Our observation of 2MASS J0249-0557 c has an SNR of $\sim$ 6 per wavelength channel.
Both brown dwarfs have an SNR of 28 per wavelength channel. We discard spectral orders in which the wavelength resolution is imprecise, as well as regions in which the telluric lines are strong (transmission $< 0.5$). The spectra used for analysis are shown in the Appendix, in Figures \ref{fig:2mjday1}, \ref{fig:2mjday2}, \ref{fig:2MIJspectra}, and \ref{fig:SIPSspectra}. The final wavelength range used for 2MASS J0249-0557c is 2.175 to 2.455 $\mu m$, and the wavelength range used for the two brown dwarfs is  2.175 to 2.471 $\mu m$.

\section{Atmospheric Retrieval}
\label{sec:retrieval}

\begin{table*}
\centering
\begin{tabular}{lcccc}
\hline
   & 2MASS J0249-0557 c & 2MASSI J0443+0002 & SIPS J2000-7523\\ \hline
  $\log(^{13}$CO/$^{12}$CO) & $\mathcal{U}$(-12,0) & $\mathcal{U}$(-12,0) & $\mathcal{U}$(-12,0)\\
  C/O & $\mathcal{U}$(0.10,1.5) & $\mathcal{U}$(0.10,1.5) & $\mathcal{U}$(0.10,1.5)\\
  $[\mathrm{M}/\mathrm{H}]$ & $\mathcal{U}$(-1.5,1.5) & $\mathcal{U}$(-1.5,1.5) & $\mathcal{U}$(-1.5,1.5) \\
  dt1 &  $\mathcal{N}$(0.053,0.021)* & $\mathcal{U}$(0.03,0.075) & $\mathcal{U}$(0.03,0.075) \\
  dt2 & $\mathcal{N}$(0.053,0.021)* & $\mathcal{U}$(0.03,0.075) & $\mathcal{U}$(0.03,0.075) \\
  dt3 &$\mathcal{N}$(0.093,0.021)* & $\mathcal{U}$(0.05, 0.18) & $\mathcal{U}$(0.05, 0.18) \\
  dt4 & $\mathcal{N}$(0.17,0.047)* & $\mathcal{U}$(0.07,0.30) & $\mathcal{U}$(0.07,0.30) \\
  dt5 & $\mathcal{N}$(0.18, 0.059)* & $\mathcal{U}$(0.08,0.34) & $\mathcal{U}$(0.08,0.34) \\
  dt6 & $\mathcal{N}$(0.29,0.17)* & $\mathcal{U}$(0.06 ,0.32) & $\mathcal{U}$(0.06 ,0.32) \\
  mass ($M_{\mathrm{Jup}}$)& $\mathcal{N}$(13, 1.0)* & $\mathcal{N}$(21, 1.0)* & $\mathcal{N}$(26, 2.0)* \\
  radius ($R_\odot$) & $\mathcal{N}$(0.15, 0.012)* & $\mathcal{N}$(0.17, 0.012)* & $\mathcal{N}$(0.18, 0.015)* \\
  rv (km/s) & $\mathcal{U}$(-20,20) & $\mathcal{U}$(-50,0) & $\mathcal{U}$(-20,20) \\
  $t_0$ (K)& $\mathcal{U}$(2000,4000) & $\mathcal{U}$(2000,4000) & $\mathcal{U}$(2000,4000) \\
  $v\sin{i}$ (km/s) & $\mathcal{U}$(0,50) & $\mathcal{U}$(0,50) & $\mathcal{U}$(0,100) \\
  $\log(g)$ (if used) & $\mathcal{U}$(3.5,5.5) & $\mathcal{U}$(3.5,5.5) & $\mathcal{U}$(3.5,5.5)\\
  sigma\_lnorm & $\mathcal{U}$(1,3) & -- & -- \\
  kzz & $\mathcal{U}$(2,10) & -- & -- \\
  fsed\_MgSiO$_3$ & $\mathcal{U}$(0,10) & -- & -- \\
  log\_X\_cloud\_base\_MgSiO$_3$ & $\mathcal{U}$($\log_{10}(0.005),\log_{10}(10)$) & -- & -- \\
\hline
\end{tabular}
\caption{Priors of parameters used in atmospheric retrieval. Priors marked with * are Gaussian in $\mathcal{N}(\mu,\sigma)$ notation, where the first value denotes the mean and the second the standard deviation. Priors without * are uniform in $\mathcal{U}(a,b)$, where the first value denotes the lower bound and second the upper bound.}
\label{tab:prior}
\end{table*}

We perform atmospheric retrievals with the python packages \texttt{petitRADTRANS} (pRT) \citep{Molliere2019pRT} and \texttt{PyMultiNest} \citep{Buchner2014Bayesian}. We used \texttt{PyMultiNest} with 400 live points, a sampling efficiency of 0.05, and constant-efficiency mode enabled, following \cite{Regt2024eso1}. We use pRT to create a spectral model that generates synthetic spectra from C/O ratio, metallicity [M/H], isotopologue ratio $^{12}$CO/$^{13}$CO, surface gravity $\log g$, radial velocity, rotational broadening $v\sin{i}$, and the pressure-temperature profile of the planet. We include the following molecular and atomic line species: $\mathrm{H}_2$O, $^{12}$CO, C$\mathrm{H}_4$, N$\mathrm{H}_3$, $^{13}$CO, $\mathrm{CO}_2$, $\mathrm{H}_2$S, Na, and K \citep{Rothman2010CO,Rothman2013C13, Azzam2016H2S, Yurchenko2017ch4, Polyansky2018h2o, Coles2019NH3, Molliere2019pRT, Allard2019Na, Hargreaves2020CH4, Costes2024c13}. We include collision-induced absorption due to $\mathrm{H}_2$-$\mathrm{H}_2$ and $\mathrm{H}_2$-He interactions as continuum sources and account for Rayleigh scattering contributions from $\mathrm{H}_2$ and He \citep{Chan1965h2, Dalgarno1962h2, Borysow1988h2he, Borysow1989h2hea, Borysow1989h2heb, Borysow2001h2h2, Borysow2002h2h2}.

We model the pressure-temperature profile with a gradient model that divides the atmosphere into six knots at pressure levels of $10^{-5}$ bar, $10^{-3.5}$ bar, $10^{-2}$ bar, $10^{-1}$ bar, 1 bar, and 10 bar. We set the temperature-pressure gradient $d\ln{T}/d\ln{P}$ at each knot, as well as the temperature of the bottom layer, as free parameters following the method used in \cite{Zhang2023cloud}. We use Gaussian priors for the pressure gradient at each knot, with values derived from the self-consistent pressure-temperature profiles from Sonora Bobcat \citep{Marley2021sonora}. 

We use a equilibrium chemistry model. It computes the chemical abundances at a given pressure, temperature, C/O ratio, and metallicity by interpolating a precomputed table that assumes chemical equilibrium \citep{Molliere2017chem, Molliere2020chem}. 
% The model enforces constant mass fractions of key species (CO, CH$_4$, and H$_2$O) above the quenching pressure level, set at $\log(P_{\mathrm{quench}}) =-5$, effectively creating an equilibrium chemistry model \citep{Zahnle2014mixing}.

Clouds are ubiquitous for gas giants like 2MASS J0249-0557 c and have a significant impact on the spectra of such objects, so we investigate the effect of clouds by adding a condensate cloud model to the retrieval process. To include clouds, we follow the setup in e.g. \cite{Zhang2021BDc13}, implemented with the \cite{Ackerman2001cloud} model. Fe and MgSiO$_3$ clouds are expected to be the primary cloud species in L dwarfs like 2MASS J0249-0557 c \citep{Morley2012cloud}. In this work, we focus on MgSiO$_3$ clouds because Fe is not expected to be the dominant aerosol composition \citep{Gao2020fecloud}. Even if Fe clouds were to form, they would condense at higher temperatures and lower altitudes than the photosphere. We add MgSiO$_3$ as a cloud species \citep{Scott1996mgsio3, Jaeger1998mgsio3}. The cloud model consists of four additional parameters: log\_X\_cloud\_base\_MgSiO$_3$, the mass fraction of the cloud species at the cloud base; fsed\_MgSiO$_3$, the settling parameter; Kzz, the eddy diffusion coefficient; and sigma\_lnorm, the width of the log-normal particle size distribution. The location of the cloud base is at the intersection of the condensation curve and the P-T profile.

We remove the continuum from the data because instrumental effects can introduce artificial slopes in the spectra. For these observations, no standard star was observed, which prevents calibration of the overall spectral shape. We also remove the continuum from the synthetic spectra to enable a direct comparison of the absorption lines, and we scale the model by a factor of $\alpha$ to match the order of magnitude of the data, where $\alpha$ is given by
\begin{equation}
    \alpha = \sum_i \frac{f_i^{'\mathrm{mod}}f_i^{\mathrm{obs}}}{\sigma_i^2}/\sum_i\frac{f_i^{'\mathrm{mod}}}{\sigma_i^2},
\end{equation}
where $f_i^{'\mathrm{mod}}$ is the model spectrum of the $i^{th}$ spectral order after it has been convolved and rebinned to the observation's resolution and wavelength grid, $f_i^{\mathrm{obs}}$ is the observed data, and $\sigma_i$ is the observation error.

Additionally, we include photometry from 2MASS (filters Ks, J, and H) and WISE (filters W1 and W2) to jointly fit a high-resolution spectrum model to the near-infrared spectrum obtained with CRIRES+ and a low-resolution spectrum model to available photometry\footnote{The retrieval results with and without jointly fitting photometry are largely consistent. For 2MASSI J0443+0002, including photometry led to higher values of $\log g$ and metallicity, while for the other two targets, the inclusion of photometry produced negligible differences.}. This allows us to better constrain the surface gravity $\log g$ and temperature structure of the object by extending the model to a wider wavelength range. We include all molecular and atomic lines from the high-resolution model, as well as FeH and TiO, which contain line features in this broader range of wavelengths \citep{Wende2010feh, McKemmish2019TiO}. We adopt parallax values of 15.1 mas, 47.6 mas, and 34.0 mas for 2MASS J0249-0557 c, 2MASSI J0443+0002, and SIPS J2000-7523, respectively \citep{Gaia2020EDR3}, and add the radii of these objects as an additional retrieval parameter. The priors of the radius parameter are derived from evolutionary models as described in Section \ref{sec:targets}.

We use a likelihood function as follows to calculate the similarity of the generated spectra to the observed data:
\begin{align}
    \ln(L) &= -\frac{1}{2}[N(\ln(2\pi)+\ln(\beta^2))\\ \notag
    &+\sum_i\frac{(f_i^{\mathrm{mod}}-f_i^{\mathrm{obs}})^2}{\beta^2\sigma_i^2}\\ \notag
    &+\sum_i\ln(\sigma_i^2)]
\end{align}
where N is the number of data points; $f_i^{\mathrm{mod}}$ is the model flux after it has been convolved, rebinned, continuum-removed, and scaled to the magnitude of the data by $\alpha$; and $\beta$ is the error inflation given by \citep{Ruffio2019rv}:
\begin{equation}
    \beta =\sqrt{\frac{1}{N}\sum_i \frac{(f_i^{\mathrm{mod}}-f_i^{\mathrm{obs}})^2}{\sigma_i^2}}.
\end{equation}
The photometry is included by adding the following to the likelihood function for each photometric observation:\begin{equation}
    \ln(L_{\mathrm{phot}}) = -\frac{1}{2}[ \frac{(F^{\mathrm{mod}}-F^{\mathrm{obs}})^2}{\epsilon^2}+\ln(2\pi\epsilon^2)]
\end{equation}
where $F^{\mathrm{mod}}$ is the model flux integrated over the photometric filter and scaled from flux to irradiance, following the methodology in \cite{Stolker2020bpicb}, $F^{\mathrm{obs}}$ is the corresponding observed $F_\lambda$, and $\epsilon$ is the uncertainty of $F^{\mathrm{obs}}$.

All the retrieval parameters and their priors are shown in Table \ref{tab:prior}.

\section{Results}
\label{sec:results}
\subsection{Atmospheric Retrieval of 2MASS J0249-0557c}
\label{sec:2MASS J0249-0557c}

\begin{figure}
    \centering
    \includegraphics[width=0.98\linewidth]{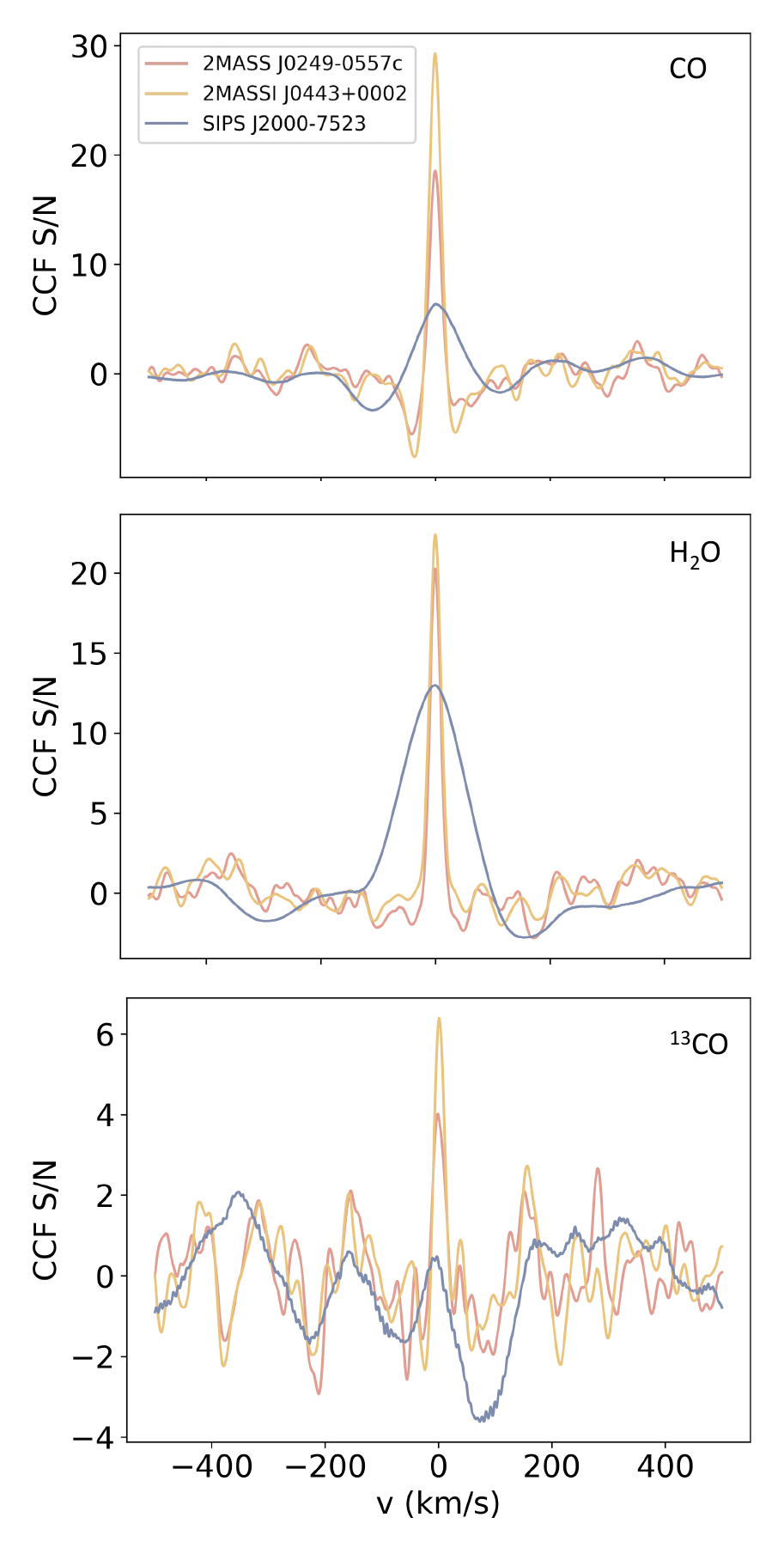}
    \caption{Cross-correlation function (CCF) of CO, H$_2$O, and  $^{13}$CO detection for the three targets. From the $^{13}$CO CCF alone, we obtain a significant detection of $^{13}$CO for 2MASSI J0443+0002, a moderate detection for 2MASS J0249-0557c, and no detection for SIPS J2000-7523.}
    \label{fig:ccfs}
\end{figure}
\begin{figure*}
    
    \centering
    \includegraphics[width=0.98\linewidth]{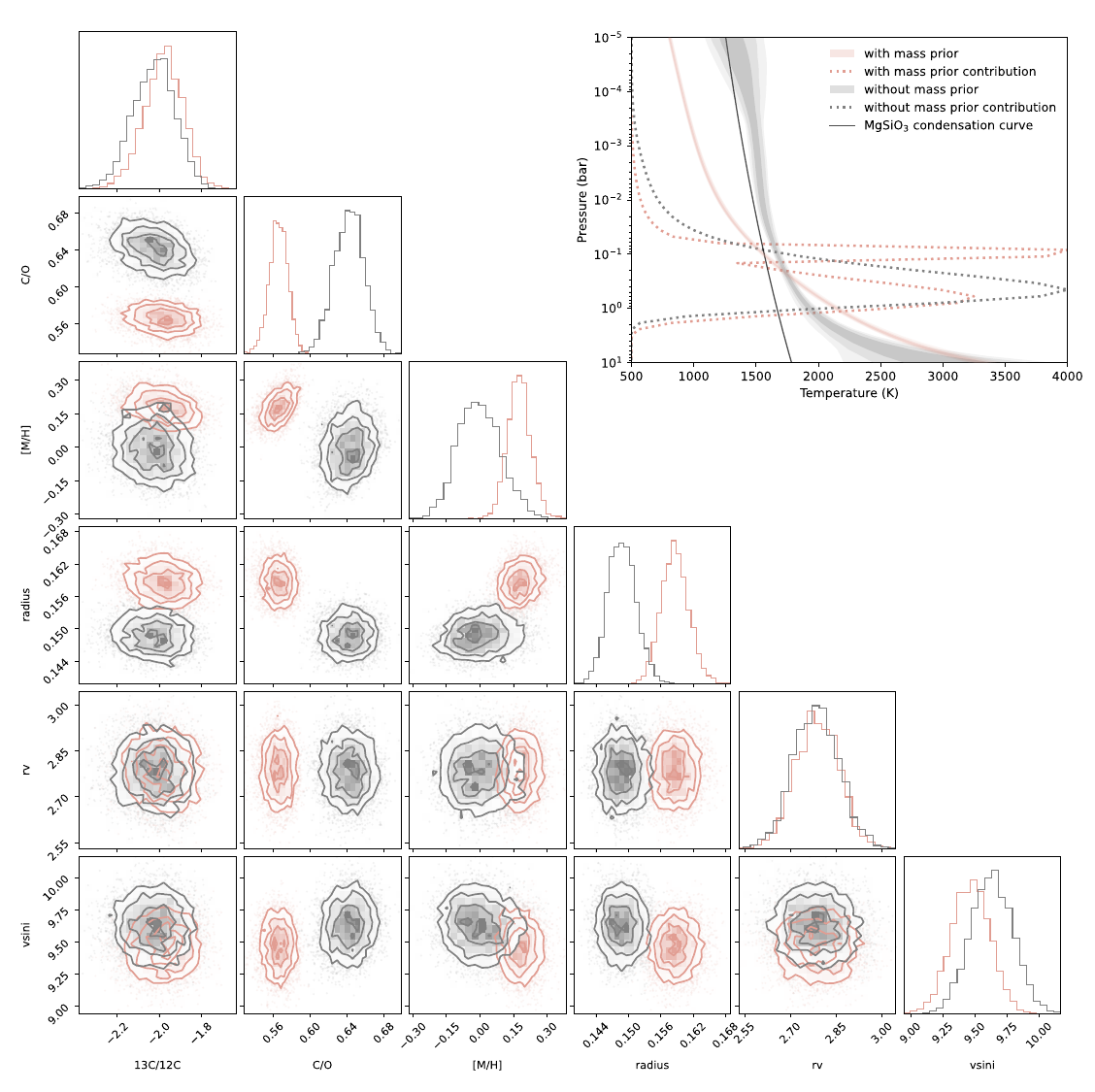}
    \caption{Comparison of using $\log(g)$ as prior for atmospheric retrieval and using mass as prior for 2MASS J0249-0557c. The result of using $\log(g)$ is shown in gray, and the result of using mass is shown in red. In the mass prior case, there is a preference for a condensate cloud model, so we included a weakly constrained condensate cloud model. In the $\log(g)$ prior retrieval, there is a preference for a cloudless model, so we did not include clouds. With the inclusion of the condensate cloud model, the contribution appears double-peaked due to the cloud deck, as indicated by the intersection of the condensation curve and the P–T profile.}
    \label{fig:cloud_logg}
\end{figure*}

\begin{figure*}
    \centering
    \includegraphics[width=0.98\linewidth]{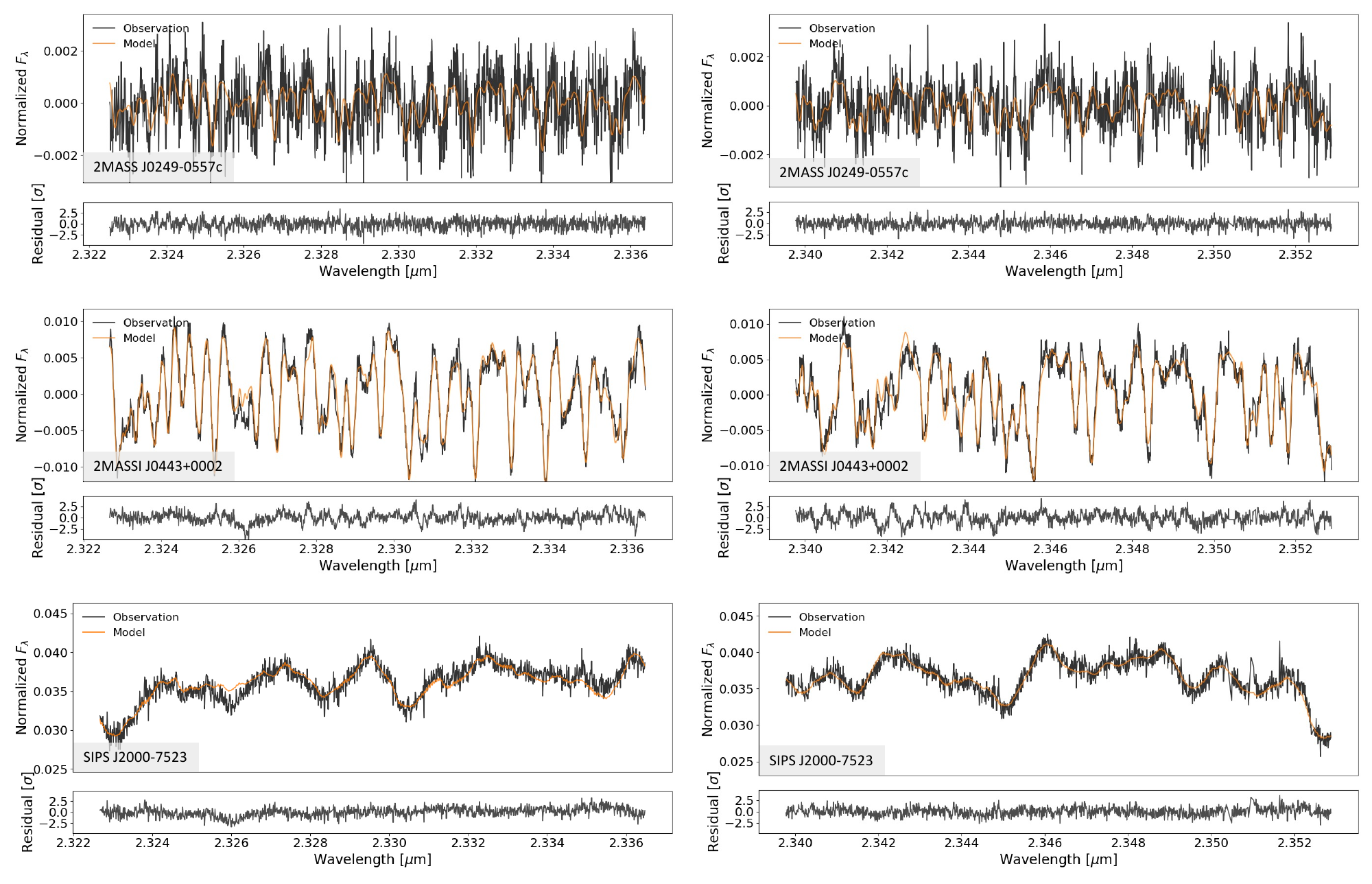}
    \caption{Two orders of high-resolution spectra with the best-fit model overplotted in yellow for each target. The observation on the first day (September 21, 2022) is chosen for 2MASS J0249-0557 c. The wavelength range is selected to include the $^{12}$CO and $^{13}$CO bandheads.}
    \label{fig:highres}
\end{figure*}

\begin{figure}
    \centering
    \includegraphics[width=0.98\linewidth]{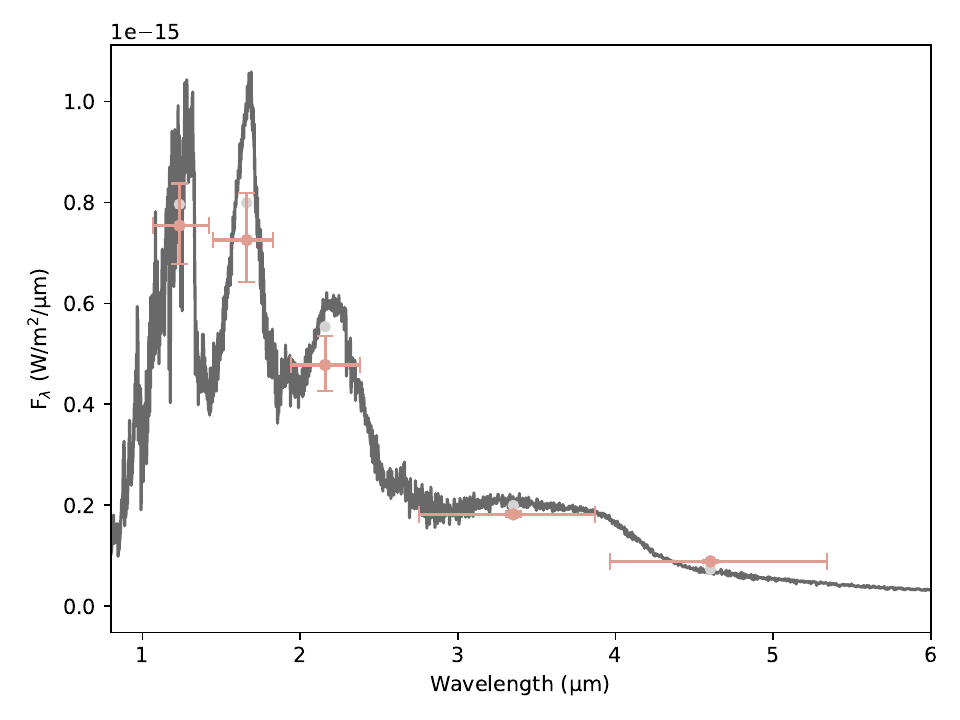}
    \caption{Best-fit low-resolution model spectrum of 2MASS J0249-0557c. The pink points with error bars show the archival photometry, while the light-gray points show the corresponding synthetic photometry of the best-fit model.}
    \label{fig:lowres}
\end{figure}

We detect H$_2$O, CO, and $^{13}$CO for 2MASS J0249-0557c, as shown in Figure \ref{fig:ccfs}. The results of the fit are shown in gray in Figure \ref{fig:cloud_logg}. The C/O ratio is $0.64\pm0.02$, the isotopologue ratio $^{12}$CO/$^{13}$CO is $105.3^{+30.7}_{-21,1}$, and the metallicity [M/H] is $0.01^{+0.10}_{-0.09}$. We retrieve a $\log(g)$ value of $3.70^{+0.13}_{-0.12}$.
However, a $\log g$ value of 3.70 and the retrieved radius value of 1.46 $R_{\mathrm{Jup}}$ together correspond to a mass of 4.31 $M_{\mathrm{Jup}}$, much lower than the expected value of $12.7\pm1.8M_{\mathrm{Jup}}$. We see that $\log(g)$ is correlated with metallicity (Figure \ref{fig:2mjlogg}), such that an underestimated $\log(g)$ may indicate that our metallicity constraint is incorrect. We therefore consider an alternative set of parameters, which includes the mass of the object instead of $\log g$, and then calculate $\log(g)$ from mass and radius. We set the mass prior using evolutionary models described in Section \ref{sec:targets}. To enforce a mass close to the expected value, we decrease the standard deviation of this object's mass from 1.8 $M_{\mathrm{Jup}}$, the value in the predicted distribution, to 1 $M_{\mathrm{Jup}}$.

Using this retrieval scheme, we also found a preference for clouds in the atmosphere of 2MASS J0249-0557c. We compare the Bayesian log-evidence of the retrieval result with clouds and the retrieval results without clouds. We found that the model with clouds is weakly preferred, with a Bayes factor $\ln(K) = 7.52$, which conventionally corresponds to a 4.3$\sigma$ significance \citep{Jeffreys1939bayes, Kass1995bayes, Buchner2014Bayesian}. However, $\sigma$ values inferred from Bayes factors are upper limits and tend to overestimate the true significance \citep{Kipping2025sigma, Thorngren2025bayes}. We report the $\sigma$ value for completeness, but emphasize that the Bayes factor provides the more reliable measure of model preference. The cloud parameters themselves remain only weakly constrained. A flexible P-T profile parametrization can partially mimic the spectral effects of cloud opacity; the limited wavelength coverage does not provide sufficient leverage to disentangle the effects of cloud opacity from adjustments in the P–T profile \citep[e.g.,][]{Zhang2021BDc13, Regt2024eso1}. We therefore do not claim a robust cloud detection. Importantly, the inferred relative abundances are consistent across cloud treatments and are robust within the adopted model.

The results of this retrieval method with the condensate cloud model are shown in red in Figure \ref{fig:2mjfinal} and Table \ref{tab:results}. We show two orders of the best-fit model overlaid on the CRIRES+ high-resolution spectra in Figure \ref{fig:highres}, and the low-resolution model with photometry overplotted in Figure \ref{fig:lowres}. The full spectra and best-fit model can be found in Figure \ref{fig:2mjday1} and \ref{fig:2mjday2}. The final C/O ratio for 2MASS J0249-0557c is $0.57\pm0.01$, the isotopologue ratio $^{12}$CO/$^{13}$CO is $95^{+23}_{-17}$, and the metallicity [M/H] is $0.18\pm0.05$. The retrieval mass and radius are $11.38\pm0.77M_{\mathrm{Jup}}$ and $1.55\pm0.03R_{\mathrm{Jup}}$. The best-fit model corresponds to a $T_{\mathrm{eff}} = 1723 K$, consistent with the results from evolutionary models in Appendix Figure \ref{fig:evolutionarymodels}.

We compare the results from a $\log(g)$ prior and a mass prior with clouds and the PT profiles in Figure \ref{fig:cloud_logg}. Because $\log(g)$ and metallicity are correlated, we can see that as $\log(g)$ is forced to a larger, more realistic value, the new metallicity value is larger. The radius now enters the $\log(g)$ calculation, whereas it was previously only used to scale the surface flux to irradiance, leading to a change in the inferred radius value. The P-T profile is less isothermal with the new prior. Changes in the metallicity alter the volume mixing ratios of molecules, which, along with changes in the P-T profile, shift the best-fit C/O ratio to a lower value.

\begin{table*}
\centering
\begin{tabular}{lccc}
\hline
   & 2MASS J0249-0557 c & 2MASSI J0443+0002 & SIPS J2000-7523 \\ \hline
  $^{12}$CO/$^{13}$CO & $95^{+23}_{-17}$ & $70\pm5$ & $254^{+61}_{-43}$ \\
  C/O & $0.57\pm0.01$ & $0.591\pm0.004$ & $0.671\pm0.005$  \\
  $[\mathrm{M}/\mathrm{H}]$ & $0.18\pm0.05$ & $-0.14\pm0.02$ & $0.24\pm0.05$ \\
  dt1 & $0.0516\pm0.0006$ & $0.056^{+0.01}_{-0.02}$ & $0.05\pm0.02$ \\
  dt2 & $0.0515\pm0.0006$ & $0.032^{+0.003}_{-0.002}$ & $0.04\pm0.01$ \\
  dt3 & $0.086\pm0.001$ & $0.0502^{+0.0003}_{-0.0002}$ & $0.052\pm0.002$ \\
  dt4 & $0.139\pm0.005$ & $0.072^{+0.003}_{-0.002}$ & $0.102\pm0.002$ \\
  dt5 & $0.150\pm0.006$ & $0.097\pm0.004$ & $0.082^{+0.003}_{-0.002}$ \\
  dt6 & $0.25\pm0.02$ & $0.063^{+0.005}_{-0.003}$ & $0.09^{+0.04}_{-0.02}$ \\
  mass ($M_{\mathrm{Jup}}$) & $11.38\pm0.77$ & $18.04\pm0.75$ & $18.80^{+2.81}_{-2.63}$ \\
  radius ($R_\odot$)& $0.159\pm0.003$ & $0.168\pm0.001$ & $0.199^{+0.002}_{-0.001}$ \\
  rv (km/s) & $2.79\pm0.07$ & $-21.65\pm0.03$ & $4.89^{+0.16}_{-0.17}$ \\
  $t_0$ (K) & $3369^{+83}_{-77}$ & $2864^{+25}_{-23}$ & $3078^{+116}_{-58}$ \\
  $v\sin{i}$ (km/s) & $9.47\pm0.15$ & $12.59^{+0.04}_{-0.05}$ & $73.33\pm0.24$\\
  sigma\_lnorm & $1.95^{+0.62}_{-0.54}$& -- & -- \\
  kzz & $5.86^{+2.30}_{-1.44}$ & -- & -- \\
  fsed\_MgSiO$_3$ & $7.78^{+1.44}_{-1.92}$ & -- & -- \\
  log\_X\_cloud\_base\_MgSiO$_3$ & $-0.53^{+0.16}_{-0.20}$ & -- & -- \\
\hline
\label{tab:results}
\end{tabular}
\caption{Final retrieval results for the three targets.}
\end{table*}

\subsection{Atmospheric Retrieval of 2MASSI J0443+0002 and SIPS J2000-7523}
\label{sec:bds}
We follow the same procedure in Section \ref{sec:retrieval} except for the following modifications:

1. For the two brown dwarfs, we used uniform priors similarly informed by Sonora Bobcat. We did not adopt Gaussian priors for the brown dwarfs, as this resulted in non-physically low gradient values.

2. For SIPS J2000-7523, we did not remove the continuum from the data and the model because this object is too fast-rotating (with $v\sin{i}\sim73\,\mathrm{km/s}$), making it difficult to remove the continuum without also removing key line features.

3. For 2MASSI J0443+0002 and SIPS J2000-7523, we also used mass and radius with Gaussian priors instead of $\log(g)$ as retrieval parameters so that we avoid underestimating $\log(g)$. We obtained mass priors from evolutionary models as described in Section \ref{sec:targets}. To enforce a mass in the brown dwarf range for 2MASSI J0443+0002, we decreased the standard deviation of the mass prior from $5 M_{\mathrm{Jup}}$, the value in the predicted distribution, to $1 M_{\mathrm{Jup}}$.

4. We did not include clouds in our models for these two targets because, as M9 brown dwarfs, they are unlikely to be cloudy.

The retrieval results for these two brown dwarfs are shown in Figures \ref{fig:2MIJfinal} and \ref{fig:SIPSfinal}, with corresponding values listed in Table \ref{tab:results}. For 2MASSI J0443+0002, the C/O ratio is $0.591\pm0.004$, the isotopologue ratio $^{12}$CO/$^{13}$CO is $70\pm5$, and the metallicity [M/H] is $-0.14\pm0.02$. For SIPS J2000-7523, the C/O ratio is $0.672\pm0.005$, and the metallicity [M/H] is $0.24\pm0.05$. As shown in Section \ref{sec:13CO}, the isotopologue ratio $^{12}$CO/$^{13}$CO of SIPS J2000-7523 is not well-constrained, as we do not have a strong detection of $^{13}$CO in this target. For 2MASSI J0443+0002, we recover a strong detection of $^{13}$CO, and we detect H$_2$O and CO for both brown dwarfs, as shown in Figure \ref{fig:ccfs}. Because SIPS J2000-7523 is fast-rotating, the CCF peak is less pronounced. However, we still have significant detections of H$_2$O and CO for this target. From the retrieved mass and radius of SIPS J200-7523, the break velocity is $v=\sqrt{GM/R}\approx131 \mathrm{km/s}$. The retrieved $v\sin i$ of $73.33 \mathrm{km/s}$ indicates that the rotational velocity will be lower than the break velocity for an inclination of less than $\sim34^\circ$. The $T_{\mathrm{eff}}$ of the best-fit model of 2MASSI J0443+0002 and SIPS J2000-7523 is 2184$K$ and 2241$K$, respectively. Both are consistent with the predictions from evolutionary models in Appendix Figure \ref{fig:evolutionarymodels}.

\subsection{Detection of $^{13}$CO}
\label{sec:13CO}

% \begin{figure}
%     \centering
%     \includegraphics[width=0.98\linewidth]{Figures/C13.pdf}
%     \caption{Cross-correlation function (CCF) of $^{13}$CO detection for the three targets. From the CCF alone, we have significant detection of $^{13}$CO for 2MASSI J0443+0002, moderate detection for 2MASS J0249-0557c, and no detection for SIPS J2000-7523.}
%     \label{fig:c13detection}
% \end{figure}

For each target, to quantify the significance of the $^{13}$CO detection, we conduct atmospheric retrieval with a new model that is the same as the one we used to create the results in section \ref{sec:retrieval} except for removing $^{13}$CO. We compare the Bayesian log-evidence of the two retrievals and found that there is a preference for the model with $^{13}$CO with a Bayes factor of 20.2, 189, and 13.1 for 2MASS J0249-0557c, 2MASSI J0443+0002, and SIPS J2000-7523, respectively. By conventional interpretation, all three Bayes factors correspond to a $>5\sigma$ significance \citep{Jeffreys1939bayes, Kass1995bayes, Buchner2014Bayesian}. However, as pointed out in Section \ref{sec:2MASS J0249-0557c}, such $\sigma$ values are upper limits and overestimates of the true significance \citep{Kipping2025sigma, Thorngren2025bayes}. The Bayes factors themselves, together with the cross-correlation function (CCF) analysis described below, provide a more reliable measure of detection significance.

To create CCFs, we subtracted all spectral features except for those of $^{13}$CO from the fitted spectra and the observed data and calculated the CCF of the two resultant spectra for each target. Figure \ref{fig:ccfs} shows the CCF functions. For 2MASS J0249-0557c and 2MASSI J0443+0002, the peak has an S/N of more than 4 and 6, respectively. Therefore, we conclude that we have a significant detection of $^{13}$CO for these two objects. For SIPS J2000-7523, the peak is not pronounced, an expected result due to its fast-rotating nature. 

\begin{figure}
    \centering  \includegraphics[width=0.98\linewidth]{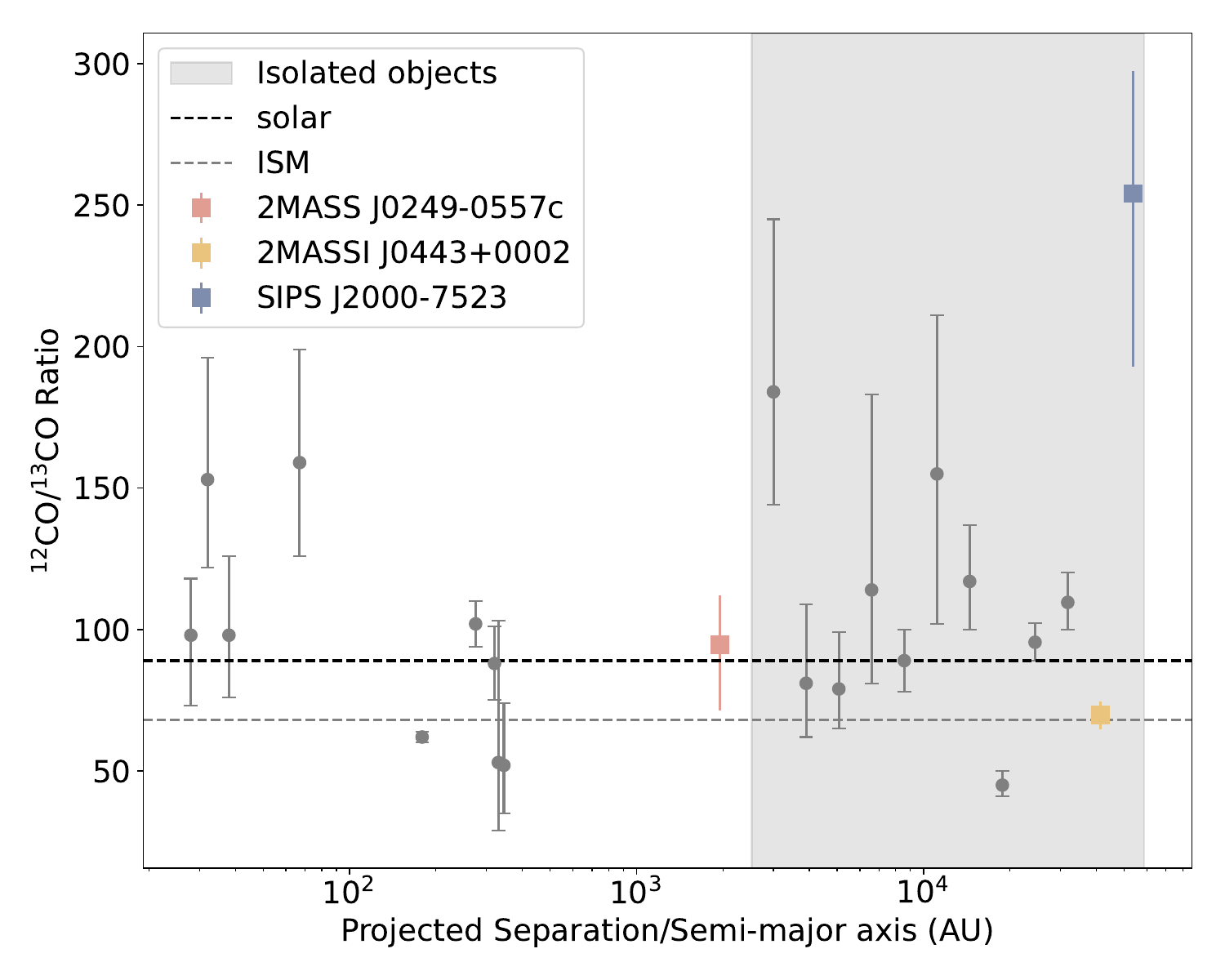}
    \caption{Substellar objects with known $^{12}$CO/$^{13}$CO, excluding the hot Jupiters. Literature values are shown in gray \citep{Gandhi2023c13, Costes2024c13, Mulder2025eso6,Gandhi2025eso5, Regt2024eso1, Xuan2024isotope, Xuan2024c13, Zhang2024yses1, Gonalezz2024eso2, Gonalezz2025eso4, Grasser2025eso8, Molliere2025PSO318, Wang2026_cd35}. Values from this work are shown in color. The solar value of 89 and ISM value of 68 are shown with dashed lines \citep{Anders1989solarc13, Langer1993ISMc13, Clayton2004stardust, Milam2005ISMc13, Meibom2007solarc13, Woods2009solarC13}. We add at least two new $^{12}$CO/$^{13}$CO values to this sample (we include SIPS J2000-7523 here for completeness, noting that we do not have a significant detection).}
    \label{fig:c13}
\end{figure}

As shown in Figure \ref{fig:c13}, we expand the sample of substellar objects with measured $^{12}$CO/$^{13}$CO ratios. The $^{13}$CO CCF for SIPS J2000-7523 is not significant, but we report two new measurements for 2MASS J0249-0557 c and 2MASSI J0443+0002.

\section{Discussion}
\label{sec:discussion}

\subsection{Caveats of Atmospheric Retrieval}

There are several caveats to the values presented by this work. The uncertainties of the retrieved values are underestimated because we did not consider correlated noise. For the two isolated brown dwarfs, the metallicity is linearly correlated with surface gravity or mass (see Appendix Figure \ref{fig:2MIJfinal} and \ref{fig:SIPSfinal}), which, in turn, is dependent on the evolutionary models. For 2MASS J0249-0557 c, this correlation is less significant.

The C/O ratio is also dependent on cloud conditions, of which we do not have tight constraints for 2MASS J0249-0557 c (see Appendix Figure \ref{fig:2mjfinal}). The expected cloud chemistry of a substellar object depends on many factors, including spectral type, $T_{\mathrm{eff}}$, and bulk elemental ratios such as Mg/Si \citep{Calamari2024mgsi}, with the cloud bases set by the intersection of condensation curves with the P–T profile \citep[e.g.][]{Visscher2010clouds}. In particular, silicate clouds are expected in giant exoplanets like 2MASS J0249-0557c (with retrieved $T_{\rm{eff}}$ of 1723 K)  instead of Fe clouds \citep{Gao2020fecloud}. Accordingly, we included silicate clouds (specifically MgSiO$_3$) in our retrieval framework of 2MASS J0249-0557c. Though we did find a preference for clouds, we did not obtain a robust detection and constraint due to the limited wavelength coverage and sensitivity of our data to cloud signatures. If unaccounted silicate clouds are present in 2MASS J0249-0557 c, then the retrieved C/O ratio is an overestimate of the true value. Furthermore, the assumption of a homogeneous cloud deck may be oversimplified. Spatially inhomogeneous or patchy clouds would also affect the C/O ratio \citep{Vos2023patchyclouds, Zhang2025patchycloud}.

\subsection{Formation History of 2MASS J0249-0557c}

\begin{figure}
    \centering
    \includegraphics[width=0.98\linewidth]{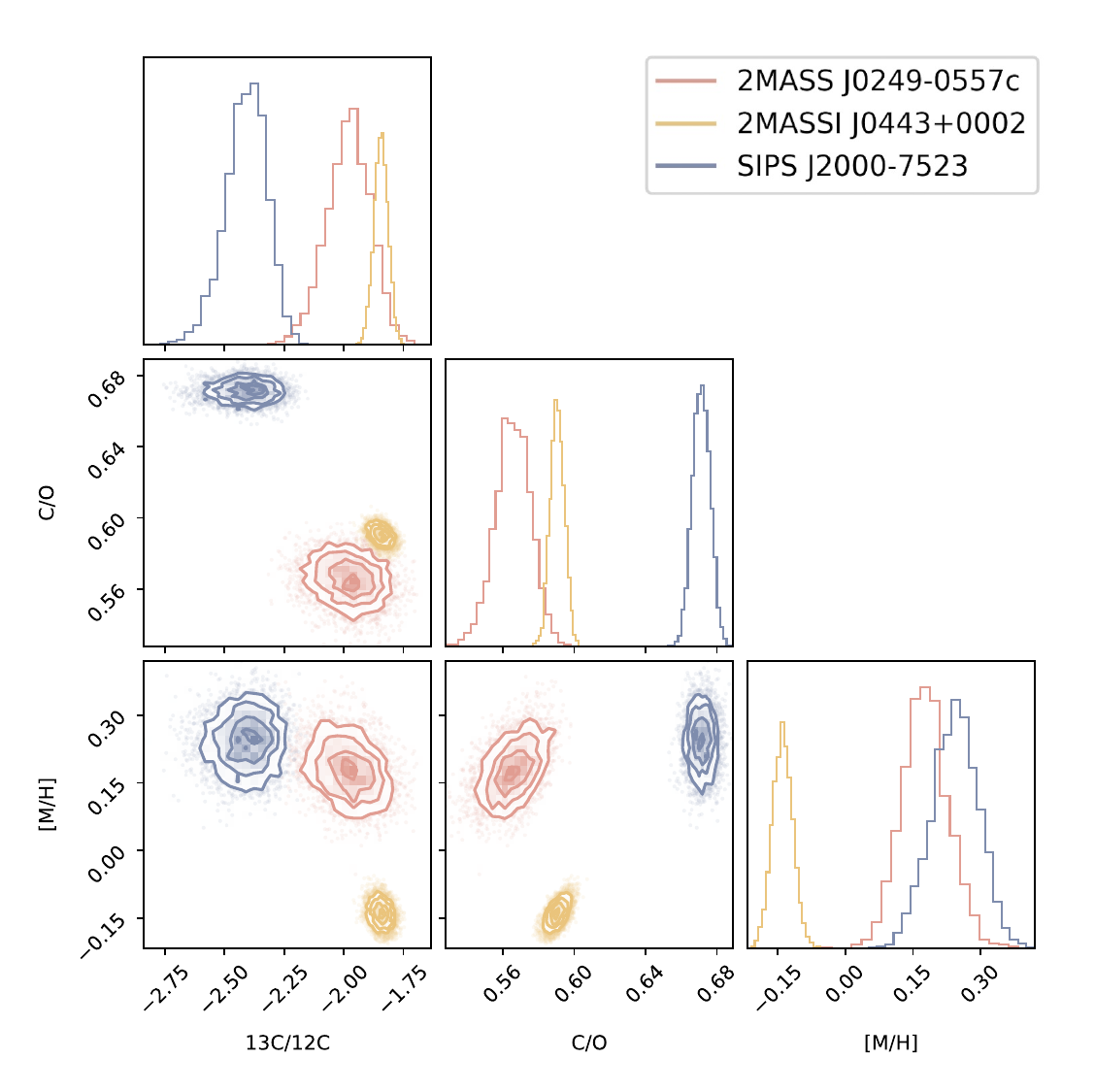}
    \caption{Corner plots of $\log(^{13}$CO/$^{12}$CO), C/O, and [M/H] of the three targets in this study. The C/O ratios of all three targets are consistent with the solar abundance range of $0.59\pm0.08$ \citep{Asplund2021solar}. The CO isotopologue ratio distribution of SIPS J2000-7523 cannot be robustly determined as we do not have a strong detection of $^{13}$CO for this object (see Section \ref{sec:13CO}).}
    \label{fig:combined_corner}
\end{figure}

2MASS J0249-0557c's C/O and $^{12}$CO/$^{13}$CO ratios are consistent with the solar level as well as the two benchmark brown dwarfs. In Figure \ref{fig:combined_corner}, the three distributions in the C/O panel are broadly consistent with the solar value range of $0.59 \pm 0.08$ \citep{Asplund2021solar}. The distributions appear distinct, but the error bars are likely underestimated because we did not consider correlated noise and model uncertainties. As shown in Figure \ref{fig:highres}, the residual noise has clearly correlated features, likely due to imperfect line lists and missing opacity sources. Additionally, cloud sequestration of oxygen can bias the C/O measurements, though this effect is partially mitigated by our use of the equilibrium chemistry model in our analysis of 2MASS J0249-0557 c. The distribution of SIPS J2000-7523 in the $^{12}$CO/$^{13}$CO appears inconsistent with the two other targets and solar values. However, this constraint is unreliable because, as described in Section \ref{sec:13CO}, we do not have a significant detection of $^{13}$CO in this object.

This consistency suggests the most likely formation pathway for 2MASS J0249-0557c is top-down, either through cloud fragmentation or gravitational instability. Gravitational instability requires enough disk material at the formation location of 2MASS J0249-0557c to create a gas giant. This is unlikely because the two brown dwarfs 2MASS J0249-0557AB are generally not massive enough to have a protoplanetary disk like this \citep[e.g.][]{Rilinger2021disk}. Therefore, cloud fragmentation is the most likely explanation for the formation of 2MASS J0249-0557c.

The consistency of the C/O ratio across the three targets and the solar value could also indicate that the planet formed beyond the CO snowline through enhanced solid-accretion \citep{Oberg2011CO}. However, the two brown dwarf hosts likely did not have enough material at 1950 AU in the protoplanetary disk to form a planet through enhanced solid-accretion. Therefore, if this were true, the planet formed around another host in its neighborhood before being ejected and later captured by its current hosts. This type of complex dynamical evolution cannot be validated or rejected without more information about the system configuration of 2MASS J0249-0557.

These conclusions regarding the formation pathway of 2MASS J0249-0557c rely on the assumption that the two isolated brown dwarfs are representative of the overall composition of the $\beta$ Pic YMG and vicinity, and thus the natal environment of 2MASS J0249-0557c. 
If, however, the overall C/O ratio of its natal environment is subsolar, this would indicate that the C/O ratio of 2MASS J0249-0557c is enhanced relative to its natal environment. If the metallicity of 2MASS J0249-0557c is lower than its environment, then it likely formed through gas accretion of oxygen-depleted gas outside of the H$_2$O snowline. If its metallicity is enhanced relative to its environment, then formation scenarios involving carbon-enhanced material, such as gas accretion near the CO and CO$_2$ snowlines or pollution from carbon-rich grains, are more likely \citep{Oberg2011CO}. On the other hand, if the overall C/O ratio of its natal environment is supersolar, then 2MASS J0249-0557c's C/O ratio would be lower than its environment and, if coupled with higher metallicity, would indicate that the atmosphere is polluted by large amounts of oxygen-rich solids \citep{Oberg2011CO, Zhang2024yses1}. Future work that expands the number of well-characterized objects in the $\beta$ Pic YMG and vicinity is needed to better constrain the chemical abundances of its natal environment, while we expect that the overall composition is solar, as discussed in Section \ref{sec:bpmg}. 

The chemical abundances of the host pair 2MASS J0249-0557AB are unknown. As a result of its binarity and close separation of 0.04'' \citep{Dupuy2018hawaii}, constraining accurate chemical abundances of the host is difficult, and would require additional assumptions that would increase the uncertainty and model dependency of the result. However, if this information could be acquired, it would facilitate a direct comparison and aid in constraining the formation pathway of 2MASS J0249-0557c.

% However, if the planet formed through this mechanism, the $^{12}$CO/$^{13}$CO ratio should be lower than the abundance in its birth environment and thus lower than that of the benchmark brown dwarfs as more $^{13}$CO than $^{12}$CO is trapped in ice. This prediction is inconsistent with our measurements. 

\subsection{Abundances of the $\beta$ Pic YMG}
\label{sec:bpmg}
\begin{table*}
\centering
\begin{tabular}{lccc}
\hline
Name & C/O  & [M/H] & Source\\
\hline
51 Eri        & $0.54\pm0.14$ & $0.03 \pm 0.08$ & \cite{Baburaj2025EriCO}\\
V* AF Lep     & $0.58^{+0.20}_{-0.15}$ & $0.39 \pm 0.09$ & Hypatia\\
HD 181327     & $0.62\pm0.08$  & $-0.08\pm0.06$ & \cite{Reggiani2024betapic}\\
% $\beta$ Pic & -- & $-0.20\pm0.11$ & \cite{Saffe2021starrv51eribpic} \\
PZ Tel & $0.28\pm0.05$ & $-0.04\pm0.07$ & \cite{Baburaj2025bpicpztel} \\
$\beta$ Pic & $0.22\pm0.06$ & $-0.25\pm0.06$ & \cite{Baburaj2025bpicpztel} \\
\hline
PSO J318      &$0.789\pm0.003$          &$0.31\pm0.01$
& \cite{Molliere2025PSO318}\\
$\beta$ Pic b & $0.35^{+0.02}_{-0.03}$ & $0.295^{+0.183}_{-0.155}$ & \cite{Reggiani2024betapic} \\
  & $0.39^{+0.10}_{-0.06}$ & $-0.12^{+0.20}_{-0.19}$& \cite{Worthen2024bpicb} ATMO \\
  & $0.43^{+0.04}_{-0.03}$ & $0.68^{+0.11}_{-0.08}$ & \cite{Collaboration2020bpicb} pRT GRAVITY + GPI  \\
AF Lep b & $0.55\pm0.10$ & $0.75\pm0.25$ & \cite{Balmer2025aflefb}, \cite{Denis2025aflepb}\\
& $0.61^{+0.05}_{-0.09}$ & $0.60^{+0.08}_{-0.13}$ & \cite{Palma2024aflepb} Restrictive Prior\\
& $0.65^{+0.07}_{-0.09}$ & $1.67^{+0.17}_{-0.21}$ & \cite{Franson2024aflepb} \\
51 Eri b & $0.38\pm0.09$ &  $ 0.26\pm0.30$ & \cite{Brown2023Erib}\\
& $0.97^{+0.09}_{-0.20}$ & $-0.04^{+0.95}_{-0.49}$ & \cite{Whiteford2023Erib} SPHERE Y, J, H and GPI K1, K2\\
& $0.65^{+0.05}_{-0.08}$ & $0.65\pm0.15$ & \cite{Balmer2025Erib}\\
% HD 197481     & --                       & $0.01 \pm 0.04$ & Hypatia & Stellar\\
% BD-13 6424    & --                      & $-0.05 \pm 0.04$ & Hypatia & Stellar\\
% V* V1005 Ori  & --                        & $0.08 \pm 0.04$ & Hypatia & Stellar\\
% PM J05019+0108& --                        & $-0.20 \pm 0.04$ & Hypatia & Stellar\\
% HD 157587     & --                        & $-0.21 \pm 0.04$ & Hypatia & Stellar\\
\hline
\end{tabular}
\caption{Known C/O ratio and metallicity of $\beta$ Pic YMG members and their sources. For sources that explicitly report [C/H], we adopt [M/H] $\approx$ [C/H], since our retrievals are most sensitive to carbon-bearing species. When only an overall metallicity parameter is reported (all such cases correspond to planetary studies, often reported as [Fe/H] or [M/H]), we adopt the overall metallicity as [M/H]. As there are many studies on the chemical abundances of the three planets, $\beta$ Pic b, 51 Eri b, and AF Lep b, we summarize three representative results for each.}
\label{tab:bpmg}
\end{table*}

\begin{figure}
    \centering   \includegraphics[width=0.98\linewidth]{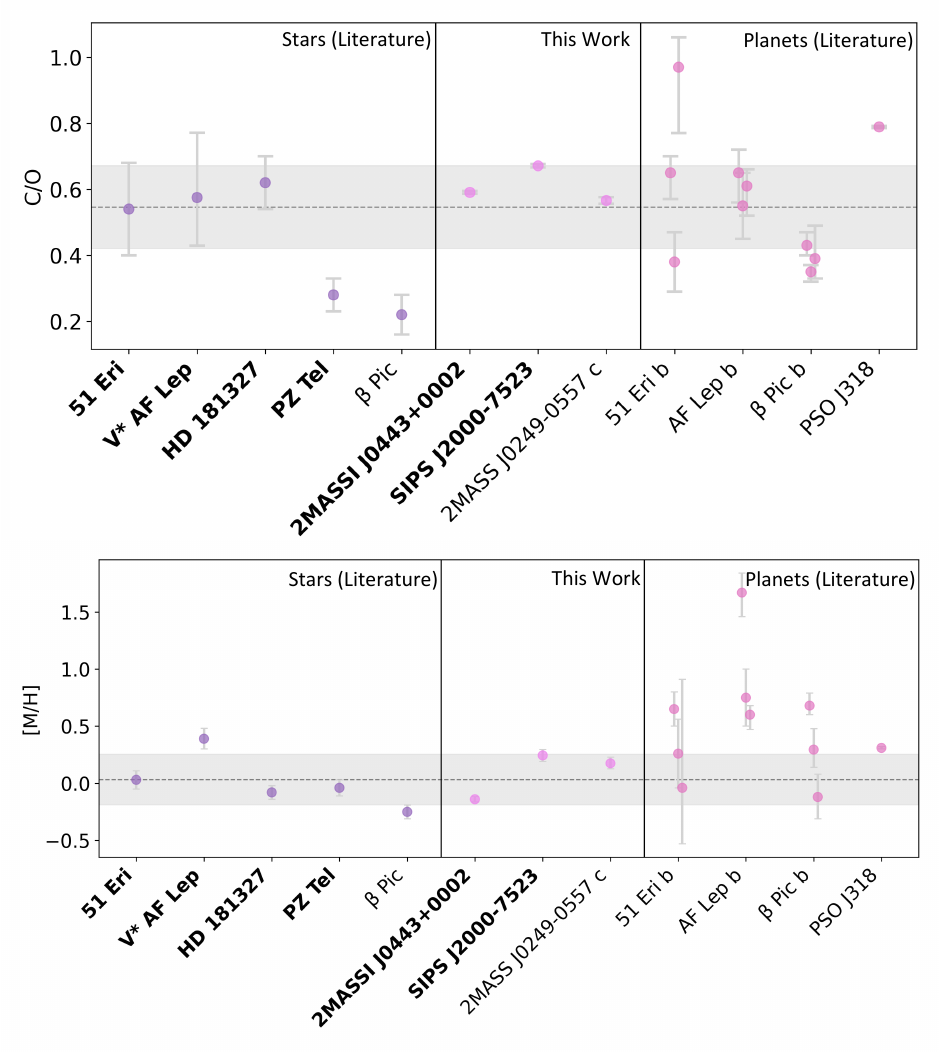}
    \caption{C/O ratio and metallicity of objects in the $\beta$ Pic Moving Group and vicinity. Values and sources are shown in Table \ref{tab:bpmg}. Each panel is divided into three regions: stellar members from literature, planetary members from literature, and this work. Benchmark members are bolded. The dotted lines are the means of benchmark members, and the gray regions cover one standard deviation above and below the mean. Similar to many literature values, the reported uncertainties in this work do not account for all possible sources of uncertainties, such as correlated noise and model dependence. However, the small error bars in our results reflect the high quality of our data.}
    \label{fig:bpmg_compare}
\end{figure}

The $\beta$ Pic YMG has four substellar members with atmospheric constraints of C/O ratio and metallicity: $\beta$ Pic b, Af Lep b, 51 Eri b, and PSO 318. We compare our retrieved C/O ratio and metallicity to these substellar members, as well as stellar members shown in Table \ref{tab:bpmg} and Figure \ref{fig:bpmg_compare}.

The reported planetary measurements from the literature are model sensitive and affected by data quality. As we can see for the three well-studied planets in Figure \ref{fig:bpmg_compare}, different studies present distinct results for C/O and metallicity that can be inconsistent with each other because they were derived from different self-consistent model grids, self-consistent models, or retrievals. Similarly, the uncertainties reported for our targets are conditional upon the specific modeling framework adopted in this study.

When we compare all the benchmark objects (isolated brown dwarfs and all stars except for the A-type star $\beta$ Pic) that are bolded in Figure \ref{fig:bpmg_compare}, we can see from the top panel that the C/O ratios are mostly consistent with each other and within the solar range of $0.59\pm0.08$ \citep{Asplund2021solar}. The average value of benchmark objects is C/O = $0.55\pm0.13$, which is shown with the dashed line and shaded region. The error bar is the standard deviation of the sample and thus does not reflect the uncertainties of individual measurements. This supports solar-like C/O abundances for the $\beta$ Pic YMG. 

The bottom panel of Figure \ref{fig:bpmg_compare} shows that the metallicity values of benchmark objects are centered around the solar value, with an average value of $0.067\pm 0.188$. The error bar is once again the standard deviation of the sample. This scatter is likely caused by the model-dependent nature of metallicity, degeneracy of metallicity with other retrieval variables, and different solar normalization standards used across different works. Overall, the average metallicity is consistent with solar metallicity. 

Planetary members can exhibit atmospheric C/O ratios and metallicities that deviate from their natal environment baselines, as seen in Figure \ref{fig:bpmg_compare}. This deviation reflects the relative abundances in gas and solids at their formation locations and can be used to determine the most viable planet formation pathway.

There are two other members of the $\beta$ Pic YMG with previously characterized $^{12}$CO/$^{13}$CO ratios. One is the isolated super-Jupiter PSO 318. The $^{12}$CO/$^{13}$CO ratio of PSO 318 has been determined to be $45^{+5}_{-4}$, but this constraint suffers from systematic effects \citep{Molliere2025PSO318}. The other is $\beta$ Pic b, for which the isotopic ratio $^{12}$CO/$^{13}$CO has been tentatively measured as $13.2^{+3.8}_{-2.2}$, with the tentative nature reflecting telluric residuals \citep{Ravet2025bpicisotope}. Additionally, $\beta$ Pic b is not isolated, so its $^{12}$CO/$^{13}$CO value cannot be used to establish the baseline of the YMG. As we do not have a robust constraint for the $^{12}$CO/$^{13}$CO ratio of SIPS J2000-7523, the only benchmark object of the $\beta$ Pic YMG with a robust constraint of the $^{12}$CO/$^{13}$CO ratio is 2MASSI J0443+0002, the value of which is $70\pm5$, consistent with the ISM value of $68\pm15$ \citep{Langer1993ISMc13, Milam2005ISMc13}. This finding is in agreement with the expectation that young objects should inherit the ISM value of the CO isotopologue ratio \citep{Zhang2021BDc13}. As time progresses, objects such as massive stars, asymptotic giant branch stars, and novae contribute $^{13}$C to the ISM \citep[e.p.][]{Romano2017cno,Romano2019cno,Romano2022A&ARvcno,Gonalezz202Mdwarfc13}, so young objects like 2MASSI J0443+0002 are expected to have lower $^{12}$CO/$^{13}$CO values than the solar value of 89 \citep[e.g.][]{Meibom2007solarc13, Woods2009solarC13}.

If we assume that 2MASS J0249-0557c formed through a top-down pathway -- its most likely formation mechanism -- then the $^{12}$CO/$^{13}$CO ratio of 2MASS J0249-0557c should also be indicative of its birth environment. The $^{12}$CO/$^{13}$CO ratio of 2MASS J0249-0557c is $95^{+23}_{-17}$, which is consistent with that of 2MASSI J0443+0002 as shown in Figure \ref{fig:combined_corner}. However, it is closer to the solar value than the ISM value of the CO isotopologue ratio. Other substellar objects have also been shown to demonstrate a depletion of $^{13}$CO relative to the ISM value, although, as is the case for 2MASS J0249-0557c, some are consistent with the ISM value within uncertainties \citep{Costes2024c13, Regt2024eso1, Gonalezz2024eso2, Grasser2025eso8}. It is unclear whether there is a fundamental scatter or measurement and model uncertainty. More precise constraints and studies on additional targets can potentially resolve this question.

The $^{12}$CO/$^{13}$CO ratio of 2MASS J0249-0557c is less well-constrained than that of 2MASSI J0443+0002, and the detection of $^{13}$CO is less significant for this target. Additionally, it is not an isolated benchmark object. Therefore, 2MASSI J0443+0002's $^{12}$CO/$^{13}$CO ratio value is a more reliable indicator of the overall $^{12}$CO/$^{13}$CO ratio of the $\beta$ Pic YMG. With only one benchmark object in the $\beta$ Pic YMG having a relatively well-constrained $^{12}$CO/$^{13}$CO ratio, it is premature to draw conclusions about the baseline CO isotopologue ratio of the entire moving group. To establish a baseline of $^{12}$CO/$^{13}$CO ratio for the $\beta$ Pic YMG, future work is needed.

\section{Conclusion}
\label{sec:conclusion}

We obtained high-resolution K-band spectra from VLT/CRIRES+ of three targets in the $\beta$ Pic YMG and vicinity and conducted atmospheric retrieval studies using high-resolution spectra and archival photometry for these targets. Of our three targets, one of them, 2MASS J0249-0557 c, is a $\sim12M_{\mathrm{Jup}}$ planet orbiting a pair of binary brown dwarfs, and the other two, 2MASSI J0443+0002 and SIPS J2000-7523, are isolated benchmark brown dwarfs.

In our retrieval process, we found that the correlation between $\log g$ and metallicity prevents us from deriving model-independent metallicity constraints. Adding mass priors from evolutionary models helps establish more reliable measurements. 

For all three targets, we found that the C/O ratio is consistent with the range for solar values. We were able to detect $^{13}$CO in the planet 2MASS J0249-0557 c and brown dwarf 2MASSI J0443+0002 with 4 S/N and 6 S/N peaks in the cross-correlation function. We found that the $^{12}$CO/$^{13}$CO ratio of 2MASSI J0443+0002 is consistent with the ISM value, whereas that of 2MASS J0249-0557 c is closer to the solar value, though still consistent with 2MASSI J0443+0002 and the ISM value within uncertainties.

The consistency of 2MASS J0249-0557 c's abundances with benchmark brown dwarfs and its wide separation from its host suggest a top-down mechanism, likely through cloud fragmentation.

Comparing the values of C/O and metallicity of the two benchmark brown dwarfs with stellar members of the $\beta$ Pic YMG with known C/O and metallicity, we were able to establish a baseline for the abundances of the $\beta$ Pic YMG: C/O = $0.599\pm0.044$ and [M/H] = $0.041\pm0.212$, both centered around solar values. The error bars represent the scatter of values for $\beta$ Pic YMG objects and do not reflect the uncertainty of each measurement. 

We also derive rigorous constraints on $^{12}$CO/$^{13}$CO for two substellar objects in the $\beta$ Pic YMG, 2MASSI J0443+0002 and SIPS J2000-7523. Future work is needed to draw conclusions about the baseline value of the CO isotopologue ratio for the YMG as a whole, given the sparsity of measurements. Establishing a baseline for both elemental and isotopic abundances within the $\beta$ Pic YMG not only informs the chemical environment of the moving group as a whole, but also provides a benchmark for investigating mechanisms of planet formation, as demonstrated with 2MASS J0249-0557 c.

\section{Acknowledgements}

Y.L. acknowledges support from the Caltech Summer Undergraduate Research Fellowships. Y.Z. acknowledges the support from the Heising-Simons Foundation 51 Pegasi b Fellowship (grant \#2023-4298). J.W.X is thankful for support from the Heising-Simons Foundation 51 Pegasi b Fellowship (grant \#2025-5887). 

This work has benefited from the use of the \textit{Grace} computing cluster at the Yale Center for Research Computing (YCRC) and the Caltech High-Performance Cluster.

\bibliography{ref.bib}

@ARTICLE{Ackerman2001cloud,
       author = {{Ackerman}, Andrew S. and {Marley}, Mark S.},
        title = "{Precipitating Condensation Clouds in Substellar Atmospheres}",
      journal = {\apj},
         year = 2001,
        month = aug,
       volume = {556},
       number = {2},
        pages = {872-884},
          doi = {10.1086/321540},
archivePrefix = {arXiv},
       eprint = {astro-ph/0103423},
 primaryClass = {astro-ph},
       adsurl = {https://ui.adsabs.harvard.edu/abs/2001ApJ...556..872A}
}

@ARTICLE{Allard2001ames-dusty,
       author = {{Allard}, France and {Hauschildt}, Peter H. and {Alexander}, David R. and {Tamanai}, Akemi and {Schweitzer}, Andreas},
        title = "{The Limiting Effects of Dust in Brown Dwarf Model Atmospheres}",
      journal = {\apj},
         year = 2001,
        month = jul,
       volume = {556},
       number = {1},
        pages = {357-372},
          doi = {10.1086/321547},
archivePrefix = {arXiv},
       eprint = {astro-ph/0104256},
 primaryClass = {astro-ph},
       adsurl = {https://ui.adsabs.harvard.edu/abs/2001ApJ...556..357A}
}

@ARTICLE{Anders1989solarc13,
       author = {{Anders}, E. and {Grevesse}, N.},
        title = "{Abundances of the elements: Meteoritic and solar}",
      journal = {\gca},
         year = 1989,
        month = jan,
       volume = {53},
       number = {1},
        pages = {197-214},
          doi = {10.1016/0016-7037(89)90286-X},
       adsurl = {https://ui.adsabs.harvard.edu/abs/1989GeCoA..53..197A}
}

@ARTICLE{Asplund2021solar,
       author = {{Asplund}, M. and {Amarsi}, A.~M. and {Grevesse}, N.},
        title = "{The chemical make-up of the Sun: A 2020 vision}",
      journal = {\aap},
         year = 2021,
        month = sep,
       volume = {653},
          eid = {A141},
        pages = {A141},
          doi = {10.1051/0004-6361/202140445},
archivePrefix = {arXiv},
       eprint = {2105.01661},
 primaryClass = {astro-ph.SR},
       adsurl = {https://ui.adsabs.harvard.edu/abs/2021A&A...653A.141A}
}

@ARTICLE{Allard2019Na,
       author = {{Allard}, N.~F. and {Spiegelman}, F. and {Leininger}, T. and {Molliere}, P.},
        title = "{New study of the line profiles of sodium perturbed by H$_{2}$}",
      journal = {\aap},
         year = 2019,
        month = aug,
       volume = {628},
          eid = {A120},
        pages = {A120},
          doi = {10.1051/0004-6361/201935593},
archivePrefix = {arXiv},
       eprint = {1908.01989},
 primaryClass = {astro-ph.SR},
       adsurl = {https://ui.adsabs.harvard.edu/abs/2019A&A...628A.120A}
}

@ARTICLE{Azzam2016H2S,
       author = {{Azzam}, Ala'a. A.~A. and {Tennyson}, Jonathan and {Yurchenko}, Sergei N. and {Naumenko}, Olga V.},
        title = "{ExoMol molecular line lists - XVI. The rotation-vibration spectrum of hot H$_{2}$S}",
      journal = {\mnras},
         year = 2016,
        month = aug,
       volume = {460},
       number = {4},
        pages = {4063-4074},
          doi = {10.1093/mnras/stw1133},
archivePrefix = {arXiv},
       eprint = {1607.00499},
 primaryClass = {astro-ph.EP},
       adsurl = {https://ui.adsabs.harvard.edu/abs/2016MNRAS.460.4063A}
}

@ARTICLE{Baraffe2015bhac15,
       author = {{Baraffe}, Isabelle and {Homeier}, Derek and {Allard}, France and {Chabrier}, Gilles},
        title = "{New evolutionary models for pre-main sequence and main sequence low-mass stars down to the hydrogen-burning limit}",
      journal = {\aap},
         year = 2015,
        month = may,
       volume = {577},
          eid = {A42},
        pages = {A42},
          doi = {10.1051/0004-6361/201425481},
archivePrefix = {arXiv},
       eprint = {1503.04107},
 primaryClass = {astro-ph.SR},
       adsurl = {https://ui.adsabs.harvard.edu/abs/2015A&A...577A..42B}
}

@ARTICLE{Baburaj2025bpicpztel,
       author = {{Baburaj}, Aneesh and {Konopacky}, Quinn M. and {Theissen}, Christopher A. and {Gerasimov}, Roman and {Hoch}, Kielan K.~W.},
        title = "{A High-Resolution Spectroscopic Survey of Directly Imaged Companion Hosts: II. Diversity in C/O Ratios among Host Stars}",
      journal = {arXiv e-prints},
         year = 2025,
        month = oct,
          eid = {arXiv:2510.17774},
        pages = {arXiv:2510.17774},
          doi = {10.48550/arXiv.2510.17774},
archivePrefix = {arXiv},
       eprint = {2510.17774},
 primaryClass = {astro-ph.EP},
       adsurl = {https://ui.adsabs.harvard.edu/abs/2025arXiv251017774B}
}

@ARTICLE{Balmer2025aflefb,
       author = {{Balmer}, William O. and {Franson}, Kyle and {Chomez}, Antoine and {Pueyo}, Laurent and {Stolker}, Tomas and {Lacour}, Sylvestre and {Nowak}, Mathias and {Nasedkin}, Evert and {Bonse}, Markus J. and {Thorngren}, Daniel and {Palma-Bifani}, Paulina and {Molli{\`e}re}, Paul and {Wang}, Jason J. and {Zhang}, Zhoujian and {Chavez}, Amanda and {Kammerer}, Jens and {Blunt}, Sarah and {Bowler}, Brendan P. and {Bonnefoy}, Mickael and {Brandner}, Wolfgang and {Charnay}, Benjamin and {Chauvin}, Gael and {Henning}, Th. and {Lagrange}, A. -M. and {Pourr{\'e}}, Nicolas and {Rickman}, Emily and {De Rosa}, Robert and {Vigan}, Arthur and {Winterhalder}, Thomas},
        title = "{VLTI/GRAVITY Observations of AF Lep b: Preference for Circular Orbits, Cloudy Atmospheres, and a Moderately Enhanced Metallicity}",
      journal = {\aj},
         year = 2025,
        month = jan,
       volume = {169},
       number = {1},
          eid = {30},
        pages = {30},
          doi = {10.3847/1538-3881/ad9265},
archivePrefix = {arXiv},
       eprint = {2411.05917},
 primaryClass = {astro-ph.EP},
       adsurl = {https://ui.adsabs.harvard.edu/abs/2025AJ....169...30B}
}

@ARTICLE{Balmer2025Erib,
       author = {{Balmer}, William O. and {Kammerer}, Jens and {Pueyo}, Laurent and {Perrin}, Marshall D. and {Girard}, Julien H. and {Leisenring}, Jarron M. and {Lawson}, Kellen and {Dennen}, Henry and {van der Marel}, Roeland P. and {Beichman}, Charles A. and {Bryden}, Geoffrey and {Llop-Sayson}, Jorge and {Valenti}, Jeff A. and {Lothringer}, Joshua D. and {Lewis}, Nikole K. and {M{\^a}lin}, Mathilde and {Rebollido}, Isabel and {Rickman}, Emily and {Hoch}, Kielan K.~W. and {Soummer}, R{\'e}mi and {Clampin}, Mark and {Mountain}, C. Matt},
        title = "{JWST-TST High Contrast: Living on the Wedge, or, NIRCam Bar Coronagraphy Reveals CO$_{2}$ in the HR 8799 and 51 Eri Exoplanets' Atmospheres}",
      journal = {\aj},
         year = 2025,
        month = apr,
       volume = {169},
       number = {4},
          eid = {209},
        pages = {209},
          doi = {10.3847/1538-3881/adb1c6},
archivePrefix = {arXiv},
       eprint = {2503.13608},
 primaryClass = {astro-ph.EP},
       adsurl = {https://ui.adsabs.harvard.edu/abs/2025AJ....169..209B}
}

@ARTICLE{Baburaj2025EriCO,
       author = {{Baburaj}, Aneesh and {Konopacky}, Quinn M. and {Theissen}, Christopher A. and {Peacock}, Sarah and {Huseby}, Lori and {Fulton}, Benjamin J. and {Gerasimov}, Roman and {Barman}, Travis S. and {Hoch}, Kielan K.~W.},
        title = "{A High-resolution Spectroscopic Survey of Directly Imaged Companion Hosts. I. Determination of Diagnostic Stellar Abundances for Planet Formation and Composition}",
      journal = {\aj},
         year = 2025,
        month = feb,
       volume = {169},
       number = {2},
          eid = {55},
        pages = {55},
          doi = {10.3847/1538-3881/ad8dfc},
archivePrefix = {arXiv},
       eprint = {2409.14239},
 primaryClass = {astro-ph.EP},
       adsurl = {https://ui.adsabs.harvard.edu/abs/2025AJ....169...55B}
}

@ARTICLE{Borysow1988h2he,
       author = {{Borysow}, Jacek and {Frommhold}, Lothar and {Birnbaum}, George},
        title = "{Collision-induced Rototranslational Absorption Spectra of H 2-He Pairs at Temperatures from 40 to 3000 K}",
      journal = {\apj},
         year = 1988,
        month = mar,
       volume = {326},
        pages = {509},
          doi = {10.1086/166112},
       adsurl = {https://ui.adsabs.harvard.edu/abs/1988ApJ...326..509B}
}

@ARTICLE{Borysow1989h2hea,
       author = {{Borysow}, Aleksandra and {Frommhold}, Lothar and {Moraldi}, Massimo},
        title = "{Collision-induced Infrared Spectra of H 2-He Pairs Involving 0 1 Vibrational Transitions and Temperatures from 18 to 7000 K}",
      journal = {\apj},
         year = 1989,
        month = jan,
       volume = {336},
        pages = {495},
          doi = {10.1086/167027},
       adsurl = {https://ui.adsabs.harvard.edu/abs/1989ApJ...336..495B}
}

@ARTICLE{Borysow1989h2heb,
       author = {{Borysow}, Aleksandra and {Frommhold}, Lothar},
        title = "{Collision-induced Infrared Spectra of H 2-He Pairs at Temperatures from 18 to 7000 K. II. Overtone and Hot Bands}",
      journal = {\apj},
         year = 1989,
        month = jun,
       volume = {341},
        pages = {549},
          doi = {10.1086/167515},
       adsurl = {https://ui.adsabs.harvard.edu/abs/1989ApJ...341..549B}
}

@ARTICLE{Borysow2001h2h2,
       author = {{Borysow}, Aleksandra and {Jorgensen}, Uffe G. and {Fu}, Yi},
        title = "{High-temperature (1000-7000 K) collision-induced absorption of H''2 pairs computed from the first principles, with application to cool and dense stellar atmospheres}",
      journal = {\jqsrt},
         year = 2001,
        month = feb,
       volume = {68},
        pages = {235-255},
          doi = {10.1016/S0022-4073(00)00023-6},
       adsurl = {https://ui.adsabs.harvard.edu/abs/2001JQSRT..68..235B}
}

@ARTICLE{Borysow2002h2h2,
       author = {{Borysow}, A.},
        title = "{Collision-induced absorption coefficients of H$_{2}$ pairs at temperatures from 60 K to 1000 K}",
      journal = {\aap},
         year = 2002,
        month = aug,
       volume = {390},
        pages = {779-782},
          doi = {10.1051/0004-6361:20020555},
       adsurl = {https://ui.adsabs.harvard.edu/abs/2002A&A...390..779B}
}

@ARTICLE{Bonse2025aflepb,
       author = {{Bonse}, Markus J. and {Gebhard}, Timothy D. and {Dannert}, Felix A. and {Absil}, Olivier and {Cantalloube}, Faustine and {Christiaens}, Valentin and {Cugno}, Gabriele and {Garvin}, Emily O. and {Hayoz}, Jean and {Kasper}, Markus and {Matthews}, Elisabeth and {Sch{\"o}lkopf}, Bernhard and {Quanz}, Sascha P.},
        title = "{Use the 4S (Signal-Safe Speckle Subtraction): Explainable Machine Learning Reveals the Giant Exoplanet AF Lep b in High-contrast Imaging Data from 2011}",
      journal = {\aj},
         year = 2025,
        month = apr,
       volume = {169},
       number = {4},
          eid = {194},
        pages = {194},
          doi = {10.3847/1538-3881/adab79},
archivePrefix = {arXiv},
       eprint = {2406.01809},
 primaryClass = {astro-ph.EP},
       adsurl = {https://ui.adsabs.harvard.edu/abs/2025AJ....169..194B}
}

@ARTICLE{Brown2023Erib,
       author = {{Brown-Sevilla}, S.~B. and {Maire}, A. -L. and {Molli{\`e}re}, P. and {Samland}, M. and {Feldt}, M. and {Brandner}, W. and {Henning}, Th. and {Gratton}, R. and {Janson}, M. and {Stolker}, T. and {Hagelberg}, J. and {Zurlo}, A. and {Cantalloube}, F. and {Boccaletti}, A. and {Bonnefoy}, M. and {Chauvin}, G. and {Desidera}, S. and {D'Orazi}, V. and {Lagrange}, A. -M. and {Langlois}, M. and {Menard}, F. and {Mesa}, D. and {Meyer}, M. and {Pavlov}, A. and {Petit}, C. and {Rochat}, S. and {Rouan}, D. and {Schmidt}, T. and {Vigan}, A. and {Weber}, L.},
        title = "{Revisiting the atmosphere of the exoplanet 51 Eridani b with VLT/SPHERE}",
      journal = {\aap},
         year = 2023,
        month = may,
       volume = {673},
          eid = {A98},
        pages = {A98},
          doi = {10.1051/0004-6361/202244826},
archivePrefix = {arXiv},
       eprint = {2211.14330},
 primaryClass = {astro-ph.EP},
       adsurl = {https://ui.adsabs.harvard.edu/abs/2023A&A...673A..98B}
}

@ARTICLE{Borisov2023starrvbpic,
       author = {{Borisov}, Sviatoslav B. and {Chilingarian}, Igor V. and {Rubtsov}, Evgenii V. and {Ledoux}, C{\'e}dric and {Melo}, Claudio and {Grishin}, Kirill A. and {Katkov}, Ivan Yu. and {Goradzhanov}, Vladimir S. and {Afanasiev}, Anton V. and {Kasparova}, Anastasia V. and {Saburova}, Anna S.},
        title = "{New Generation Stellar Spectral Libraries in the Optical and Near-infrared. I. The Recalibrated UVES-POP Library for Stellar Population Synthesis}",
      journal = {\apjs},
         year = 2023,
        month = may,
       volume = {266},
       number = {1},
          eid = {11},
        pages = {11},
          doi = {10.3847/1538-4365/acc321},
archivePrefix = {arXiv},
       eprint = {2211.09130},
 primaryClass = {astro-ph.IM},
       adsurl = {https://ui.adsabs.harvard.edu/abs/2023ApJS..266...11B}
}

@ARTICLE{Bryan2020vsini,
       author = {{Bryan}, Marta L. and {Ginzburg}, Sivan and {Chiang}, Eugene and {Morley}, Caroline and {Bowler}, Brendan P. and {Xuan}, Jerry W. and {Knutson}, Heather A.},
        title = "{As the Worlds Turn: Constraining Spin Evolution in the Planetary-mass Regime}",
      journal = {\apj},
         year = 2020,
        month = dec,
       volume = {905},
       number = {1},
          eid = {37},
        pages = {37},
          doi = {10.3847/1538-4357/abc0ef},
archivePrefix = {arXiv},
       eprint = {2010.07315},
 primaryClass = {astro-ph.EP},
       adsurl = {https://ui.adsabs.harvard.edu/abs/2020ApJ...905...37B}
}

@ARTICLE{Boss1997instability,
       author = {{Boss}, A.~P.},
        title = "{Giant planet formation by gravitational instability.}",
      journal = {Science},
         year = 1997,
        month = jan,
       volume = {276},
        pages = {1836-1839},
          doi = {10.1126/science.276.5320.1836},
       adsurl = {https://ui.adsabs.harvard.edu/abs/1997Sci...276.1836B}
}

@ARTICLE{Bowler2023starrv51eri,
       author = {{Bowler}, Brendan P. and {Tran}, Quang H. and {Zhang}, Zhoujian and {Morgan}, Marvin and {Ashok}, Katelyn B. and {Blunt}, Sarah and {Bryan}, Marta L. and {Evans}, Analis E. and {Franson}, Kyle and {Huber}, Daniel and {Nagpal}, Vighnesh and {Wu}, Ya-Lin and {Zhou}, Yifan},
        title = "{Rotation Periods, Inclinations, and Obliquities of Cool Stars Hosting Directly Imaged Substellar Companions: Spin-Orbit Misalignments Are Common}",
      journal = {\aj},
         year = 2023,
        month = apr,
       volume = {165},
       number = {4},
          eid = {164},
        pages = {164},
          doi = {10.3847/1538-3881/acbd34},
archivePrefix = {arXiv},
       eprint = {2301.04692},
 primaryClass = {astro-ph.EP},
       adsurl = {https://ui.adsabs.harvard.edu/abs/2023AJ....165..164B}
}

@ARTICLE{Buchner2014Bayesian,
       author = {{Buchner}, J. and {Georgakakis}, A. and {Nandra}, K. and {Hsu}, L. and {Rangel}, C. and {Brightman}, M. and {Merloni}, A. and {Salvato}, M. and {Donley}, J. and {Kocevski}, D.},
        title = "{X-ray spectral modelling of the AGN obscuring region in the CDFS: Bayesian model selection and catalogue}",
      journal = {\aap},
         year = 2014,
        month = apr,
       volume = {564},
          eid = {A125},
        pages = {A125},
          doi = {10.1051/0004-6361/201322971},
archivePrefix = {arXiv},
       eprint = {1402.0004},
 primaryClass = {astro-ph.HE},
       adsurl = {https://ui.adsabs.harvard.edu/abs/2014A&A...564A.125B}
}

@ARTICLE{Wang2026_cd35,
       author = {{Wang}, Gavin and {Xuan}, Jerry W. and {Gonz{\'a}lez Picos}, Dar{\'\i}o and {Zhang}, Zhoujian and {Zhang}, Yapeng and {Mawet}, Dimitri and {Hsu}, Chih-Chun and {Wang}, Jason J. and {Blake}, Geoffrey A. and {Ruffio}, Jean-Baptiste and {Horstman}, Katelyn and {Sappey}, Ben and {Xin}, Yinzi and {Finnerty}, Luke and {Echeverri}, Daniel and {Jovanovic}, Nemanja and {Baker}, Ashley and {Bartos}, Randall and {Calvin}, Benjamin and {Cetre}, Sylvain and {Delorme}, Jacques-Robert and {Doppmann}, Gregory W. and {Fitzgerald}, Michael P. and {Liberman}, Joshua and {L{\'o}pez}, Ronald A. and {Morris}, Evan and {Pezzato-Rovner}, Jacklyn and {Phillips}, Caprice L. and {Schofield}, Tobias and {Skemer}, Andrew and {Wallace}, J. Kent and {Wang}, Ji},
        title = "{Chemical and Isotopic Homogeneity between the L Dwarf CD-35 2722 B and Its Early M Host Star}",
      journal = {\apj},
         year = 2026,
        month = feb,
       volume = {997},
       number = {2},
          eid = {195},
        pages = {195},
          doi = {10.3847/1538-4357/ae232f},
       adsurl = {https://ui.adsabs.harvard.edu/abs/2026ApJ...997..195W}
}

@ARTICLE{Calamari2024mgsi,
       author = {{Calamari}, Emily and {Faherty}, Jacqueline K. and {Visscher}, Channon and {Gemma}, Marina E. and {Burningham}, Ben and {Rothermich}, Austin},
        title = "{Predicting Cloud Conditions in Substellar Mass Objects Using Ultracool Dwarf Companions}",
      journal = {\apj},
         year = 2024,
        month = mar,
       volume = {963},
       number = {1},
          eid = {67},
        pages = {67},
          doi = {10.3847/1538-4357/ad1f6d},
archivePrefix = {arXiv},
       eprint = {2401.11038},
 primaryClass = {astro-ph.SR},
       adsurl = {https://ui.adsabs.harvard.edu/abs/2024ApJ...963...67C}
}

@ARTICLE{Chinchilla2021Halpha,
       author = {{Chinchilla}, P. and {B{\'e}jar}, V.~J.~S. and {Lodieu}, N. and {Zapatero Osorio}, M.~R. and {Gauza}, B.},
        title = "{Strong H{\ensuremath{\alpha}} emission in the young planetary mass companion 2MASS J0249-0557 c}",
      journal = {\aap},
         year = 2021,
        month = jan,
       volume = {645},
          eid = {A17},
        pages = {A17},
          doi = {10.1051/0004-6361/202038731},
archivePrefix = {arXiv},
       eprint = {2011.10002},
 primaryClass = {astro-ph.EP},
       adsurl = {https://ui.adsabs.harvard.edu/abs/2021A&A...645A..17C}
}

@ARTICLE{Clayton2004stardust,
       author = {{Clayton}, Donald D. and {Nittler}, Larry R.},
        title = "{Astrophysics with Presolar Stardust}",
      journal = {\araa},
         year = 2004,
        month = sep,
       volume = {42},
       number = {1},
        pages = {39-78},
          doi = {10.1146/annurev.astro.42.053102.134022},
       adsurl = {https://ui.adsabs.harvard.edu/abs/2004ARA&A..42...39C}
}

@ARTICLE{Chabrier2023atmo,
       author = {{Chabrier}, Gilles and {Baraffe}, Isabelle and {Phillips}, Mark and {Debras}, Florian},
        title = "{Impact of a new H/He equation of state on the evolution of massive brown dwarfs. New determination of the hydrogen burning limit}",
      journal = {\aap},
         year = 2023,
        month = mar,
       volume = {671},
          eid = {A119},
        pages = {A119},
          doi = {10.1051/0004-6361/202243832},
archivePrefix = {arXiv},
       eprint = {2212.07153},
 primaryClass = {astro-ph.SR},
       adsurl = {https://ui.adsabs.harvard.edu/abs/2023A&A...671A.119C}
}

@ARTICLE{Coles2019NH3,
       author = {{Coles}, Phillip A. and {Yurchenko}, Sergei N. and {Tennyson}, Jonathan},
        title = "{ExoMol molecular line lists - XXXV. A rotation-vibration line list for hot ammonia}",
      journal = {\mnras},
         year = 2019,
        month = dec,
       volume = {490},
       number = {4},
        pages = {4638-4647},
          doi = {10.1093/mnras/stz2778},
archivePrefix = {arXiv},
       eprint = {1911.10369},
 primaryClass = {astro-ph.SR},
       adsurl = {https://ui.adsabs.harvard.edu/abs/2019MNRAS.490.4638C}
}

@ARTICLE{Collaboration2020bpicb,
       author = {{GRAVITY Collaboration} and {Nowak}, M. and {Lacour}, S. and {Molli{\`e}re}, P. and {Wang}, J. and {Charnay}, B. and {van Dishoeck}, E.~F. and {Abuter}, R. and {Amorim}, A. and {Berger}, J.~P. and {Beust}, H. and {Bonnefoy}, M. and {Bonnet}, H. and {Brandner}, W. and {Buron}, A. and {Cantalloube}, F. and {Collin}, C. and {Chapron}, F. and {Cl{\'e}net}, Y. and {Coud{\'e} Du Foresto}, V. and {de Zeeuw}, P.~T. and {Dembet}, R. and {Dexter}, J. and {Duvert}, G. and {Eckart}, A. and {Eisenhauer}, F. and {F{\"o}rster Schreiber}, N.~M. and {F{\'e}dou}, P. and {Garcia Lopez}, R. and {Gao}, F. and {Gendron}, E. and {Genzel}, R. and {Gillessen}, S. and {Hau{\ss}mann}, F. and {Henning}, T. and {Hippler}, S. and {Hubert}, Z. and {Jocou}, L. and {Kervella}, P. and {Lagrange}, A. -M. and {Lapeyr{\`e}re}, V. and {Le Bouquin}, J. -B. and {L{\'e}na}, P. and {Maire}, A. -L. and {Ott}, T. and {Paumard}, T. and {Paladini}, C. and {Perraut}, K. and {Perrin}, G. and {Pueyo}, L. and {Pfuhl}, O. and {Rabien}, S. and {Rau}, C. and {Rodr{\'\i}guez-Coira}, G. and {Rousset}, G. and {Scheithauer}, S. and {Shangguan}, J. and {Straub}, O. and {Straubmeier}, C. and {Sturm}, E. and {Tacconi}, L.~J. and {Vincent}, F. and {Widmann}, F. and {Wieprecht}, E. and {Wiezorrek}, E. and {Woillez}, J. and {Yazici}, S. and {Ziegler}, D.},
        title = "{Peering into the formation history of {\ensuremath{\beta}} Pictoris b with VLTI/GRAVITY long-baseline interferometry}",
      journal = {\aap},
         year = 2020,
        month = jan,
       volume = {633},
          eid = {A110},
        pages = {A110},
          doi = {10.1051/0004-6361/201936898},
archivePrefix = {arXiv},
       eprint = {1912.04651},
 primaryClass = {astro-ph.EP},
       adsurl = {https://ui.adsabs.harvard.edu/abs/2020A&A...633A.110G}
}

@ARTICLE{Costes2024c13,
       author = {{Costes}, J.~C. and {Xuan}, J.~W. and {Vigan}, A. and {Wang}, J. and {D'Orazi}, V. and {Molli{\`e}re}, P. and {Baker}, A. and {Bartos}, R. and {Blake}, G.~A. and {Calvin}, B. and {Cetre}, S. and {Delorme}, J. and {Doppmann}, G. and {Echeveri}, D. and {Finnerty}, L. and {Fitzgerald}, M.~P. and {Hsu}, C. and {Jovanovic}, N. and {Lopez}, R. and {Mawet}, D. and {Morris}, E. and {Pezzato}, J. and {Phillips}, C.~L. and {Ruffio}, J. and {Sappey}, B. and {Schneeberger}, A. and {Schofield}, T. and {Skemer}, A.~J. and {Wallace}, J.~K. and {Wang}, J.},
        title = "{Fresh view of the hot brown dwarf HD 984 B through high-resolution spectroscopy}",
      journal = {\aap},
         year = 2024,
        month = jun,
       volume = {686},
          eid = {A294},
        pages = {A294},
          doi = {10.1051/0004-6361/202348370},
archivePrefix = {arXiv},
       eprint = {2404.11523},
 primaryClass = {astro-ph.SR},
       adsurl = {https://ui.adsabs.harvard.edu/abs/2024A&A...686A.294C}
}

@ARTICLE{Chan1965h2,
       author = {{Chan}, Y.~M. and {Dalgarno}, A.},
        title = "{The refractive index of helium}",
      journal = {Proceedings of the Physical Society},
         year = 1965,
        month = feb,
       volume = {85},
       number = {2},
        pages = {227-230},
          doi = {10.1088/0370-1328/85/2/304},
       adsurl = {https://ui.adsabs.harvard.edu/abs/1965PPS....85..227C}
}

@ARTICLE{Dalgarno1962h2,
       author = {{Dalgarno}, A. and {Williams}, D.~A.},
        title = "{Rayleigh Scattering by Molecular Hydrogen.}",
      journal = {\apj},
         year = 1962,
        month = sep,
       volume = {136},
        pages = {690-692},
          doi = {10.1086/147428},
       adsurl = {https://ui.adsabs.harvard.edu/abs/1962ApJ...136..690D}
}

@ARTICLE{DeRosa2023aflepb,
       author = {{De Rosa}, Robert J. and {Nielsen}, Eric L. and {Wahhaj}, Zahed and {Ruffio}, Jean-Baptiste and {Kalas}, Paul G. and {Peck}, Anne E. and {Hirsch}, Lea A. and {Roberson}, William},
        title = "{Direct imaging discovery of a super-Jovian around the young Sun-like star AF Leporis}",
      journal = {\aap},
         year = 2023,
        month = apr,
       volume = {672},
          eid = {A94},
        pages = {A94},
          doi = {10.1051/0004-6361/202345877},
archivePrefix = {arXiv},
       eprint = {2302.06332},
 primaryClass = {astro-ph.EP},
       adsurl = {https://ui.adsabs.harvard.edu/abs/2023A&A...672A..94D}
}

@ARTICLE{Denis2025aflepb,
       author = {{Denis}, A. and {Vigan}, A. and {Costes}, J. and {Chauvin}, G. and {Radcliffe}, A. and {Ravet}, M. and {Balmer}, W. and {Palma-Bifani}, P. and {Petrus}, S. and {Parmentier}, V. and {Martos}, S. and {Simonnin}, A. and {Bonnefoy}, M. and {Cadet}, R. and {Forveille}, T. and {Charnay}, B. and {Kiefer}, F. and {Lagrange}, A. -M. and {Chiavassa}, A. and {Stolker}, T. and {Lavail}, A. and {Godoy}, N. and {Janson}, M. and {Pourcelot}, R. and {Delorme}, P. and {Rickman}, E. and {Cont}, D. and {Reiners}, A. and {De Rosa}, R. and {Anwand-Heerwart}, H. and {Charles}, Y. and {Costille}, A. and {El Morsy}, M. and {Garcia}, J. and {Houll{\'e}}, M. and {Lopez}, M. and {Murray}, G. and {Muslimov}, E. and {Otten}, G.~P.~P.~L. and {Paufique}, J. and {Phillips}, M. and {Seemann}, U. and {Viret}, A. and {Zins}, G.},
        title = "{Characterization of AF Lep b at high spectral resolution with VLT/HiRISE}",
      journal = {\aap},
         year = 2025,
        month = apr,
       volume = {696},
          eid = {A6},
        pages = {A6},
          doi = {10.1051/0004-6361/202453108},
archivePrefix = {arXiv},
       eprint = {2502.19558},
 primaryClass = {astro-ph.EP},
       adsurl = {https://ui.adsabs.harvard.edu/abs/2025A&A...696A...6D}
}

@ARTICLE{Deshpande2012MIJvsini,
       author = {{Deshpande}, R. and {Mart{\'\i}n}, E.~L. and {Montgomery}, M.~M. and {Zapatero Osorio}, M.~R. and {Rodler}, F. and {del Burgo}, C. and {Phan Bao}, N. and {Lyubchik}, Y. and {Tata}, R. and {Bouy}, H. and {Pavlenko}, Y.},
        title = "{Intermediate Resolution Near-infrared Spectroscopy of 36 Late M Dwarfs}",
      journal = {\aj},
         year = 2012,
        month = oct,
       volume = {144},
       number = {4},
          eid = {99},
        pages = {99},
          doi = {10.1088/0004-6256/144/4/99},
archivePrefix = {arXiv},
       eprint = {1207.2781},
 primaryClass = {astro-ph.SR},
       adsurl = {https://ui.adsabs.harvard.edu/abs/2012AJ....144...99D}
}

@ARTICLE{Dorn2014crires,
       author = {{Dorn}, R.~J. and {Anglada-Escude}, G. and {Baade}, D. and {Bristow}, P. and {Follert}, R. and {Gojak}, D. and {Grunhut}, J. and {Hatzes}, A. and {Heiter}, U. and {Hilker}, M. and {Ives}, D.~J. and {Jung}, Y. and {K{\"a}ufl}, H. -U. and {Kerber}, F. and {Klein}, B. and {Lizon}, J. -L. and {Lockhart}, M. and {L{\"o}winger}, T. and {Marquart}, T. and {Oliva}, E. and {Origlia}, L. and {Pasquini}, L. and {Paufique}, J. and {Piskunov}, N. and {Pozna}, E. and {Reiners}, A. and {Smette}, A. and {Smoker}, J. and {Seemann}, U. and {Stempels}, E. and {Valenti}, E.},
        title = "{CRIRES+: Exploring the Cold Universe at High Spectral Resolution}",
      journal = {The Messenger},
         year = 2014,
        month = jun,
       volume = {156},
        pages = {7-11},
       adsurl = {https://ui.adsabs.harvard.edu/abs/2014Msngr.156....7D}
}

@ARTICLE{Dorn2023crires,
       author = {{Dorn}, R.~J. and {Bristow}, P. and {Smoker}, J.~V. and {Rodler}, F. and {Lavail}, A. and {Accardo}, M. and {van den Ancker}, M. and {Baade}, D. and {Baruffolo}, A. and {Courtney-Barrer}, B. and {Blanco}, L. and {Brucalassi}, A. and {Cumani}, C. and {Follert}, R. and {Haimerl}, A. and {Hatzes}, A. and {Haug}, M. and {Heiter}, U. and {Hinterschuster}, R. and {Hubin}, N. and {Ives}, D.~J. and {Jung}, Y. and {Jones}, M. and {Kaeufl}, H. -U. and {Kirchbauer}, J. -P. and {Klein}, B. and {Kochukhov}, O. and {Korhonen}, H.~H. and {K{\"o}hler}, J. and {Lizon}, J. -L. and {Moins}, C. and {Molina-Conde}, I. and {Marquart}, T. and {Neeser}, M. and {Oliva}, E. and {Pallanca}, L. and {Pasquini}, L. and {Paufique}, J. and {Piskunov}, N. and {Reiners}, A. and {Schneller}, D. and {Schmutzer}, R. and {Seemann}, U. and {Slumstrup}, D. and {Smette}, A. and {Stegmeier}, J. and {Stempels}, E. and {Tordo}, S. and {Valenti}, E. and {Valenzuela}, J.~J. and {Vernet}, J. and {Vinther}, J. and {Wehrhahn}, A.},
        title = "{CRIRES$^{+}$ on sky at the ESO Very Large Telescope. Observing the Universe at infrared wavelengths and high spectral resolution}",
      journal = {\aap},
         year = 2023,
        month = mar,
       volume = {671},
          eid = {A24},
        pages = {A24},
          doi = {10.1051/0004-6361/202245217},
archivePrefix = {arXiv},
       eprint = {2301.08048},
 primaryClass = {astro-ph.IM},
       adsurl = {https://ui.adsabs.harvard.edu/abs/2023A&A...671A..24D}
}

@ARTICLE{Dupuy2017evolution,
       author = {{Dupuy}, Trent J. and {Liu}, Michael C.},
        title = "{Individual Dynamical Masses of Ultracool Dwarfs}",
      journal = {\apjs},
         year = 2017,
        month = aug,
       volume = {231},
       number = {2},
          eid = {15},
        pages = {15},
          doi = {10.3847/1538-4365/aa5e4c},
archivePrefix = {arXiv},
       eprint = {1703.05775},
 primaryClass = {astro-ph.SR},
       adsurl = {https://ui.adsabs.harvard.edu/abs/2017ApJS..231...15D}
}

@ARTICLE{Dupuy2018hawaii,
       author = {{Dupuy}, Trent J. and {Liu}, Michael C. and {Allers}, Katelyn N. and {Biller}, Beth A. and {Kratter}, Kaitlin M. and {Mann}, Andrew W. and {Shkolnik}, Evgenya L. and {Kraus}, Adam L. and {Best}, William M.~J.},
        title = "{The Hawaii Infrared Parallax Program. III. 2MASS J0249-0557 c: A Wide Planetary-mass Companion to a Low-mass Binary in the {\ensuremath{\beta}} Pic Moving Group}",
      journal = {\aj},
         year = 2018,
        month = aug,
       volume = {156},
       number = {2},
          eid = {57},
        pages = {57},
          doi = {10.3847/1538-3881/aacbc2},
archivePrefix = {arXiv},
       eprint = {1807.05235},
 primaryClass = {astro-ph.EP},
       adsurl = {https://ui.adsabs.harvard.edu/abs/2018AJ....156...57D}
}

@ARTICLE{Filippazzo2015mass,
       author = {{Filippazzo}, Joseph C. and {Rice}, Emily L. and {Faherty}, Jacqueline and {Cruz}, Kelle L. and {Van Gordon}, Mollie M. and {Looper}, Dagny L.},
        title = "{Fundamental Parameters and Spectral Energy Distributions of Young and Field Age Objects with Masses Spanning the Stellar to Planetary Regime}",
      journal = {\apj},
         year = 2015,
        month = sep,
       volume = {810},
       number = {2},
          eid = {158},
        pages = {158},
          doi = {10.1088/0004-637X/810/2/158},
archivePrefix = {arXiv},
       eprint = {1508.01767},
 primaryClass = {astro-ph.SR},
       adsurl = {https://ui.adsabs.harvard.edu/abs/2015ApJ...810..158F}
}

@ARTICLE{Franson2024aflepb,
       author = {{Franson}, Kyle and {Balmer}, William O. and {Bowler}, Brendan P. and {Pueyo}, Laurent and {Zhou}, Yifan and {Rickman}, Emily and {Zhang}, Zhoujian and {Mukherjee}, Sagnick and {Pearce}, Tim D. and {Bardalez Gagliuffi}, Daniella C. and {Biddle}, Lauren I. and {Brandt}, Timothy D. and {Bowens-Rubin}, Rachel and {Crepp}, Justin R. and {Davidson}, James W. and {Faherty}, Jacqueline and {Ginski}, Christian and {Horch}, Elliott P. and {Morgan}, Marvin and {Morley}, Caroline V. and {Perrin}, Marshall D. and {Sanghi}, Aniket and {Salama}, Ma{\"\i}ssa and {Theissen}, Christopher A. and {Tran}, Quang H. and {Wolf}, Trevor N.},
        title = "{JWST/NIRCam 4{\textendash}5 {\ensuremath{\mu}}m Imaging of the Giant Planet AF Lep b}",
      journal = {\apjl},
         year = 2024,
        month = oct,
       volume = {974},
       number = {1},
          eid = {L11},
        pages = {L11},
          doi = {10.3847/2041-8213/ad736a},
archivePrefix = {arXiv},
       eprint = {2406.09528},
 primaryClass = {astro-ph.EP},
       adsurl = {https://ui.adsabs.harvard.edu/abs/2024ApJ...974L..11F}
}

@dataset{Gaia2020EDR3,
       author = {{Gaia Collaboration}},
        title = "{VizieR Online Data Catalog: Gaia EDR3 (Gaia Collaboration, 2020)}",
 howpublished = {VizieR On-line Data Catalog: I/350.  Originally published in: 2021A\&A...649A...1G},
         year = 2020,
        month = nov,
          eid = {I/350},
          doi = {10.26093/cds/vizier.1350},
       adsurl = {https://ui.adsabs.harvard.edu/abs/2020yCat.1350....0G}
}

@ARTICLE{Gagne2015banyan,
       author = {{Gagn{\'e}}, Jonathan and {Faherty}, Jacqueline K. and {Cruz}, Kelle L. and {Lafreni{\'e}re}, David and {Doyon}, Ren{\'e} and {Malo}, Lison and {Burgasser}, Adam J. and {Naud}, Marie-Eve and {Artigau}, {\'E}tienne and {Bouchard}, Sandie and {Gizis}, John E. and {Albert}, Lo{\"\i}c},
        title = "{BANYAN. VII. A New Population of Young Substellar Candidate Members of Nearby Moving Groups from the BASS Survey}",
      journal = {\apjs},
         year = 2015,
        month = aug,
       volume = {219},
       number = {2},
          eid = {33},
        pages = {33},
          doi = {10.1088/0067-0049/219/2/33},
archivePrefix = {arXiv},
       eprint = {1506.07712},
 primaryClass = {astro-ph.SR},
       adsurl = {https://ui.adsabs.harvard.edu/abs/2015ApJS..219...33G}
}

@ARTICLE{Gandhi2023c13,
       author = {{Gandhi}, Siddharth and {de Regt}, Sam and {Snellen}, Ignas and {Zhang}, Yapeng and {Rugers}, Benson and {van Leur}, Niels and {Bosschaart}, Quincy},
        title = "{JWST Measurements of $^{13}$C, $^{18}$O, and $^{17}$O in the Atmosphere of Super-Jupiter VHS 1256 b}",
      journal = {\apjl},
         year = 2023,
        month = nov,
       volume = {957},
       number = {2},
          eid = {L36},
        pages = {L36},
          doi = {10.3847/2041-8213/ad07e2},
archivePrefix = {arXiv},
       eprint = {2311.05349},
 primaryClass = {astro-ph.EP},
       adsurl = {https://ui.adsabs.harvard.edu/abs/2023ApJ...957L..36G}
}

@ARTICLE{Gandhi2025eso5,
       author = {{Gandhi}, Siddharth and {de Regt}, Sam and {Snellen}, Ignas and {Palma-Bifani}, Paulina and {Abdoulwahab}, Idriss and {Chauvin}, Ga{\"e}l and {Gonz{\'a}lez Picos}, Dar{\'\i}o and {Zhang}, Yapeng and {Landman}, Rico and {Stolker}, Tomas and {Kesseli}, Aurora and {Mulder}, Willeke and {Chomez}, Antoine and {Lagrange}, Anne-Marie and {Zurlo}, Alice},
        title = "{The ESO SupJup Survey V: exploring atmospheric variability and orbit of the super-Jupiter AB Pictoris b with CRIRES+}",
      journal = {\mnras},
         year = 2025,
        month = feb,
       volume = {537},
       number = {1},
        pages = {134-153},
          doi = {10.1093/mnras/staf004},
archivePrefix = {arXiv},
       eprint = {2501.05114},
 primaryClass = {astro-ph.EP},
       adsurl = {https://ui.adsabs.harvard.edu/abs/2025MNRAS.537..134G}
}

@ARTICLE{Gao2020fecloud,
       author = {{Gao}, Peter and {Thorngren}, Daniel P. and {Lee}, Elspeth K.~H. and {Fortney}, Jonathan J. and {Morley}, Caroline V. and {Wakeford}, Hannah R. and {Powell}, Diana K. and {Stevenson}, Kevin B. and {Zhang}, Xi},
        title = "{Aerosol composition of hot giant exoplanets dominated by silicates and hydrocarbon hazes}",
      journal = {Nature Astronomy},
         year = 2020,
        month = may,
       volume = {4},
        pages = {951-956},
          doi = {10.1038/s41550-020-1114-3},
archivePrefix = {arXiv},
       eprint = {2005.11939},
 primaryClass = {astro-ph.EP},
       adsurl = {https://ui.adsabs.harvard.edu/abs/2020NatAs...4..951G}
}

@ARTICLE{Gagnne2021movinggroup,
       author = {{Gagn{\'e}}, Jonathan and {Faherty}, Jacqueline K. and {Moranta}, Leslie and {Popinchalk}, Mark},
        title = "{A Number of nearby Moving Groups May Be Fragments of Dissolving Open Clusters}",
      journal = {\apjl},
         year = 2021,
        month = jul,
       volume = {915},
       number = {2},
          eid = {L29},
        pages = {L29},
          doi = {10.3847/2041-8213/ac0e9a},
archivePrefix = {arXiv},
       eprint = {2106.11873},
 primaryClass = {astro-ph.GA},
       adsurl = {https://ui.adsabs.harvard.edu/abs/2021ApJ...915L..29G}
}

@ARTICLE{Grasser2025eso8,
       author = {{Grasser}, N. and {Snellen}, I.~A.~G. and {de Regt}, S. and {Gonz{\'a}lez Picos}, D. and {Zhang}, Y. and {Stolker}, T. and {Gandhi}, S. and {Nasedkin}, E. and {Landman}, R. and {Kesseli}, A.~Y. and {Mulder}, W.},
        title = "{The ESO SupJup Survey: VIII. Chemical fingerprints of young L dwarf twins}",
      journal = {\aap},
         year = 2025,
        month = jun,
       volume = {698},
          eid = {A252},
        pages = {A252},
          doi = {10.1051/0004-6361/202554195},
archivePrefix = {arXiv},
       eprint = {2507.02706},
 primaryClass = {astro-ph.EP},
       adsurl = {https://ui.adsabs.harvard.edu/abs/2025A&A...698A.252G}
}

@ARTICLE{Gonalezz2024eso2,
       author = {{Gonz{\'a}lez Picos}, D. and {Snellen}, I.~A.~G. and {de Regt}, S. and {Landman}, R. and {Zhang}, Y. and {Gandhi}, S. and {Ginski}, C. and {Kesseli}, A.~Y. and {Molli{\`e}re}, P. and {Stolker}, T.},
        title = "{The ESO SupJup Survey: II. The $^{12}$C/$^{13}$C isotope ratios of three young brown dwarfs with CRIRES$^{+}$}",
      journal = {\aap},
         year = 2024,
        month = sep,
       volume = {689},
          eid = {A212},
        pages = {A212},
          doi = {10.1051/0004-6361/202450028},
archivePrefix = {arXiv},
       eprint = {2407.07678},
 primaryClass = {astro-ph.EP},
       adsurl = {https://ui.adsabs.harvard.edu/abs/2024A&A...689A.212G}
}

@ARTICLE{Gonalezz2025eso4,
       author = {{Gonz{\'a}lez Picos}, D. and {Snellen}, I.~A.~G. and {de Regt}, S. and {Landman}, R. and {Zhang}, Y. and {Gandhi}, S. and {S{\'a}nchez-L{\'o}pez}, A.},
        title = "{The ESO SupJup Survey: IV. Unveiling the carbon isotope ratio of GQ Lup B and its host star}",
      journal = {\aap},
         year = 2025,
        month = jan,
       volume = {693},
          eid = {A298},
        pages = {A298},
          doi = {10.1051/0004-6361/202451936},
archivePrefix = {arXiv},
       eprint = {2501.01789},
 primaryClass = {astro-ph.EP},
       adsurl = {https://ui.adsabs.harvard.edu/abs/2025A&A...693A.298G}
}

@ARTICLE{Gonalezz202Mdwarfc13,
       author = {{Gonz{\'a}lez Picos}, Dar{\'\i}o and {Snellen}, Ignas and {de Regt}, Sam},
        title = "{Chemical evolution imprints in the rare isotopes of nearby M dwarfs}",
      journal = {Nature Astronomy},
         year = 2025,
        month = aug,
          doi = {10.1038/s41550-025-02641-4},
archivePrefix = {arXiv},
       eprint = {2508.18424},
 primaryClass = {astro-ph.SR},
       adsurl = {https://ui.adsabs.harvard.edu/abs/2025NatAs.tmp..172G}
}

@ARTICLE{Hayoz2025aflepb,
       author = {{Hayoz}, J. and {Bonse}, M.~J. and {Dannert}, F. and {Garvin}, E.~O. and {Cugno}, G. and {Patapis}, P. and {Gebhard}, T.~D. and {Balmer}, W.~O. and {De Rosa}, R.~J. and {Agudo Berbel}, A. and {Cao}, Y. and {Orban de Xivry}, G. and {Stolker}, T. and {Davies}, R. and {Absil}, O. and {Schmid}, H.~M. and {Quanz}, S.~P. and {Agapito}, G. and {Baruffolo}, A. and {Black}, M. and {Bonaglia}, M. and {Briguglio}, R. and {Carbonaro}, L. and {Cresci}, G. and {Dallilar}, Y. and {Deysenroth}, M. and {Di Antonio}, I. and {Di Cianno}, A. and {Di Rico}, G. and {Doelman}, D. and {Dolci}, M. and {Eisenhauer}, F. and {Esposito}, S. and {Fantinel}, D. and {Ferruzzi}, D. and {Feuchtgruber}, H. and {F{\"o}rster Schreiber}, N.~M. and {Gao}, X. and {Genzel}, R. and {Gillessen}, S. and {Glauser}, A.~M. and {Grani}, P. and {Hartl}, M. and {Henry}, D. and {Huber}, H. and {Keller}, C. and {Kenworthy}, M. and {Kravchenko}, K. and {Lightfoot}, J. and {Lunney}, D. and {Lutz}, D. and {MacIntosh}, M. and {Mannucci}, F. and {Ott}, T. and {Pearson}, D. and {Puglisi}, A. and {Rabien}, S. and {Rau}, C. and {Riccardi}, A. and {Salasnich}, B. and {Shimizu}, T. and {Snik}, F. and {Sturm}, E. and {Tacconi}, L. and {Taylor}, W. and {Valentini}, A. and {Waring}, C. and {Wiezorrek}, E. and {Xompero}, M.},
        title = "{High-contrast spectroscopy with the new VLT/ERIS instrument: Molecular maps and radial velocity of the gas giant AF Lep b}",
      journal = {\aap},
         year = 2025,
        month = jun,
       volume = {698},
          eid = {A87},
        pages = {A87},
          doi = {10.1051/0004-6361/202453297},
archivePrefix = {arXiv},
       eprint = {2502.19961},
 primaryClass = {astro-ph.EP},
       adsurl = {https://ui.adsabs.harvard.edu/abs/2025A&A...698A..87H}
}

@ARTICLE{Hargreaves2020CH4,
       author = {{Hargreaves}, Robert J. and {Gordon}, Iouli E. and {Rey}, Michael and {Nikitin}, Andrei V. and {Tyuterev}, Vladimir G. and {Kochanov}, Roman V. and {Rothman}, Laurence S.},
        title = "{An Accurate, Extensive, and Practical Line List of Methane for the HITEMP Database}",
      journal = {\apjs},
         year = 2020,
        month = apr,
       volume = {247},
       number = {2},
          eid = {55},
        pages = {55},
          doi = {10.3847/1538-4365/ab7a1a},
archivePrefix = {arXiv},
       eprint = {2001.05037},
 primaryClass = {astro-ph.EP},
       adsurl = {https://ui.adsabs.harvard.edu/abs/2020ApJS..247...55H}
}

@ARTICLE{Hsu2024pds70b,
       author = {{Hsu}, Chih-Chun and {Wang}, Jason J. and {Blake}, Geoffrey A. and {Xuan}, Jerry W. and {Zhang}, Yapeng and {Ruffio}, Jean-Baptiste and {Horstman}, Katelyn and {Cronin}, Julianne and {Sappey}, Ben and {Xin}, Yinzi and {Finnerty}, Luke and {Echeverri}, Daniel and {Mawet}, Dimitri and {Jovanovic}, Nemanja and {Do {\'O}}, Clarissa R. and {Baker}, Ashley and {Bartos}, Randall and {Calvin}, Benjamin and {Cetre}, Sylvain and {Delorme}, Jacques-Robert and {Doppmann}, Gregory W. and {Fitzgerald}, Michael P. and {Liberman}, Joshua and {L{\'o}pez}, Ronald A. and {Morris}, Evan and {Pezzato-Rovner}, Jacklyn and {Schofield}, Tobias and {Skemer}, Andrew and {Wallace}, J. Kent and {Wang}, Ji},
        title = "{PDS 70b Shows Stellar-like Carbon-to-oxygen Ratio}",
      journal = {\apjl},
         year = 2024,
        month = dec,
       volume = {977},
       number = {2},
          eid = {L47},
        pages = {L47},
          doi = {10.3847/2041-8213/ad95e8},
archivePrefix = {arXiv},
       eprint = {2411.15117},
 primaryClass = {astro-ph.EP},
       adsurl = {https://ui.adsabs.harvard.edu/abs/2024ApJ...977L..47H}
}

@BOOK{Jeffreys1939bayes,
       author = {{Jeffreys}, Harold},
        title = "{Theory of Probability}",
         year = 1939,
       adsurl = {https://ui.adsabs.harvard.edu/abs/1939thpr.book.....J}
}

@ARTICLE{Janson2025bpicb,
       author = {{Janson}, Markus and {Wehrung-Montpezat}, Jonas and {Wehrhahn}, Ansgar and {Brandeker}, Alexis and {Viswanath}, Gayathri and {Molli{\`e}re}, Paul and {Stolker}, Thomas},
        title = "{Deep high-resolution L band spectroscopy in the {\ensuremath{\beta}} Pictoris planetary system}",
      journal = {\aap},
         year = 2025,
        month = feb,
       volume = {694},
          eid = {A63},
        pages = {A63},
          doi = {10.1051/0004-6361/202452411},
archivePrefix = {arXiv},
       eprint = {2501.08445},
 primaryClass = {astro-ph.EP},
       adsurl = {https://ui.adsabs.harvard.edu/abs/2025A&A...694A..63J}
}

@ARTICLE{Jaeger1998mgsio3,
       author = {{Jaeger}, C. and {Molster}, F.~J. and {Dorschner}, J. and {Henning}, Th. and {Mutschke}, H. and {Waters}, L.~B.~F.~M.},
        title = "{Steps toward interstellar silicate mineralogy. IV. The crystalline revolution}",
      journal = {\aap},
         year = 1998,
        month = nov,
       volume = {339},
        pages = {904-916},
       adsurl = {https://ui.adsabs.harvard.edu/abs/1998A&A...339..904J}
}

@ARTICLE{Kammerer2024bpicb,
       author = {{Kammerer}, Jens and {Lawson}, Kellen and {Perrin}, Marshall D. and {Rebollido}, Isabel and {Stark}, Christopher C. and {Stolker}, Tomas and {Girard}, Julien H. and {Pueyo}, Laurent and {Balmer}, William O. and {Worthen}, Kadin and {Chen}, Christine and {van der Marel}, Roeland P. and {Lewis}, Nikole K. and {Ward-Duong}, Kimberly and {Valenti}, Jeff A. and {Clampin}, Mark and {Mountain}, C. Matt},
        title = "{JWST-TST High Contrast: JWST/NIRCam Observations of the Young Giant Planet {\ensuremath{\beta}} Pic b}",
      journal = {\aj},
         year = 2024,
        month = aug,
       volume = {168},
       number = {2},
          eid = {51},
        pages = {51},
          doi = {10.3847/1538-3881/ad4ffe},
archivePrefix = {arXiv},
       eprint = {2405.18422},
 primaryClass = {astro-ph.EP},
       adsurl = {https://ui.adsabs.harvard.edu/abs/2024AJ....168...51K}
}

@article{Kass1995bayes,
author = {Robert E. Kass and Adrian E. Raftery},
title = {Bayes Factors},
journal = {Journal of the American Statistical Association},
volume = {90},
number = {430},
pages = {773--795},
year = {1995},
publisher = {ASA Website},
doi = {10.1080/01621459.1995.10476572},
URL = { https://www.tandfonline.com/doi/abs/10.1080/01621459.1995.10476572
},
eprint = {  https://www.tandfonline.com/doi/pdf/10.1080/01621459.1995.10476572
}
}

@ARTICLE{Kipping2025sigma,
       author = {{Kipping}, David and {Benneke}, Bj{\"o}rn},
        title = "{Exoplaneteers Keep Overestimating Sigma Significances}",
      journal = {arXiv e-prints},
         year = 2025,
        month = jun,
          eid = {arXiv:2506.05392},
        pages = {arXiv:2506.05392},
          doi = {10.48550/arXiv.2506.05392},
archivePrefix = {arXiv},
       eprint = {2506.05392},
 primaryClass = {astro-ph.IM},
       adsurl = {https://ui.adsabs.harvard.edu/abs/2025arXiv250605392K}
}

@ARTICLE{Kratter2016instability,
       author = {{Kratter}, Kaitlin and {Lodato}, Giuseppe},
        title = "{Gravitational Instabilities in Circumstellar Disks}",
      journal = {\araa},
         year = 2016,
        month = sep,
       volume = {54},
        pages = {271-311},
          doi = {10.1146/annurev-astro-081915-023307},
archivePrefix = {arXiv},
       eprint = {1603.01280},
 primaryClass = {astro-ph.SR},
       adsurl = {https://ui.adsabs.harvard.edu/abs/2016ARA&A..54..271K}
}

@ARTICLE{Langer1993ISMc13,
       author = {{Langer}, William D. and {Penzias}, Arno A.},
        title = "{12C/ 13C Isotope Ratio in the Local Interstellar Medium from Observations of 13C 18O in Molecular Clouds}",
      journal = {\apj},
         year = 1993,
        month = may,
       volume = {408},
        pages = {539},
          doi = {10.1086/172611},
       adsurl = {https://ui.adsabs.harvard.edu/abs/1993ApJ...408..539L}
}

@ARTICLE{Landman2024bpicb,
       author = {{Landman}, R. and {Stolker}, T. and {Snellen}, I.~A.~G. and {Costes}, J. and {de Regt}, S. and {Zhang}, Y. and {Gandhi}, S. and {Molliere}, P. and {Kesseli}, A. and {Vigan}, A. and {Sanchez-L{\'o}pez}, A.},
        title = "{{\ensuremath{\beta}} Pictoris b through the eyes of the upgraded CRIRES+. Atmospheric composition, spin rotation, and radial velocity}",
      journal = {\aap},
         year = 2024,
        month = feb,
       volume = {682},
          eid = {A48},
        pages = {A48},
          doi = {10.1051/0004-6361/202347846},
archivePrefix = {arXiv},
       eprint = {2311.13527},
 primaryClass = {astro-ph.EP},
       adsurl = {https://ui.adsabs.harvard.edu/abs/2024A&A...682A..48L}
}

@ARTICLE{Lambrechts2012accretion,
       author = {{Lambrechts}, M. and {Johansen}, A.},
        title = "{Rapid growth of gas-giant cores by pebble accretion}",
      journal = {\aap},
         year = 2012,
        month = aug,
       volume = {544},
          eid = {A32},
        pages = {A32},
          doi = {10.1051/0004-6361/201219127},
archivePrefix = {arXiv},
       eprint = {1205.3030},
 primaryClass = {astro-ph.EP},
       adsurl = {https://ui.adsabs.harvard.edu/abs/2012A&A...544A..32L}
}

@ARTICLE{Luhman2024bpmg,
       author = {{Luhman}, K.~L.},
        title = "{A Census of the {\ensuremath{\beta}} Pic Moving Group and Other Nearby Associations with Gaia}",
      journal = {\aj},
         year = 2024,
        month = oct,
       volume = {168},
       number = {4},
          eid = {159},
        pages = {159},
          doi = {10.3847/1538-3881/ad697d},
archivePrefix = {arXiv},
       eprint = {2409.06092},
 primaryClass = {astro-ph.GA},
       adsurl = {https://ui.adsabs.harvard.edu/abs/2024AJ....168..159L}
}

@ARTICLE{Madhusudhan2019atmospheres,
       author = {{Madhusudhan}, Nikku},
        title = "{Exoplanetary Atmospheres: Key Insights, Challenges, and Prospects}",
      journal = {\araa},
         year = 2019,
        month = aug,
       volume = {57},
        pages = {617-663},
          doi = {10.1146/annurev-astro-081817-051846},
archivePrefix = {arXiv},
       eprint = {1904.03190},
 primaryClass = {astro-ph.EP},
       adsurl = {https://ui.adsabs.harvard.edu/abs/2019ARA&A..57..617M}
}

@ARTICLE{Maldonado2022starrvaflep,
       author = {{Maldonado}, J. and {Colombo}, S. and {Petralia}, A. and {Benatti}, S. and {Desidera}, S. and {Malavolta}, L. and {Lanza}, A.~F. and {Damasso}, M. and {Micela}, G. and {Mallonn}, M. and {Messina}, S. and {Sozzetti}, A. and {Stelzer}, B. and {Biazzo}, K. and {Gratton}, R. and {Maggio}, A. and {Nardiello}, D. and {Scandariato}, G. and {Affer}, L. and {Baratella}, M. and {Claudi}, R. and {Molinari}, E. and {Bignamini}, A. and {Covino}, E. and {Pagano}, I. and {Piotto}, G. and {Poretti}, E. and {Cosentino}, R. and {Carleo}, I.},
        title = "{The GAPS programme at TNG. XXXIV. Activity-rotation, flux-flux relationships, and active-region evolution through stellar age}",
      journal = {\aap},
         year = 2022,
        month = jul,
       volume = {663},
          eid = {A142},
        pages = {A142},
          doi = {10.1051/0004-6361/202243360},
archivePrefix = {arXiv},
       eprint = {2204.12206},
 primaryClass = {astro-ph.SR},
       adsurl = {https://ui.adsabs.harvard.edu/abs/2022A&A...663A.142M}
}

@ARTICLE{Marley2021sonora,
       author = {{Marley}, Mark S. and {Saumon}, Didier and {Visscher}, Channon and {Lupu}, Roxana and {Freedman}, Richard and {Morley}, Caroline and {Fortney}, Jonathan J. and {Seay}, Christopher and {Smith}, Adam J.~R.~W. and {Teal}, D.~J. and {Wang}, Ruoyan},
        title = "{The Sonora Brown Dwarf Atmosphere and Evolution Models. I. Model Description and Application to Cloudless Atmospheres in Rainout Chemical Equilibrium}",
      journal = {\apj},
         year = 2021,
        month = oct,
       volume = {920},
       number = {2},
          eid = {85},
        pages = {85},
          doi = {10.3847/1538-4357/ac141d},
archivePrefix = {arXiv},
       eprint = {2107.07434},
 primaryClass = {astro-ph.SR},
       adsurl = {https://ui.adsabs.harvard.edu/abs/2021ApJ...920...85M}
}

@INPROCEEDINGS{Mawet2016kpic,
       author = {{Mawet}, D. and {Wizinowich}, P. and {Dekany}, R. and {Chun}, M. and {Hall}, D. and {Cetre}, S. and {Guyon}, O. and {Wallace}, J.~K. and {Bowler}, B. and {Liu}, M. and {Ruane}, G. and {Serabyn}, E. and {Bartos}, R. and {Wang}, J. and {Vasisht}, G. and {Fitzgerald}, M. and {Skemer}, A. and {Ireland}, M. and {Fucik}, J. and {Fortney}, J. and {Crossfield}, I. and {Hu}, R. and {Benneke}, B.},
        title = "{Keck Planet Imager and Characterizer: concept and phased implementation}",
    booktitle = {Adaptive Optics Systems V},
         year = 2016,
       editor = {{Marchetti}, Enrico and {Close}, Laird M. and {V{\'e}ran}, Jean-Pierre},
       series = {Society of Photo-Optical Instrumentation Engineers (SPIE) Conference Series},
       volume = {9909},
        month = jul,
          eid = {99090D},
        pages = {99090D},
          doi = {10.1117/12.2233658},
       adsurl = {https://ui.adsabs.harvard.edu/abs/2016SPIE.9909E..0DM}
}

@ARTICLE{Meibom2007solarc13,
       author = {{Meibom}, Anders and {Krot}, Alexander N. and {Robert}, Fran{\c{c}}ois and {Mostefaoui}, Smail and {Russell}, Sara S. and {Petaev}, Michael I. and {Gounelle}, Matthieu},
        title = "{Nitrogen and Carbon Isotopic Composition of the Sun Inferred from a High-Temperature Solar Nebular Condensate}",
      journal = {\apjl},
         year = 2007,
        month = feb,
       volume = {656},
       number = {1},
        pages = {L33-L36},
          doi = {10.1086/512052},
       adsurl = {https://ui.adsabs.harvard.edu/abs/2007ApJ...656L..33M}
}

@ARTICLE{McKemmish2019TiO,
       author = {{McKemmish}, Laura K. and {Masseron}, Thomas and {Hoeijmakers}, H. Jens and {P{\'e}rez-Mesa}, V{\'\i}ctor and {Grimm}, Simon L. and {Yurchenko}, Sergei N. and {Tennyson}, Jonathan},
        title = "{ExoMol molecular line lists - XXXIII. The spectrum of Titanium Oxide}",
      journal = {\mnras},
         year = 2019,
        month = sep,
       volume = {488},
       number = {2},
        pages = {2836-2854},
          doi = {10.1093/mnras/stz1818},
archivePrefix = {arXiv},
       eprint = {1905.04587},
 primaryClass = {astro-ph.SR},
       adsurl = {https://ui.adsabs.harvard.edu/abs/2019MNRAS.488.2836M}
}

@ARTICLE{Miret2020betapic,
       author = {{Miret-Roig}, N. and {Galli}, P.~A.~B. and {Brandner}, W. and {Bouy}, H. and {Barrado}, D. and {Olivares}, J. and {Antoja}, T. and {Romero-G{\'o}mez}, M. and {Figueras}, F. and {Lillo-Box}, J.},
        title = "{Dynamical traceback age of the {\ensuremath{\beta}} Pictoris moving group}",
      journal = {\aap},
         year = 2020,
        month = oct,
       volume = {642},
          eid = {A179},
        pages = {A179},
          doi = {10.1051/0004-6361/202038765},
archivePrefix = {arXiv},
       eprint = {2007.10997},
 primaryClass = {astro-ph.GA},
       adsurl = {https://ui.adsabs.harvard.edu/abs/2020A&A...642A.179M}
}

@ARTICLE{Molliere2019pRT,
       author = {{Molli{\`e}re}, P. and {Wardenier}, J.~P. and {van Boekel}, R. and {Henning}, Th. and {Molaverdikhani}, K. and {Snellen}, I.~A.~G.},
        title = "{petitRADTRANS. A Python radiative transfer package for exoplanet characterization and retrieval}",
      journal = {\aap},
         year = 2019,
        month = jul,
       volume = {627},
          eid = {A67},
        pages = {A67},
          doi = {10.1051/0004-6361/201935470},
archivePrefix = {arXiv},
       eprint = {1904.11504},
 primaryClass = {astro-ph.EP},
       adsurl = {https://ui.adsabs.harvard.edu/abs/2019A&A...627A..67M}
}

@ARTICLE{Morley2012cloud,
       author = {{Morley}, Caroline V. and {Fortney}, Jonathan J. and {Marley}, Mark S. and {Visscher}, Channon and {Saumon}, Didier and {Leggett}, S.~K.},
        title = "{Neglected Clouds in T and Y Dwarf Atmospheres}",
      journal = {\apj},
         year = 2012,
        month = sep,
       volume = {756},
       number = {2},
          eid = {172},
        pages = {172},
          doi = {10.1088/0004-637X/756/2/172},
archivePrefix = {arXiv},
       eprint = {1206.4313},
 primaryClass = {astro-ph.SR},
       adsurl = {https://ui.adsabs.harvard.edu/abs/2012ApJ...756..172M}
}

@ARTICLE{Milam2005ISMc13,
       author = {{Milam}, S.~N. and {Savage}, C. and {Brewster}, M.~A. and {Ziurys}, L.~M. and {Wyckoff}, S.},
        title = "{The $^{12}$C/$^{13}$C Isotope Gradient Derived from Millimeter Transitions of CN: The Case for Galactic Chemical Evolution}",
      journal = {\apj},
         year = 2005,
        month = dec,
       volume = {634},
       number = {2},
        pages = {1126-1132},
          doi = {10.1086/497123},
       adsurl = {https://ui.adsabs.harvard.edu/abs/2005ApJ...634.1126M}
}

@ARTICLE{Molliere2017chem,
       author = {{Molli{\`e}re}, P. and {van Boekel}, R. and {Bouwman}, J. and {Henning}, Th. and {Lagage}, P. -O. and {Min}, M.},
        title = "{Observing transiting planets with JWST. Prime targets and their synthetic spectral observations}",
      journal = {\aap},
         year = 2017,
        month = apr,
       volume = {600},
          eid = {A10},
        pages = {A10},
          doi = {10.1051/0004-6361/201629800},
archivePrefix = {arXiv},
       eprint = {1611.08608},
 primaryClass = {astro-ph.EP},
       adsurl = {https://ui.adsabs.harvard.edu/abs/2017A&A...600A..10M}
}

@ARTICLE{Molliere2020chem,
       author = {{Molli{\`e}re}, P. and {Stolker}, T. and {Lacour}, S. and {Otten}, G.~P.~P.~L. and {Shangguan}, J. and {Charnay}, B. and {Molyarova}, T. and {Nowak}, M. and {Henning}, Th. and {Marleau}, G. -D. and {Semenov}, D.~A. and {van Dishoeck}, E. and {Eisenhauer}, F. and {Garcia}, P. and {Garcia Lopez}, R. and {Girard}, J.~H. and {Greenbaum}, A.~Z. and {Hinkley}, S. and {Kervella}, P. and {Kreidberg}, L. and {Maire}, A. -L. and {Nasedkin}, E. and {Pueyo}, L. and {Snellen}, I.~A.~G. and {Vigan}, A. and {Wang}, J. and {de Zeeuw}, P.~T. and {Zurlo}, A.},
        title = "{Retrieving scattering clouds and disequilibrium chemistry in the atmosphere of HR 8799e}",
      journal = {\aap},
         year = 2020,
        month = aug,
       volume = {640},
          eid = {A131},
        pages = {A131},
          doi = {10.1051/0004-6361/202038325},
archivePrefix = {arXiv},
       eprint = {2006.09394},
 primaryClass = {astro-ph.EP},
       adsurl = {https://ui.adsabs.harvard.edu/abs/2020A&A...640A.131M}
}

@ARTICLE{Molliere2025PSO318,
       author = {{Molli{\`e}re}, P. and {K{\"u}hnle}, H. and {Matthews}, E.~C. and {Henning}, Th. and {Min}, M. and {Patapis}, P. and {Lagage}, P. -O. and {Waters}, L.~B.~F.~M. and {G{\"u}del}, M. and {J{\"a}ger}, Cornelia and {Zhang}, Z. and {Decin}, L. and {Biller}, B.~A. and {Absil}, O. and {Argyriou}, I. and {Barrado}, D. and {Cossou}, C. and {Glasse}, A. and {Olofsson}, G. and {Pye}, J.~P. and {Rouan}, D. and {Samland}, M. and {Scheithauer}, S. and {Tremblin}, P. and {Whiteford}, N. and {van Dishoeck}, E.~F. and {{\"O}stlin}, G. and {Ray}, T.},
        title = "{Evidence for SiO cloud nucleation in the rogue planet PSO J318}",
      journal = {arXiv e-prints},
         year = 2025,
        month = jul,
          eid = {arXiv:2507.18691},
        pages = {arXiv:2507.18691},
          doi = {10.48550/arXiv.2507.18691},
archivePrefix = {arXiv},
       eprint = {2507.18691},
 primaryClass = {astro-ph.EP},
       adsurl = {https://ui.adsabs.harvard.edu/abs/2025arXiv250718691M}
}

@ARTICLE{Mulder2025eso6,
       author = {{Mulder}, W. and {de Regt}, S. and {Landman}, R. and {Picos}, D. Gonz{\'a}lez and {Snellen}, I.~A.~G. and {Zhang}, Y. and {Gandhi}, S. and {Ginski}, C. and {Kesseli}, A.~Y. and {Nasedkin}, E. and {Stolker}, T.},
        title = "{The ESO SupJup Survey: VI. $^{12}$C/$^{13}$C isotope ratio comparison of three L-type brown dwarfs}",
      journal = {\aap},
         year = 2025,
        month = feb,
       volume = {694},
          eid = {A164},
        pages = {A164},
          doi = {10.1051/0004-6361/202452859},
       adsurl = {https://ui.adsabs.harvard.edu/abs/2025A&A...694A.164M}
}

@ARTICLE{Oberg2011CO,
       author = {{{\"O}berg}, Karin I. and {Murray-Clay}, Ruth and {Bergin}, Edwin A.},
        title = "{The Effects of Snowlines on C/O in Planetary Atmospheres}",
      journal = {\apjl},
         year = 2011,
        month = dec,
       volume = {743},
       number = {1},
          eid = {L16},
        pages = {L16},
          doi = {10.1088/2041-8205/743/1/L16},
archivePrefix = {arXiv},
       eprint = {1110.5567},
 primaryClass = {astro-ph.GA},
       adsurl = {https://ui.adsabs.harvard.edu/abs/2011ApJ...743L..16O}
}

@ARTICLE{Palma2024aflepb,
       author = {{Palma-Bifani}, P. and {Chauvin}, G. and {Borja}, D. and {Bonnefoy}, M. and {Petrus}, S. and {Mesa}, D. and {De Rosa}, R.~J. and {Gratton}, R. and {Baudoz}, P. and {Boccaletti}, A. and {Charnay}, B. and {Desgrange}, C. and {Tremblin}, P. and {Vigan}, A.},
        title = "{Atmospheric properties of AF Lep b with forward modeling}",
      journal = {\aap},
         year = 2024,
        month = mar,
       volume = {683},
          eid = {A214},
        pages = {A214},
          doi = {10.1051/0004-6361/202347653},
archivePrefix = {arXiv},
       eprint = {2401.05491},
 primaryClass = {astro-ph.EP},
       adsurl = {https://ui.adsabs.harvard.edu/abs/2024A&A...683A.214P}
}

@ARTICLE{Parker2024bpicb,
       author = {{Parker}, Luke T. and {Birkby}, Jayne L. and {Landman}, Rico and {Wardenier}, Joost P. and {Young}, Mitchell E. and {Vaughan}, Sophia R. and {van Sluijs}, Lennart and {Brogi}, Matteo and {Parmentier}, Vivien and {Line}, Michael R.},
        title = "{Into the red: an M-band study of the chemistry and rotation of {\ensuremath{\beta}} Pictoris b at high spectral resolution}",
      journal = {\mnras},
         year = 2024,
        month = jun,
       volume = {531},
       number = {2},
        pages = {2356-2378},
          doi = {10.1093/mnras/stae1277},
archivePrefix = {arXiv},
       eprint = {2405.08867},
 primaryClass = {astro-ph.EP},
       adsurl = {https://ui.adsabs.harvard.edu/abs/2024MNRAS.531.2356P}
}

@ARTICLE{Phillips2020atmo,
       author = {{Phillips}, M.~W. and {Tremblin}, P. and {Baraffe}, I. and {Chabrier}, G. and {Allard}, N.~F. and {Spiegelman}, F. and {Goyal}, J.~M. and {Drummond}, B. and {H{\'e}brard}, E.},
        title = "{A new set of atmosphere and evolution models for cool T-Y brown dwarfs and giant exoplanets}",
      journal = {\aap},
         year = 2020,
        month = may,
       volume = {637},
          eid = {A38},
        pages = {A38},
          doi = {10.1051/0004-6361/201937381},
archivePrefix = {arXiv},
       eprint = {2003.13717},
 primaryClass = {astro-ph.SR},
       adsurl = {https://ui.adsabs.harvard.edu/abs/2020A&A...637A..38P}
}

@ARTICLE{Polyansky2018h2o,
       author = {{Polyansky}, Oleg L. and {Kyuberis}, Aleksandra A. and {Zobov}, Nikolai F. and {Tennyson}, Jonathan and {Yurchenko}, Sergei N. and {Lodi}, Lorenzo},
        title = "{ExoMol molecular line lists XXX: a complete high-accuracy line list for water}",
      journal = {\mnras},
         year = 2018,
        month = oct,
       volume = {480},
       number = {2},
        pages = {2597-2608},
          doi = {10.1093/mnras/sty1877},
archivePrefix = {arXiv},
       eprint = {1807.04529},
 primaryClass = {astro-ph.EP},
       adsurl = {https://ui.adsabs.harvard.edu/abs/2018MNRAS.480.2597P}
}

@article{Pollack1996accretion,
title = {Formation of the Giant Planets by Concurrent Accretion of Solids and Gas},
journal = {Icarus},
volume = {124},
number = {1},
pages = {62-85},
year = {1996},
issn = {0019-1035},
doi = {https://doi.org/10.1006/icar.1996.0190},
url = {https://www.sciencedirect.com/science/article/pii/S0019103596901906},
author = {James B. Pollack and Olenka Hubickyj and Peter Bodenheimer and Jack J. Lissauer and Morris Podolak and Yuval Greenzweig},
}

@ARTICLE{Ravet2025bpicisotope,
       author = {{Ravet}, M. and {Bonnefoy}, M. and {Chauvin}, G. and {Lacour}, S. and {Nowak}, M. and {Charnay}, B. and {Tremblin}, P. and {Homeier}, D. and {Morley}, C. and {Fortney}, J. and {Denis}, A. and {Petrus}, S. and {Palma-Bifani}, P. and {Landman}, R. and {Parker}, L.~T. and {Houll{\'e}}, M. and {Chomez}, A. and {Worthen}, K. and {Kiefer}, F. and {Marleau}, G. -D. and {Zhang}, Z. and {Birkby}, J.~L. and {Millour}, F. and {Lagrange}, A. -M. and {Vigan}, A. and {Otten}, G.~P.~P.~L. and {Shangguan}, J.},
        title = "{Multi-modal atmospheric characterization of $β$ Pictoris b: Adding high-resolution continuum spectra from GRAVITY}",
      journal = {arXiv e-prints},
         year = 2025,
        month = sep,
          eid = {arXiv:2509.25338},
        pages = {arXiv:2509.25338},
archivePrefix = {arXiv},
       eprint = {2509.25338},
 primaryClass = {astro-ph.EP},
       adsurl = {https://ui.adsabs.harvard.edu/abs/2025arXiv250925338R}
}

@ARTICLE{Reggiani2024betapic,
       author = {{Reggiani}, Henrique and {Galarza}, Jhon Yana and {Schlaufman}, Kevin C. and {Sing}, David K. and {Healy}, Brian F. and {McWilliam}, Andrew and {Lothringer}, Joshua D. and {Pueyo}, Laurent},
        title = "{Insight into the Formation of {\ensuremath{\beta}} Pic b through the Composition of Its Parent Protoplanetary Disk as Revealed by the {\ensuremath{\beta}} Pic Moving Group Member HD 181327}",
      journal = {\aj},
         year = 2024,
        month = jan,
       volume = {167},
       number = {1},
          eid = {45},
        pages = {45},
          doi = {10.3847/1538-3881/ad0f93},
archivePrefix = {arXiv},
       eprint = {2311.12210},
 primaryClass = {astro-ph.SR},
       adsurl = {https://ui.adsabs.harvard.edu/abs/2024AJ....167...45R}
}

@ARTICLE{Regt2024eso1,
       author = {{de Regt}, S. and {Gandhi}, S. and {Snellen}, I.~A.~G. and {Zhang}, Y. and {Ginski}, C. and {Gonz{\'a}lez Picos}, D. and {Kesseli}, A.~Y. and {Landman}, R. and {Molli{\`e}re}, P. and {Nasedkin}, E. and {S{\'a}nchez-L{\'o}pez}, A. and {Stolker}, T.},
        title = "{The ESO SupJup Survey. I. Chemical and isotopic characterisation of the late L-dwarf DENIS J0255-4700 with CRIRES$^{+}$}",
      journal = {\aap},
         year = 2024,
        month = aug,
       volume = {688},
          eid = {A116},
        pages = {A116},
          doi = {10.1051/0004-6361/202348508},
archivePrefix = {arXiv},
       eprint = {2405.10841},
 primaryClass = {astro-ph.EP},
       adsurl = {https://ui.adsabs.harvard.edu/abs/2024A&A...688A.116D}
}

@ARTICLE{Regt2025eso7,
       author = {{de Regt}, S. and {Snellen}, I.~A.~G. and {Allard}, N.~F. and {Gonz{\'a}lez Picos}, D. and {Gandhi}, S. and {Grasser}, N. and {Landman}, R. and {Molli{\`e}re}, P. and {Nasedkin}, E. and {Stolker}, T. and {Zhang}, Y.},
        title = "{The ESO SupJup Survey: VII. Clouds and line asymmetries in CRIRES$^{+}$ J-band spectra of the Luhman 16 binary}",
      journal = {\aap},
         year = 2025,
        month = apr,
       volume = {696},
          eid = {A225},
        pages = {A225},
          doi = {10.1051/0004-6361/202453190},
archivePrefix = {arXiv},
       eprint = {2503.21266},
 primaryClass = {astro-ph.EP},
       adsurl = {https://ui.adsabs.harvard.edu/abs/2025A&A...696A.225D}
}

@ARTICLE{Rilinger2021disk,
       author = {{Rilinger}, Anneliese M. and {Espaillat}, Catherine C.},
        title = "{Disk Masses and Dust Evolution of Protoplanetary Disks around Brown Dwarfs}",
      journal = {\apj},
         year = 2021,
        month = nov,
       volume = {921},
       number = {2},
          eid = {182},
        pages = {182},
          doi = {10.3847/1538-4357/ac09e5},
archivePrefix = {arXiv},
       eprint = {2106.05247},
 primaryClass = {astro-ph.SR},
       adsurl = {https://ui.adsabs.harvard.edu/abs/2021ApJ...921..182R}
}

@ARTICLE{Ruffio2019rv,
       author = {{Ruffio}, Jean-Baptiste and {Macintosh}, Bruce and {Konopacky}, Quinn M. and {Barman}, Travis and {De Rosa}, Robert J. and {Wang}, Jason J. and {Wilcomb}, Kielan K. and {Czekala}, Ian and {Marois}, Christian},
        title = "{Radial Velocity Measurements of HR 8799 b and c with Medium Resolution Spectroscopy}",
      journal = {\aj},
         year = 2019,
        month = nov,
       volume = {158},
       number = {5},
          eid = {200},
        pages = {200},
          doi = {10.3847/1538-3881/ab4594},
archivePrefix = {arXiv},
       eprint = {1909.07571},
 primaryClass = {astro-ph.EP},
       adsurl = {https://ui.adsabs.harvard.edu/abs/2019AJ....158..200R}
}

\appendix 
\renewcommand{\thefigure}{A.\arabic{figure}}
\setcounter{figure}{0}
\section{2MASS J0249-0557c Plots}
This section contains the retrieval results and the best-fit model of 2MASS J0249-0557c.
% \begin{figure*}
%     \centering
%     \includegraphics[width=0.98\linewidth]{Figures/allmodels_mrtefflogg_2MASSJ0249-0557c.pdf}
%     \caption{Predictions from evolutionary models for 2MASS J0249-0557c}
%     \label{fig:2mjevolutionary}
% \end{figure*}

\begin{figure*}
    \centering
    \includegraphics[width=0.98\linewidth]{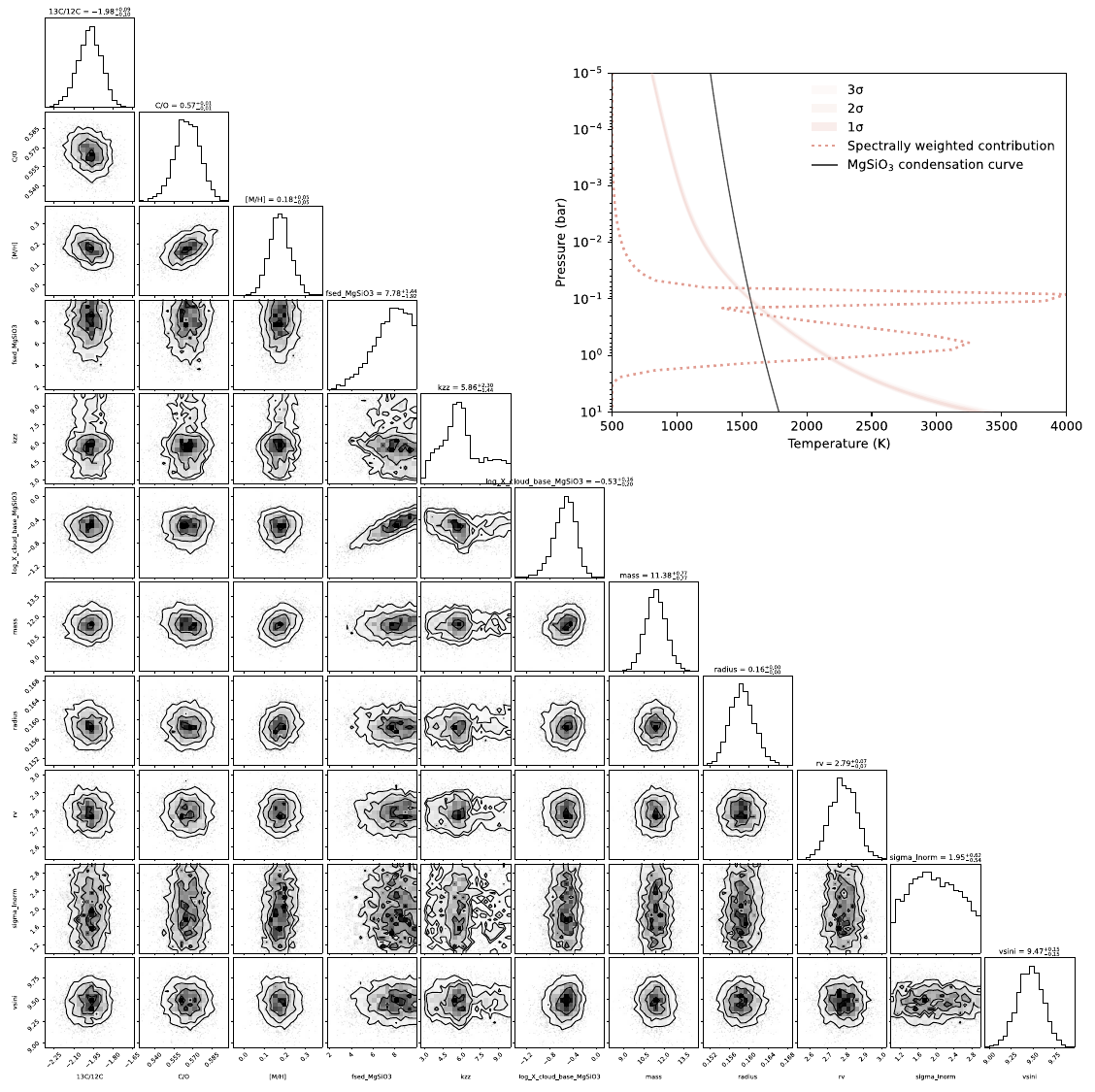}
    \caption{Retrieval result and PT profile of 2MASS J0249-0557c with mass prior and condensate cloud model. This constitutes the final result that we report.}
    \label{fig:2mjfinal}
\end{figure*}

\begin{figure*}
    \centering
    \includegraphics[width=0.98\linewidth]{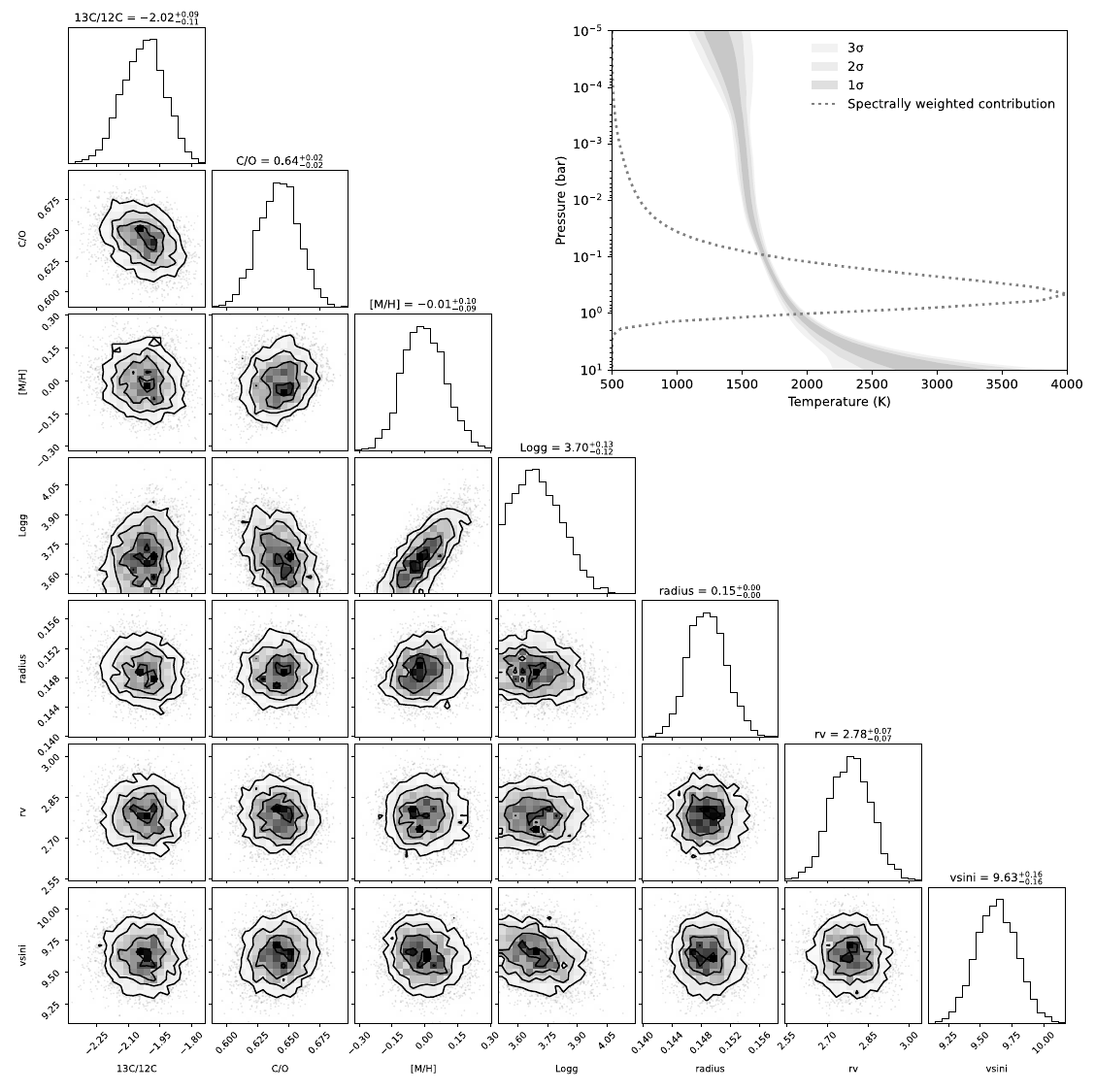}
    \caption{Retrieval result and PT profile of 2MASS J0249-0557c with $\log(g)$ prior and no clouds.}
    \label{fig:2mjlogg}
\end{figure*}

\begin{figure*}
    \centering
    \includegraphics[width=0.98\linewidth]{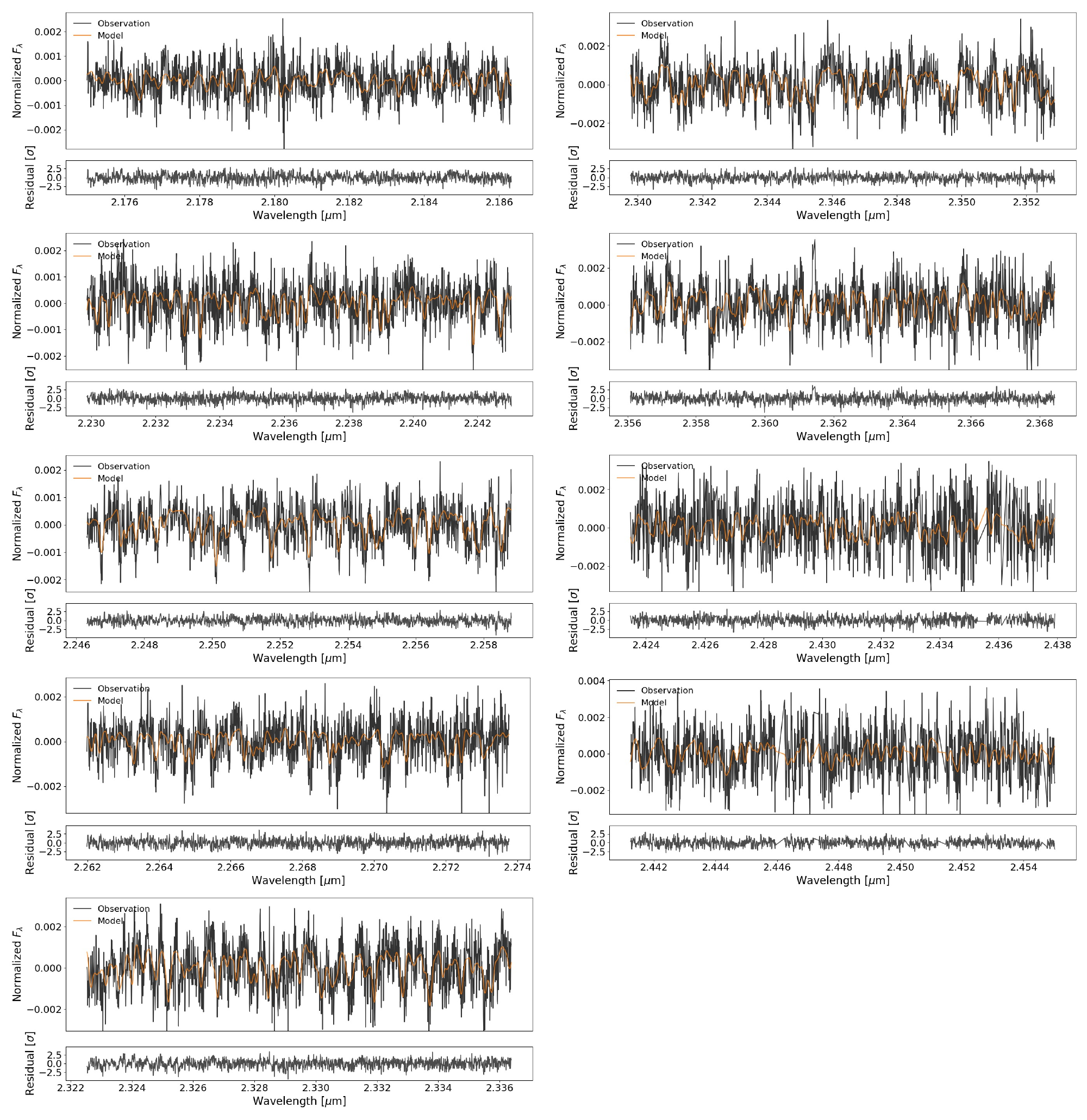}
    \caption{Spectra and best-fit model of 2MASS J0249-0557c on September 21, 2022.}
    \label{fig:2mjday1}
\end{figure*}

\begin{figure*}
    \centering
    \includegraphics[width=0.98\linewidth]{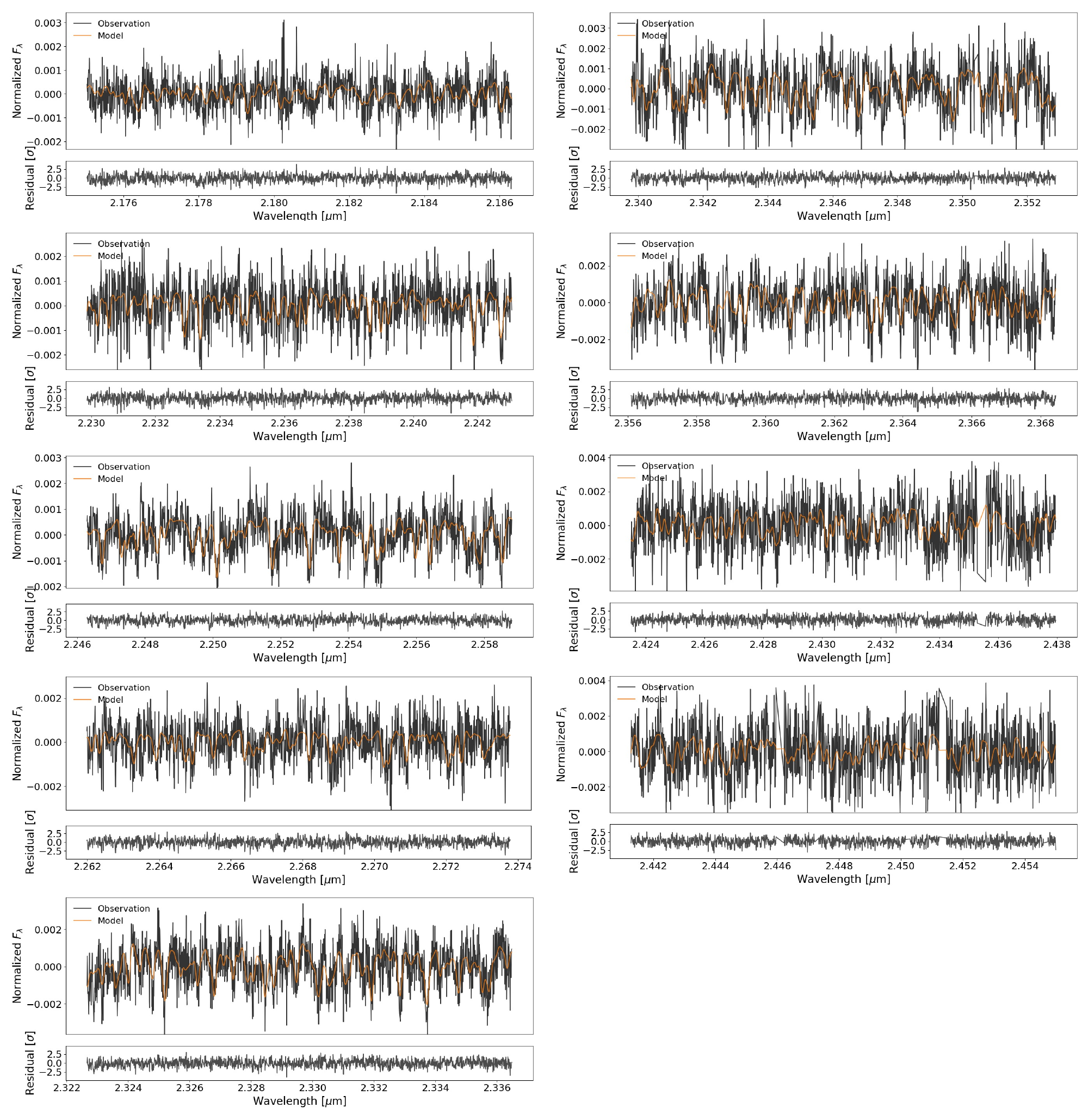}
    \caption{Spectra and best-fit model of 2MASS J0249-0557c on September 22, 2022.}
    \label{fig:2mjday2}
\end{figure*}

\section{2MASS J0443+0002 Plots}
This section contains the retrieval result and the best-fit model of 2MASS J0443+0002.
% \begin{figure*}
%     \centering
%     \includegraphics[width=0.98\linewidth]{Figures/allmodels_mrtefflogg_2MASSIJ0443376+000205.pdf}
%     \caption{Predictions from evolutionary models for 2MASS J0443+0002}
%     \label{fig:2mijevolutionary}
% \end{figure*}

\begin{figure*}
    \centering
    \includegraphics[width=0.98\linewidth]{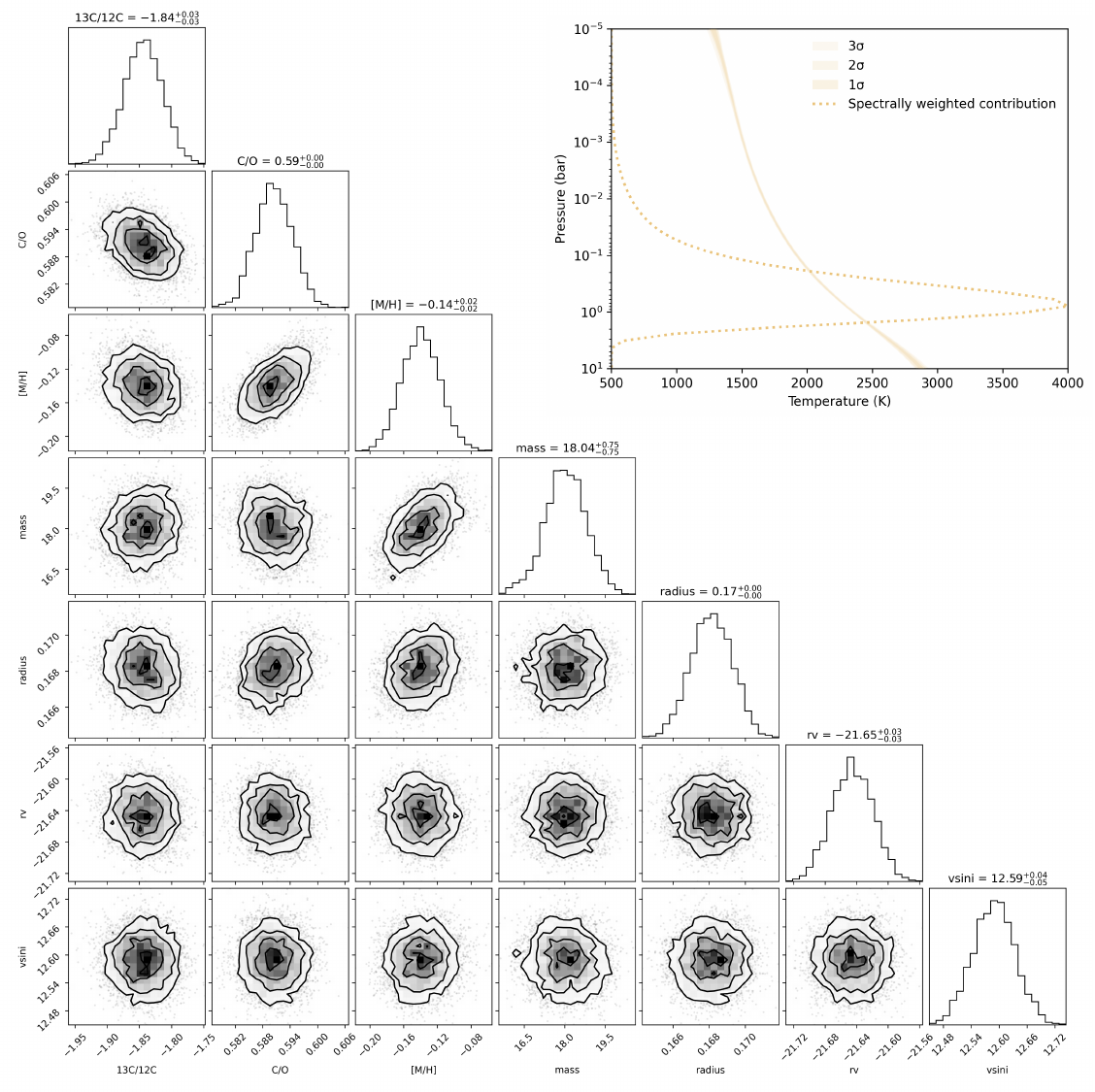}
    \caption{Retrieval result and PT profile of 2MASSI J0443+0002 with mass prior. This constitutes the final result that we report and analyze.}
    \label{fig:2MIJfinal}
\end{figure*}

% \begin{figure*}
%     \centering
%     \includegraphics[width=0.98\linewidth]{Figures/2MIJ/logg_2MIJ0443376_corner.pdf}
%     \caption{Retrieval result and PT profile of 2MASSI J0443+0002 with $\log(g)$ prior}
%     \label{fig:2MIJlogg}
% \end{figure*}

\begin{figure*}
    \centering
    \includegraphics[width=0.98\linewidth]{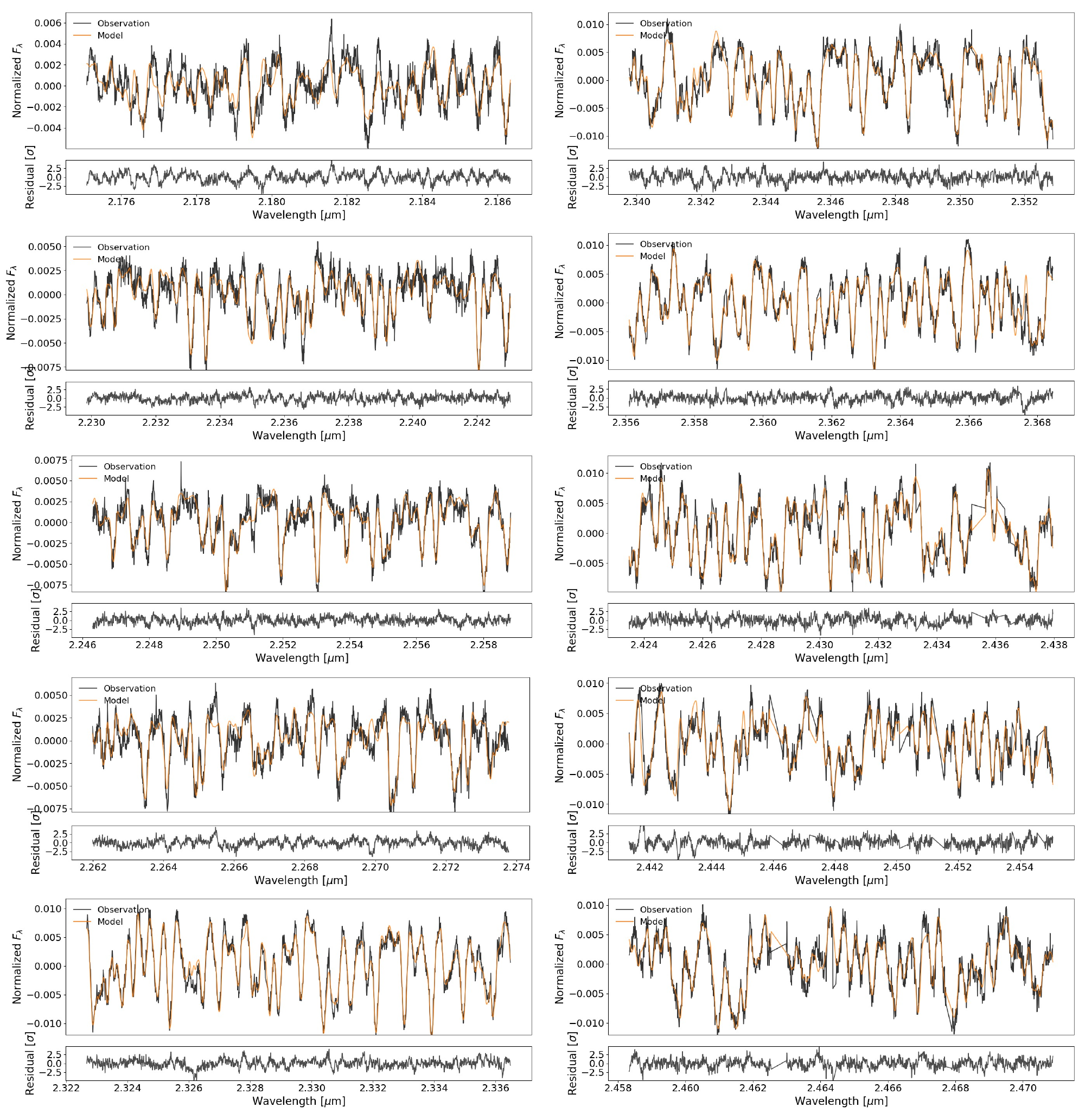}
    \caption{Spectra and best-fit model of 2MASSI J0443+0002.}
    \label{fig:2MIJspectra}
\end{figure*}

\begin{figure}
    \centering
    \includegraphics[width=0.98\linewidth]{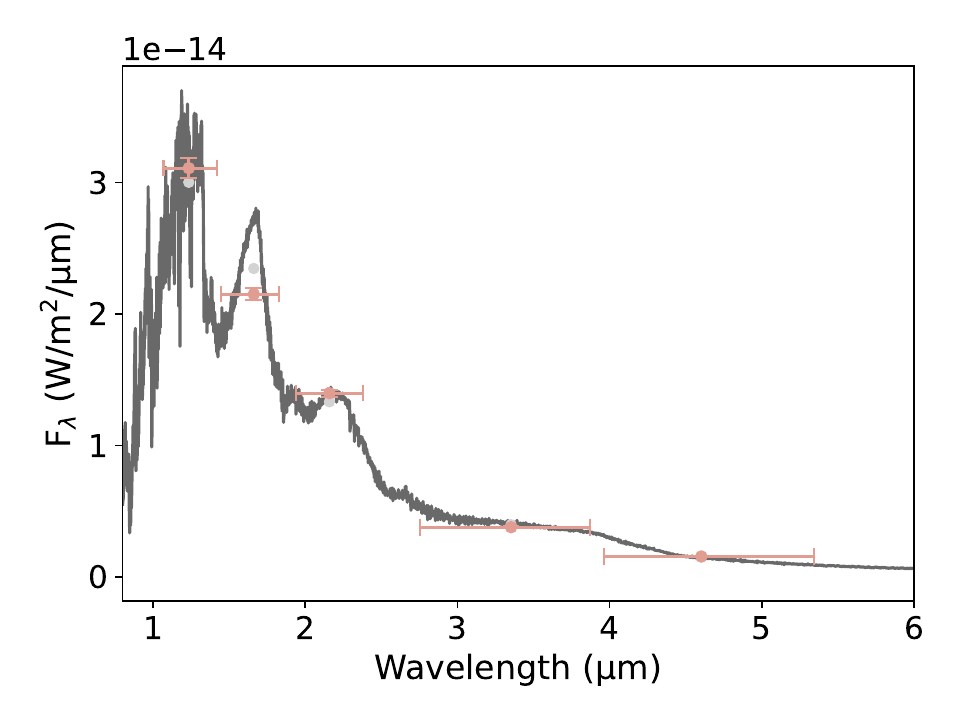}
    \caption{Best-fit low-resolution model spectrum of 2MASSI J0443+0002. The pink points with error bars show the archival photometry, while the light-gray points show the corresponding synthetic photometry of the best-fit model.}
    \label{fig:lowres}
\end{figure}

\section{SIPS J2000-7523 Plots}
This section contains the retrieval results and the best-fit model of SIPS J2000-7523.
% \begin{figure*}
%     \centering
%     \includegraphics[width=0.98\linewidth]{Figures/allmodels_mrtefflogg_SIPSJ2000-7523.pdf}
%     \caption{Predictions from evolutionary models for SIPS J2000-7523}
%     \label{fig:SIPSevolutionary}
% \end{figure*}

\begin{figure*}
    \centering
    \includegraphics[width=0.98\linewidth]{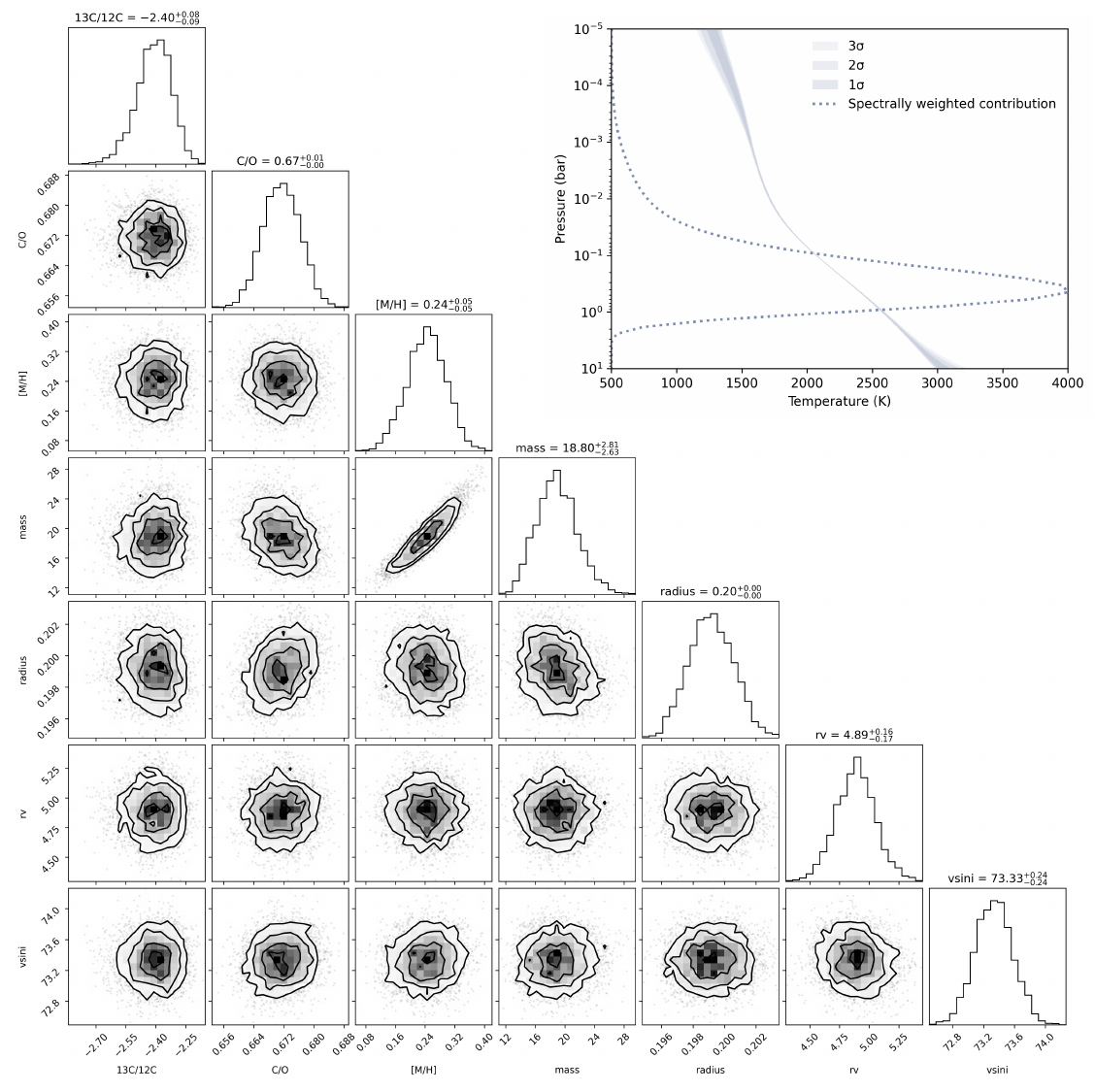}
    \caption{Retrieval result and PT profile of SIPS J2000-7523 with mass prior. This constitutes the final result that we report and analyze.}
    \label{fig:SIPSfinal}
\end{figure*}

% \begin{figure*}
%     \centering
%     \includegraphics[width=0.98\linewidth]{Figures/SIPS/logg_SIPSJ2000_corner.pdf}
%     \caption{Retrieval result and PT profile of SIPS J2000-7523 with $\log(g)$ prior}
%     \label{fig:SIPSlogg}
% \end{figure*}

\begin{figure*}
    \centering
    \includegraphics[width=0.98\linewidth]{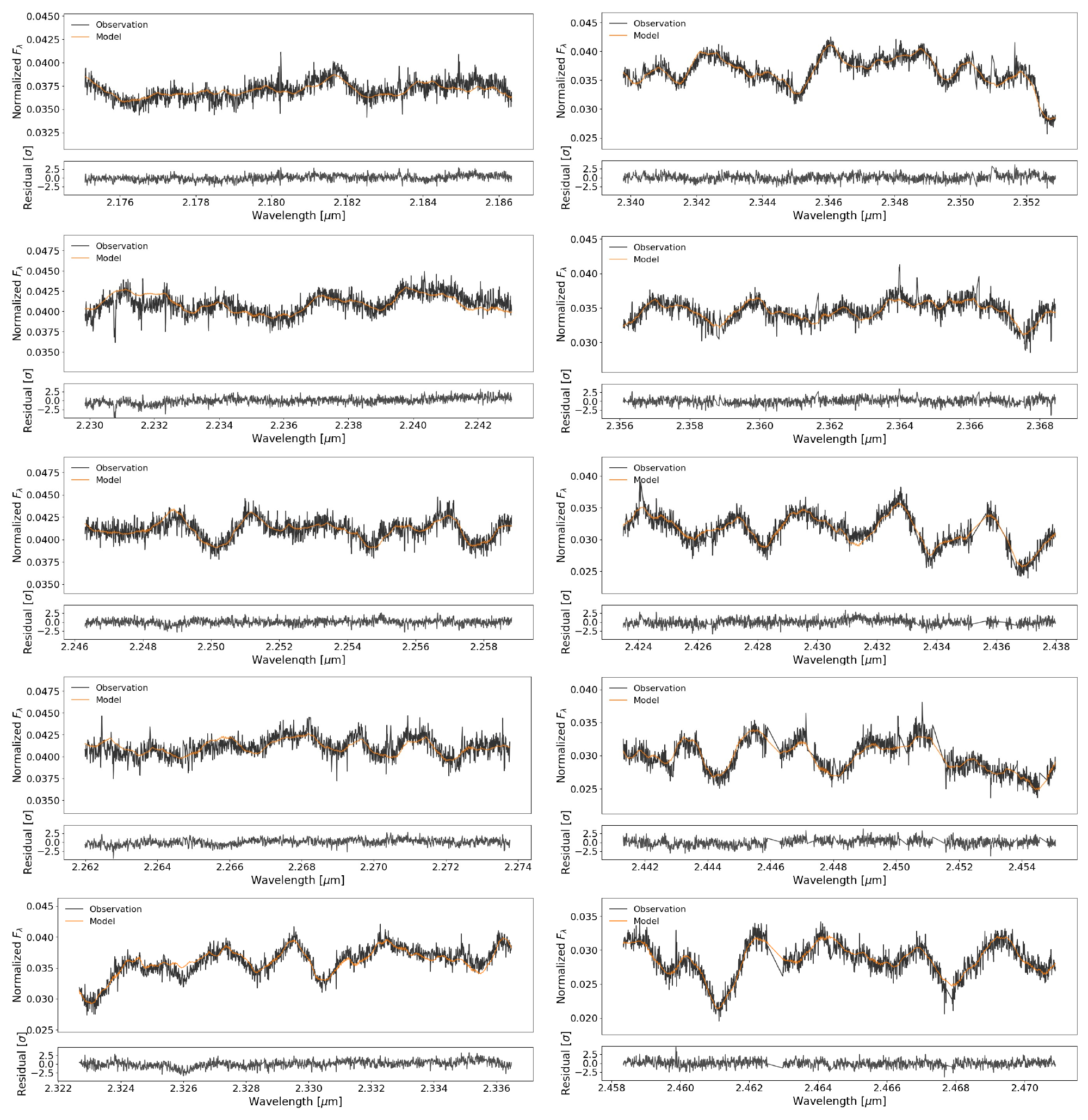}
    \caption{Spectra and best-fit model of SIPS J2000-7523.}
    \label{fig:SIPSspectra}
\end{figure*}

\begin{figure}
    \centering
    \includegraphics[width=0.98\linewidth]{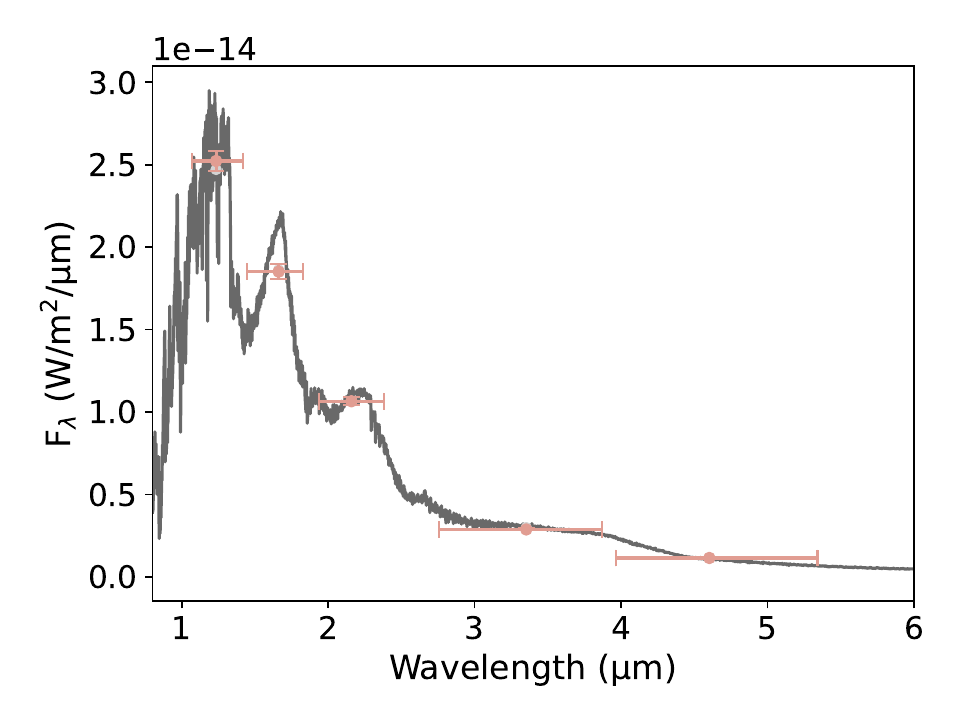}
    \caption{Best-fit low-resolution model spectrum of SIPS J2000-7523. The pink points with error bars show the archival photometry, while the light-gray points show the corresponding synthetic photometry of the best-fit model.}
    \label{fig:lowres}
\end{figure}

\section{Evolutionary Models}
This section contains the predictions of mass, radius, log(g), and effective temperature of the three targets from evolutionary models.
\begin{figure*}
    \centering
    \includegraphics[width=0.98\linewidth]{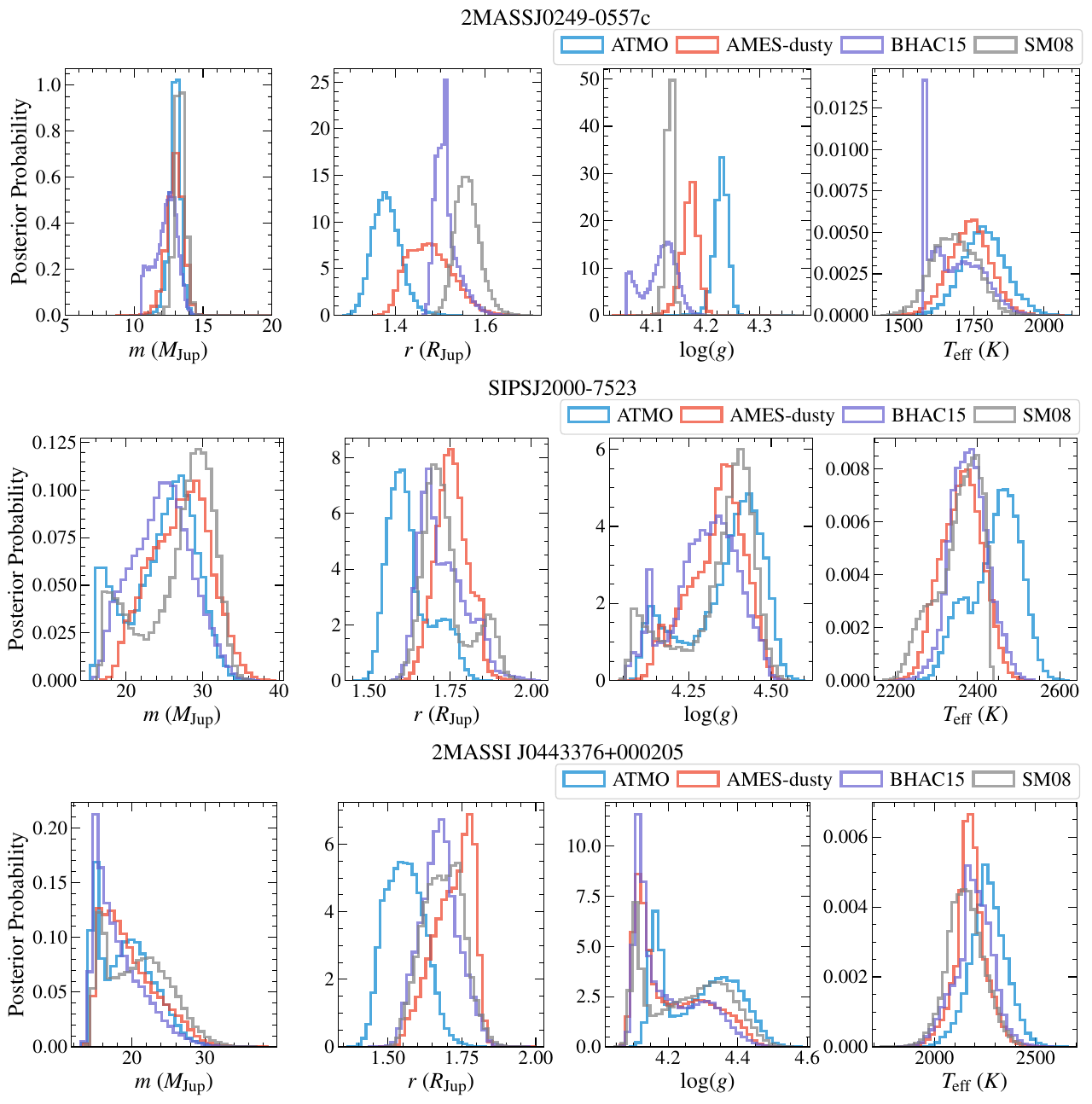}
    \caption{Predictions from evolutionary models for the three targets.}
    \label{fig:evolutionarymodels}
\end{figure*}

\end{document}